\newif{\ifarxiv}
\newif{\ifdraft}
\newif{\ifremarks}

\arxivtrue  
\remarkstrue  

\ifremarks
\newcommand{\remarktb}[1]{{\renewcommand{\bfdefault}{b}{\color[RGB]{0,150,0}{\textbf{#1}}}}}
\newcommand{\remarkfc}[1]{{\renewcommand{\bfdefault}{b}{\color[RGB]{0,0,150}{\textbf{#1}}}}}
\newcommand{\remarkpv}[1]{{\renewcommand{\bfdefault}{b}{\color[RGB]{150,0,0}{\textbf{#1}}}}}
\fi
\providecommand{\remarktb}[1]{\ignorespaces}
\providecommand{\remarkfc}[1]{\ignorespaces}
\providecommand{\remarkpv}[1]{\ignorespaces}

\documentclass[12pt,a4paper]{article}
\pdfoutput=1

\usepackage{amsmath}
\usepackage{amssymb}
\usepackage{longtable}
\usepackage{caption}
\usepackage[nosort]{cite}
\ifdraft\usepackage{showkeys}\fi 
\usepackage[bookmarks=true,colorlinks=true]{hyperref}
\usepackage{graphicx}
\usepackage[usenames,dvipsnames,svgnames,table]{xcolor}
\usepackage{float}
\usepackage[normalem]{ulem}
\usepackage{booktabs} 
\usepackage{listings}
\usepackage{siunitx}
\usepackage[mode=buildnew]{standalone} 
\usepackage{graphbox}
\usepackage{multirow}
\usepackage{xspace}
\usepackage[mathbold,autobold,greekcaps,greeklower]{mathfixs}
\providecommand{\mathbold}{\mathbf}

\usepackage{lmodern}
\usepackage[T1]{fontenc}
\usepackage{microtype}

\newcommand{\BMNworks}{Kristjansen:2002bb,Constable:2002hw,Beisert:2002bb,Constable:2002vq,Beisert:2002ff}

\usepackage[left=2.5cm,right=2.5cm,top=2.8cm,bottom=2.8cm]{geometry}

\ifdraft\else\fi

\allowdisplaybreaks
\usepackage{enumitem}

\definecolor{mathgreen}{RGB}{0,90,39}
\definecolor{mypurple}{rgb}{0.6,0,0.8}

\newcommand{\beq}{\begin{equation}}
\newcommand{\eeq}{\end{equation}}
\newcommand{\bba}{\begin{align}}
\newcommand{\eea}{\end{align}}
\newcommand{\bbn}{\begin{align*}}
\newcommand{\een}{\end{align*}}
\newcommand{\nn}{\nonumber}

\newenvironment{myeqnarray}{\arraycolsep0pt\begin{eqnarray}}{\end{eqnarray}\ignorespacesafterend}
\newenvironment{myeqnarray*}{\arraycolsep0pt\begin{eqnarray*}}{\end{eqnarray*}\ignorespacesafterend}

\def\[{\begin{equation}}
\def\]{\end{equation}}
\def\<{\begin{myeqnarray}}
\def\>{\end{myeqnarray}}

\numberwithin{equation}{section}

\def\etal.{et\penalty50\ al.}
\newcommand*{\eg}{e.\,g.\@\xspace}
\newcommand*{\ie}{i.\,e.\@\xspace}

\usepackage[font=small,labelfont=bf,width=0.88\textwidth]{caption}

\providecommand{\hypersetup}[1]{}
\providecommand{\hyperref}[2][]{#2}
\providecommand{\texorpdfstring}[2]{#1}
\providecommand{\pdfbookmark}[3][]{}

\hypersetup{plainpages=false}
\hypersetup{pdfpagemode=UseOutlines}
\hypersetup{bookmarksnumbered=true}
\hypersetup{bookmarksopen=true}
\hypersetup{pdfstartview=FitH}
\hypersetup{citecolor=[rgb]{0 .4 0}}
\hypersetup{urlcolor=[rgb]{.4 0 0}}
\hypersetup{linkcolor=[rgb]{0 0 .4}}
\hypersetup{citebordercolor={.6 .9 .6}}
\hypersetup{urlbordercolor={.7 .8 1}}
\hypersetup{linkbordercolor={1 .7 .7}}
\hypersetup{pdfborder={0 0 .5}}

\newcommand{\namedref}[2]{\hyperref[#2]{#1~\ref*{#2}}}
\newcommand{\secref}[1]{\namedref{Section}{#1}}
\newcommand{\appref}[1]{\namedref{Appendix}{#1}}
\newcommand{\tabref}[1]{\namedref{Table}{#1}}
\newcommand{\figref}[1]{\namedref{Figure}{#1}}

\makeatletter
\def\mr@ignsp#1 {\ifx\:#1\@empty\else #1\expandafter\mr@ignsp\fi}%
\newcommand{\multiref}[1]{\begingroup
\xdef\mr@no@sparg{\expandafter\mr@ignsp#1 \: }%
\def\mr@comma{}%
\@for\mr@refs:=\mr@no@sparg\do{\mr@comma\def\mr@comma{,}\ref{\mr@refs}}%
\endgroup}
\makeatother
\renewcommand{\eqref}[1]{(\multiref{#1})}

\makeatletter
\let\@myabstract\@empty
\let\@keywords\@empty
\let\@subject\@empty
\providecommand{\affiliation}[1]{\gdef\@affiliation{#1}}
\providecommand{\myabstract}[1]{\gdef\@myabstract{#1}}
\providecommand{\keywords}[1]{\gdef\@keywords{#1}}
\providecommand{\subject}[1]{\gdef\@subject{#1}}
\def\thetitle{\@title}
\def\theauthor{\@author}
\def\theaffiliation{\@affiliation}
\def\theabstract{\@myabstract}
\def\thesubject{\@subject}
\def\thedate{\@date}
\def\thekeywords{\@keywords}
\makeatother
\AtBeginDocument{
\hypersetup{pdftitle={\thetitle}}%
\hypersetup{pdfauthor={\theauthor}}%
\hypersetup{pdfsubject={\thesubject}}%
\hypersetup{pdfkeywords={\thekeywords}}%
}

\newcommand{\sfrac}[2]{{\textstyle\frac{#1}{#2}}}

\newcommand{\order}[1]{\mathcal{O}(#1)}

\newcommand{\superN}{\mathcal{N}}

\newcommand{\Nc}{N\subrm{c}}
\newcommand{\Csphere}{{}^\bullet\kern-1.2pt C}
\newcommand{\Ctorus}{{}^\circ\kern-1.2pt C}

\newcommand{\oct}{\mathbb{O}}

\newcommand{\tr}{\operatorname{tr}}

\newcommand{\aut}{\operatorname{Aut}}
\newcommand{\op}[1]{\mathcal{#1}}

\newcommand{\subrm}[1]{_{\text{#1}}}

\newcommand{\grp}[1]{\mathrm{#1}}

\newcommand{\mathematica}{\textsc{Mathematica}\@\xspace}

\newcommand{\filename}[1]{\texttt{#1}}

\newcommand{\brk}[1]{(#1)}
\newcommand{\lrbrk}[1]{\left(#1\right)}
\newcommand{\bigbrk}[1]{\bigl(#1\bigr)}
\newcommand{\Bigbrk}[1]{\Bigl(#1\Bigr)}

\newcommand{\biggbrk}[1]{\biggl(#1\biggr)}

\newcommand{\sbrk}[1]{[#1]}
\newcommand{\lrsbrk}[1]{\left[#1\right]}

\newcommand{\Bigsbrk}[1]{\Bigl[#1\Bigr]}
\newcommand{\biggsbrk}[1]{\biggl[#1\biggr]}

\newcommand{\vev}[1]{\langle#1\rangle}
\newcommand{\lrvev}[1]{\left\langle#1\right\rangle}

\newcommand{\Bigvev}[1]{\Bigl\langle#1\Bigr\rangle}
\newcommand{\biggvev}[1]{\biggl\langle#1\biggr\rangle}

\newcommand{\abs}[1]{|#1|}

\title{Octagons I: Combinatorics\texorpdfstring{\\}{ }and Non-Planar Resummations}

\author{%
Till Bargheer\texorpdfstring{$^{a,b}$}{},
Frank Coronado\texorpdfstring{$^{c,d}$}{},
Pedro Vieira\texorpdfstring{$^{c,d}$}{}}

\myabstract{We explain how the 't~Hooft expansion of
correlators of half-BPS operators can be resummed in a large-charge
limit in $\superN=4$ super Yang--Mills theory. The full
correlator in the limit is given
by a non-trivial function of two variables: One variable is the charge
of the BPS operators divided by the square root of the number~$\Nc$ of
colors; the other variable is the \emph{octagon} that contains all the
't~Hooft coupling and spacetime dependence. At each genus $g$
in the large~$\Nc$ expansion, this function is a polynomial of degree
$2g+2$ in the octagon. We find several dual matrix model
representations of the correlators in the large-charge limit.
Amusingly, the number of colors in these matrix models is
formally taken to zero in the relevant limit.}

\subject{Mathematical Physics}


\begin{document}

\pdfbookmark[1]{Title Page}{title}

\thispagestyle{empty}
\setcounter{page}{0}

\mbox{}
\vfill

\begin{center}

{\Large\textbf{\mathversion{bold}\thetitle}\par}

\vspace{1cm}

\textsc{\theauthor}

\bigskip

\begingroup
\footnotesize\itshape

$^{a}$Institut f\"ur Theoretische Physik, Leibniz Universit\"at Hannover,\\
Appelstra{\ss}e 2, 30167 Hannover, Germany

\medskip

$^{b}$DESY Theory Group, DESY Hamburg,\\
Notkestra\ss e 85, D-22603 Hamburg, Germany\\

\medskip

$^{c}$Perimeter Institute for Theoretical Physics,\\
31 Caroline St N Waterloo, Ontario N2L 2Y5, Canada

\medskip

$^{d}$Instituto de F\'isica Te\'orica, UNESP - Univ. Estadual Paulista,\\
ICTP South American Institute for Fundamental Research,\\
Rua Dr. Bento Teobaldo Ferraz 271, 01140-070, S\~ao Paulo, SP, Brasil

\endgroup

\bigskip

\newcommand{\email}[1]{\href{mailto:#1}{#1}}

{\small\ttfamily
\email{till.bargheer@desy.de},
\email{fcoronado@perimeterinstitute.ca},
\email{pedrogvieira@gmail.com}}
\par

\vspace{1cm}

\textbf{Abstract}\vspace{5mm}

\begin{minipage}{12cm}
\theabstract
\end{minipage}

\end{center}

\vfill
\vfill

\newpage

\providecommand{\microtypesetup}[1]{}
\microtypesetup{protrusion=false}
\setcounter{tocdepth}{2}
\pdfbookmark[1]{\contentsname}{contents}
\tableofcontents
\microtypesetup{protrusion=true}



\section{Introduction}

\begin{figure}[t]
\center \includegraphics[scale=.55]{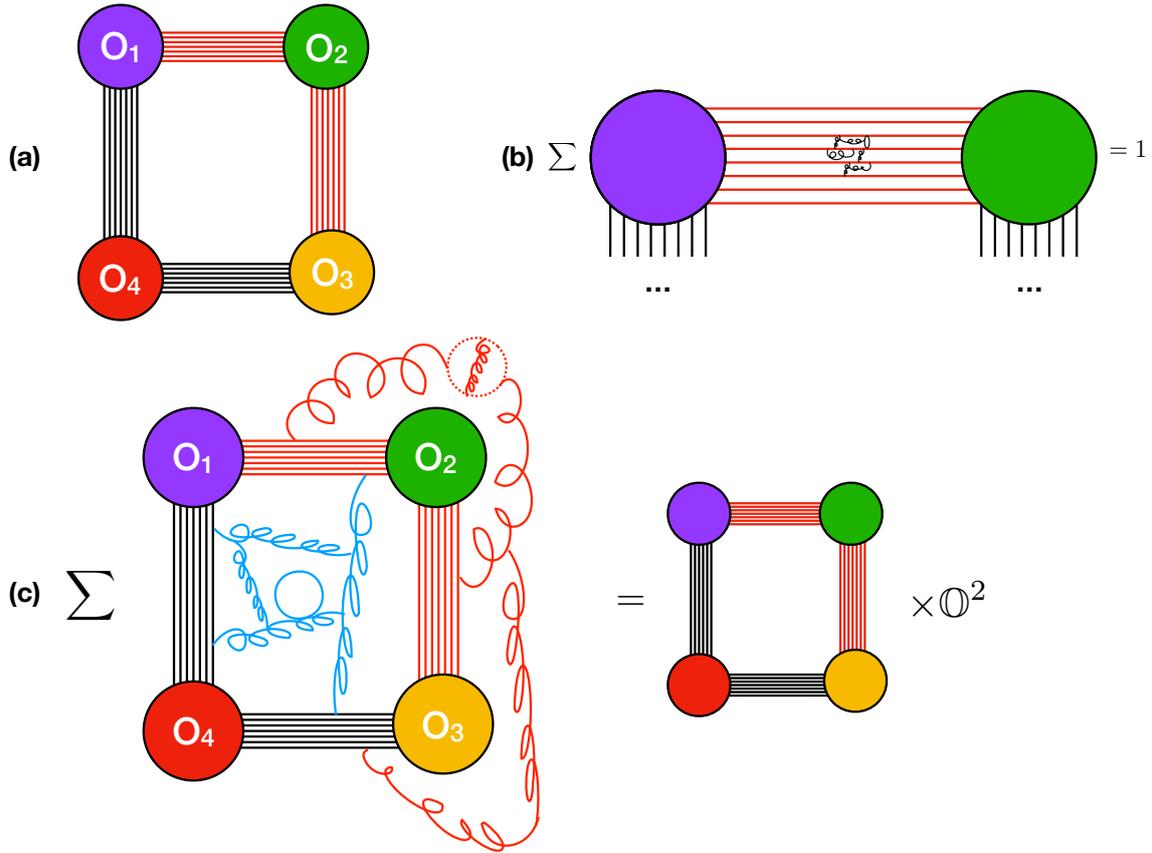}
\caption{Large cyclic operators at genus zero.
(a) The tree-level result is given by a single Feynman graph of
rectangular form. For large operators, this rectangle creates a big
frame.
(b) The frame, when big, does not receive loop corrections because of
supersymmetry. Indeed these loop corrections are indistinguishable
from those arising in the two-point function of BPS operators, a
protected quantity. (c) The inside and the outside of the frame, on
the other hand, are loop corrected. The sum of all quantum corrections
to the inside (or outside) define a function $\oct$ of the 't~Hooft
coupling and of the four-point cross ratio.}
\label{fig:intro}
\end{figure}

In this work, we will consider correlation functions of single-trace half-BPS
operators in~$\mathcal{N}=4$ super Yang--Mills theory. Each of
these operators creates a closed string state, so these correlation
functions describe closed-string scattering in $\grp{AdS}_5\times\grp{S}^5$.

We will focus on four-point correlation functions in an
interesting limit of very large BPS operators with carefully chosen
polarizations, where the closed string scattering process factorizes
into several copies of an off-shell open string partition function
$\oct$ that was determined exactly
in~\cite{Coronado:2018ypq,Coronado:2018cxj}
at any value of the 't~Hooft coupling and further simplified into an
infinite determinant representation in~\cite{Kostov:2019stn}.

The dimensions of the operators we will consider scale with
the rank of the $U(\Nc)$ gauge group as $\sqrt{\Nc}$, reminiscent of
inspiring earlier studies~\cite{\BMNworks} in the plane-wave
Berenstein--Maldacena--Nastase (BMN) limit~\cite{Berenstein:2002jq}. The
motivation for this particular limit is similar to the one
considered in those works: It will allow us to re-sum the large
$\Nc$ 't~Hooft expansion. We now have a much stronger control over the
't~Hooft coupling behavior due to integrability and bootstrap
techniques that were not yet available at the time, so it seems rather
timely to revive those explorations in light of these newer
technologies.

A key difference compared to the earlier BMN-related works~\cite{\BMNworks}
is that in those studies there was typically a single
R-charge that was taken to be large, while for the present work it is crucial that
the operators correspond to closed strings rotating in different $\grp{S}^5$
equators.
To be precise, we will take two operators to be two \textit{different}
BMN highest-weight states
\begin{equation}
\op{O}_{2}= \tr\brk{{\color{red}X}^{2k}\,}(z)
\,,\quad
\op{O}_{4}= \tr\brk{Z^{2k}}(\infty)
\,,
\label{eq:24}
\end{equation}
and two other operators to be two \textit{equal}  BMN descendants
\begin{equation}
\op{O}_{1} = \tr\brk{\bar{Z}^{k}\,{\color{red}\bar{X}}^{k}}(0)+\texttt{permutations}
\,,\quad
\op{O}_{3} = \tr\brk{\bar{Z}^{k}\,{\color{red}\bar{X}}^{k}}(1)+\texttt{permutations}
\,,
\label{eq:13}
\end{equation}
where $X$ and $Z$ are two complex scalars in $\mathcal{N}=4$.
This choice of two highest-weight states and two BMN descendents might seem
asymmetric and unorthodox but is actually quite important,
both technically and physically.

The technical simplification can already be seen at tree level in the
planar limit: Because of R-charge conservation, there is only a single
Feynman diagram computing the four-point correlation function! The
correlator is simply given by a product of $4k$ propagators, with $k$ parallel propagators
connecting each pair of consecutive operators $\op{O}_i$ and
$\op{O}_{i+1}$, thus drawing a square frame as depicted in~\figref{fig:intro}(a).

\begin{figure}[t]
\centering
\includegraphics[scale=.55,trim=1.3cm 1.5cm 0.4cm 1cm,clip]{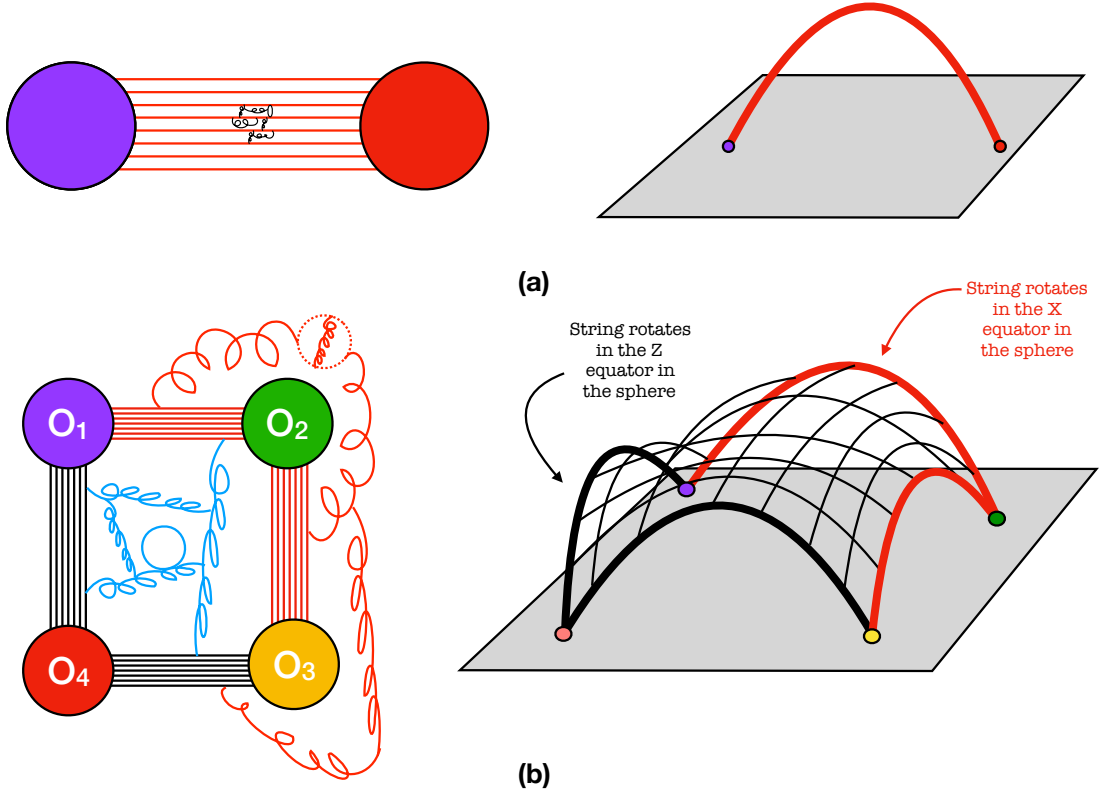}
\caption{Each bundle of propagators in the field theory can be thought
of as a very heavy BMN string geodesic which is protected by supersymmetry. The
octagon is the sum of all Feynman diagrams inside (or outside) the
tree level frame or, equivalently, the top (or bottom) of the folded
string stretching between four consecutive geodesics. At genus zero,
the folded string has a top and bottom and hence the result is given
by~$\oct^2$ in agreement with the field theory result.}
\label{fig:StringFigure}
\end{figure}

Beyond tree level -- but still
at genus zero -- we decorate this correlator by all possible
Feynman loops. The diagrams inside individual propagator bundles
connecting two operators -- as depicted in~\figref{fig:intro}(b) --
cancel out by supersymmetry, so they do not correct the correlator.
After all, those diagrams do not \emph{know} they belong to a
four-point function rather than a protected two-point function of BPS operators.
The diagrams inside the square -- represented in~\figref{fig:intro}(c) --
do probe all four operators and hence lead to a non-trivial
function $\oct$ that depends on the 't~Hooft
coupling $\lambda$ and on the conformal cross ratios formed by the four operators.
This function $\oct$ was studied in detail
in~\cite{Coronado:2018ypq,Coronado:2018cxj}. The diagrams outside the square
contribute by the same amount as the diagrams inside, hence the full
genus-zero result is simply given by
(throughout this work, $g$ denotes the genus)
\begin{equation}
\vev{ \op{O}_1\dots \op{O}_4}_{g=0}
= \vev{ \op{O}_1\dots \op{O}_4}_{\lambda=0,g=0} \times  \oct^2
\,.
\end{equation}
Note that if it were not for large $k$, the decoupling between outside
and inside would be absent. Indeed, for any finite $k$ and large
enough
loop order, diagrams can communicate all the way from the inside to the outside.%
\footnote{In the language of hexagonalization~\cite{Fleury:2016ykk,Eden:2016xvg,Eden:2017ozn,Fleury:2017eph,Bargheer:2017nne,Bargheer:2018jvq},
the two faces decouple because mirror-particle propagation across
large propagator bundles is suppressed. }

\begin{figure}
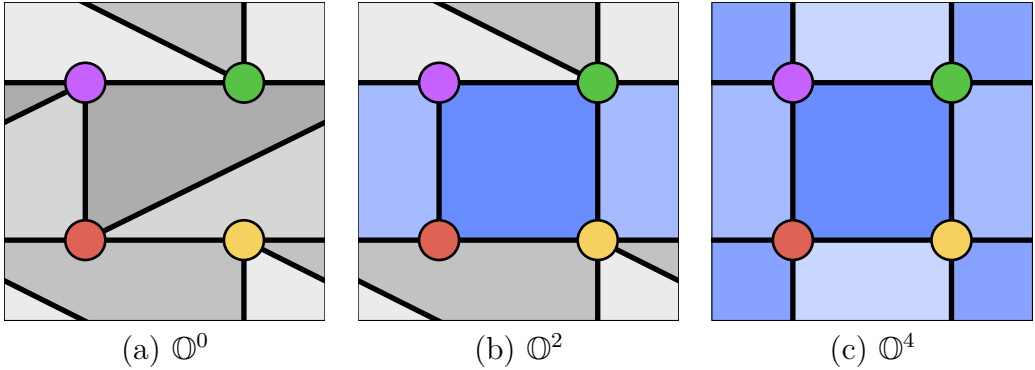

\centering
\begin{tabular}{ccc}
\includegraphics{FigGraphOct0Torus} &
\includegraphics{FigGraphOct2Torus} &
\includegraphics{FigGraphOct4Torus}
\\
(a) $\oct^0$ &
(b) $\oct^2$ &
(c) $\oct^4$
\end{tabular}
\caption{Four operators (circles) are inserted on a torus (the top and
bottom sides as well as the left and right sides of the square are
identified). Wick contractions organize into \emph{skeleton graphs},
where each edge (black line) is a bundle of parallel propagators. For
our choice of operators, all faces of all skeleton graphs are
octagons. Shown in shades of gray are octagons that touch only two or
three of the operators, these are protected by supersymmetry. Octagons
that touch all four operators (blue shades) are non-trivial functions
of the 't~Hooft coupling and conformal cross ratios.}
\label{fig:threegraphs}
\end{figure}

In dual string theory terms, each bundle of propagators connecting
consecutive BPS operators can be thought of as a heavy geodesic
connecting points $x_i$
and $x_{i+1}$ on the AdS boundary, as represented
in~\figref{fig:StringFigure}(a). Because there are so many propagators
$k$ in each bundle, these
geodesics are very heavy and will not move away from their classical
configuration. The four classical geodesics will be connected by a folded string,
as depicted in~\figref{fig:StringFigure}(b). The
fold lines are given by the heavy geodesics, which effectively
decouple the top and bottom of the folded string. The two sides of the
folded string are the string counterpart
to the gauge-theory Feynman diagrams inside and outside of the
square.
In contrast to the heavy geodesics, they do vibrate quantum
mechanically, each of them thus defining a
full-fledged open string partition function%
\footnote{The boundary conditions for this open string
partition function say that the string should end on
the BMN classical geodesics in the bulk. This is somewhat unusual -- typically
the boundary conditions are such that the worldsheet ends at the
boundary of $\grp{AdS}$. To properly define the boundary conditions for
this open string partition function, we also need to specify how the
four classical geodesics rotate in the sphere. There are $k$ units of R-charge of
type $X$ ($Z$) connecting $\op{O}_2$ ($\op{O}_4$) with its
cyclic neighbours, so the geodesics emanating from operator
$\op{O}_2$ ($\op{O}_4$) rotate in the $X\bar X$ ($Z\bar Z$)
equator of $S^5$ with $k$ units of angular momentum, see~\figref{fig:intro}(a).
The full open string will thus interpolate between these two different
BMN geodesic behaviors. At large 't~Hooft
coupling, the open string surfaces become classical, and the open
partition function should be given by the area of a minimal surface ending on
the four BMN geodesics. Reference~\cite{Dorn:2009hs} is an inspiring
related paper where a slightly different class of folded strings
were considered,
corresponding to null squares with further movement in the sphere.}
-- this open
string partition function is the string definition of the function
$\oct$.

\begin{figure}
\centering
\includegraphics[width=\textwidth,trim=0.2cm 10.5cm 0.6cm 0.9cm,clip]{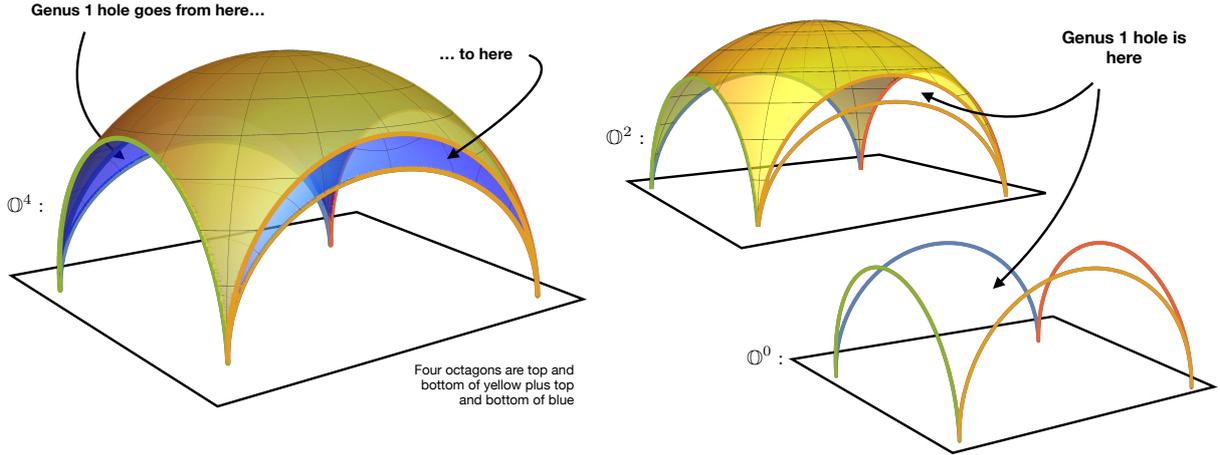}
\caption{AdS embeddings of the three graphs
in~\figref{fig:threegraphs}. Each edge in~\figref{fig:threegraphs}
represents a bundle of $\order{\sqrt{\Nc}}$ propagators, and therefore
becomes a heavy BMN geodesic connecting two operators. These geodesics
are folds of the worldsheet that connect adjacent octagons. The BPS
octagons have no extent in AdS, they curl up along the BMN geodesics.
The non-BPS octagons are extended objects that touch all four operators.}
\label{fig:threegraphsAdS}
\end{figure}

This concludes the genus-zero considerations. This paper's main focus
is on the higher-genus picture.
We will explain the general structure
of the correlator in detail in the next
section. The upshot is that \emph{(i)} the leading term in the
large-$k$ limit at fixed genus $g$ is proportional to $k^{4g}$, and
\emph{(ii)} we can can stack from zero to $2g+2$ folded strings on top of
each other to construct a genus-$g$ surface.%
\footnote{For example, if
we remove the folded string from~\figref{fig:StringFigure}(b), we are
left with the four geodesics with a hole in the middle -- a genus $1$
surface, see~\figref{fig:threegraphsAdS}.}
Since each fold joins two open strings, the number of open
string surfaces ought to be even, and thus the full correlation
function, \ie the full closed-string partition function -- in the
limit of large $k$ -- will be simply given by a polynomial $P_{g+1}$
of degree $g+1$ in
the square of the open string partition function $\oct$,%
\footnote{The
dots in this formula contain higher-genus terms, but also, for each genus, including
the terms presented here, smaller powers of $k$, subleading in the
large $k$ limit we are interested in.}
\begin{equation}
\vev{ \op{O}_1\dots \op{O}_4}
=\vev{ \op{O}_1\dots \op{O}_4}_{\lambda=0,g= 0}
\times \lrsbrk{
\oct^2 + \frac{k^4}{\Nc^2} P_2(\oct^2) + \frac{k^8}{\Nc^4} P_3(\oct^2) +\dots
}\,.
\label{eq:polynomials}
\end{equation}
Resumming the full large $\Nc$ expansion,
at any value of the 't~Hooft coupling, thus amounts to finding the
function of two variables
\begin{equation}
\mathcal{A}(\zeta,\oct) \equiv \lim_{\Nc\to \infty}
\left.
\frac{\langle \op{O}_1\dots \op{O}_4\rangle}{\langle \op{O}_1\dots \op{O}_4\rangle_{\lambda=0,g= 0}}
\right|_{k=\zeta\sqrt{\Nc}}
\,.
\label{eq:Afunction}
\end{equation}
Note that this correlation function $\mathcal{A}$ depends very
non-trivially on the conformal cross ratios and on the 't~Hooft
coupling of the theory through the octagon function~$\oct$ computed
in~\cite{Coronado:2018ypq,Coronado:2018cxj}. The main result of this
paper is a representation of the function $\mathcal{A}$ and of the associated
polynomials $P_{g+1}$ in~\eqref{eq:polynomials} in terms of a matrix
model, where the octagon function~$\oct$ enters as an effective quartic
coupling.


\section{A Matrix Model for Large Operators}
\label{sec:matrix-model-large}

The basis of our computation is the (planar and non-planar) hexagonalization
prescription for correlation
functions~\cite{Fleury:2016ykk,Eden:2016xvg,Eden:2017ozn,Fleury:2017eph,Bargheer:2017nne,Bargheer:2018jvq}.
The starting point of that prescription
is a sum over all Wick
contractions of the free gauge theory. We organize this sum by first
summing over ``skeleton graphs'' of the desired genus. Each edge in a
skeleton graph represents a bundle of one or more parallel propagators.%
\footnote{Because they represent Wick contractions of single-trace
operators, the incident edges at each vertex (operator) have a
well-defined cyclic ordering. Graphs with this property are
called \emph{ribbon graphs} (or fat graphs).
See~\appref{sec:generating-maximal-graphs} for more details.}
For each skeleton graph, we then sum over all possible ways of
distributing propagators on the edges of the graph (that are
compatible with the charges of the operators).

\begin{figure}[t]
\centering
\includegraphics[scale=.55,trim=0 8.0cm 1.3cm 0,clip]{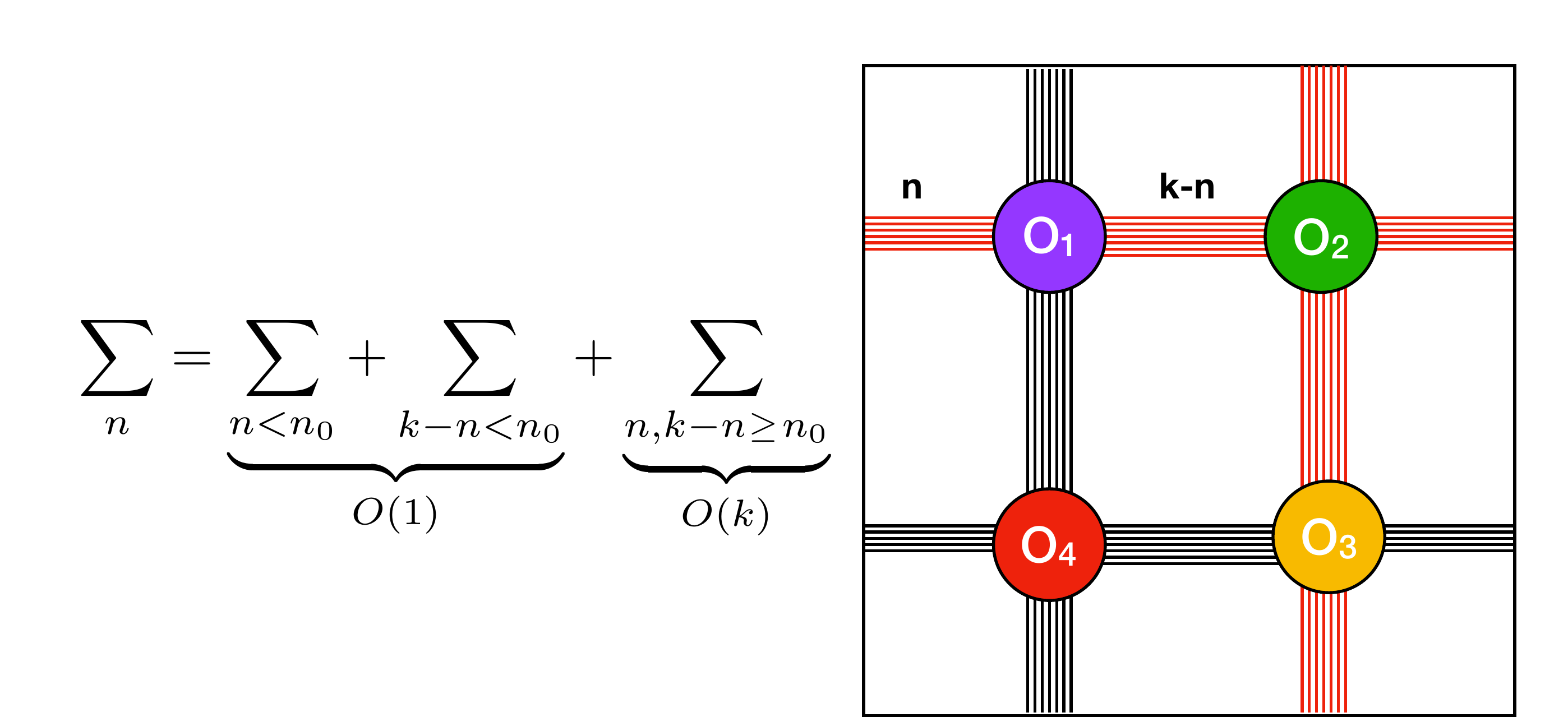}
\caption{Octopus principle. Configurations where some available
propagator bundles are not massively occupied are strongly suppressed
in the large charge limit. All skeleton graphs are thus maximal
graphs, where no further propagator bundle can be added
without increasing the genus at hand.}
\label{fig:octopus}
\end{figure}

\begin{figure}[t]
\center \includegraphics[scale=.55]{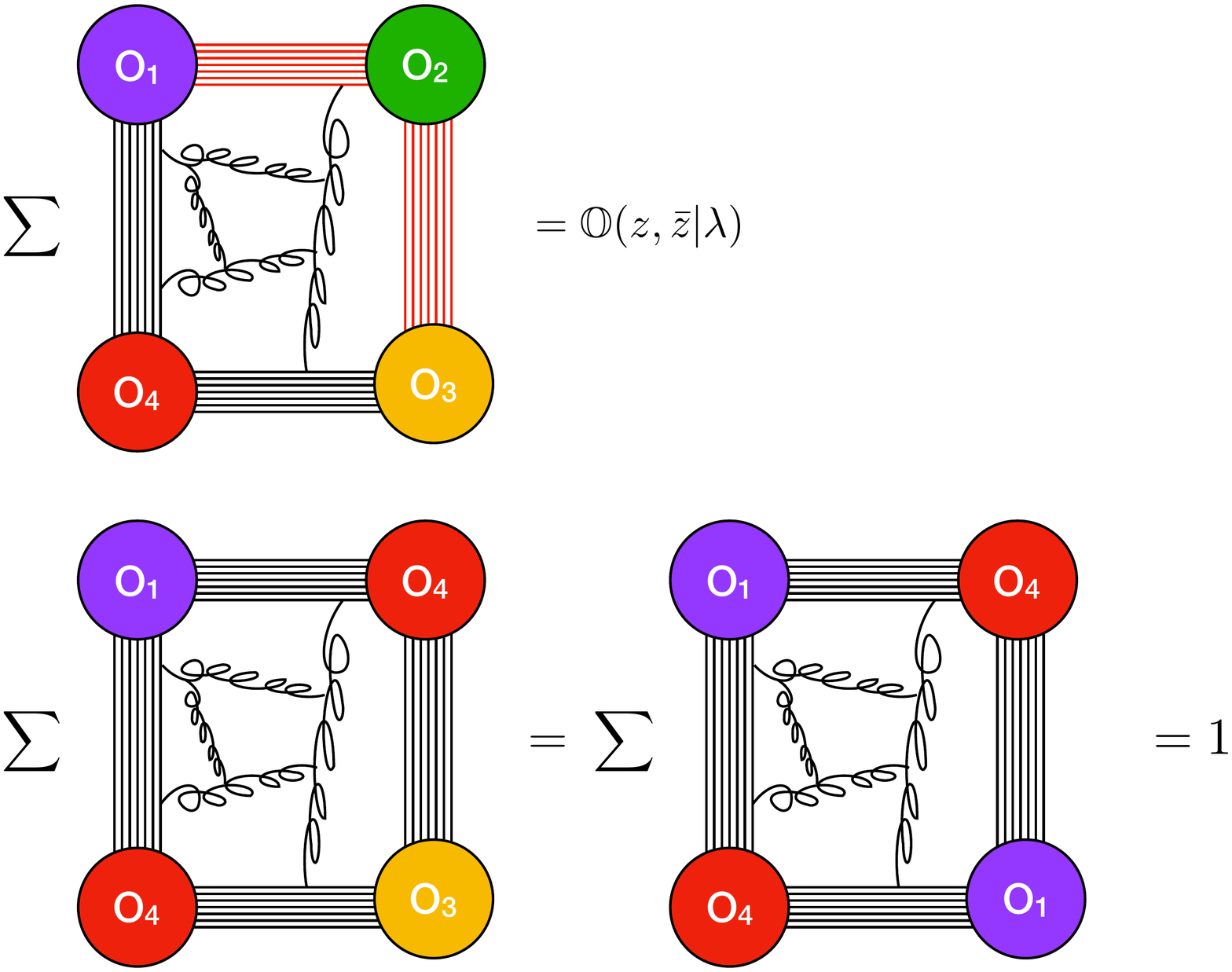}
\caption{The three types of squares all large-$k$
skeleton graphs are made of, see~\protect\eqref{eq:squaretypes}. Squares can connect all four
operators (top figure), and all such squares equal the non-trivial function
$\oct$ of the 't~Hooft coupling $\lambda$ and the cross ratios
parametrized by $z$ and $\bar z$. Squares may also only connect three
operators (bottom left) or two operators (bottom right), such squares
are protected by supersymmetry and hence do not receive loop corrections.}
\label{fig:squareTypes}
\end{figure}

We saw in the introduction that, for our choice of operators, there is
only a single skeleton graph
at genus zero. At higher genus, there are
several contributing diagrams.
For example, the top row of~\figref{fig:threegraphs} shows three
different genus-one graphs contributing to the four-point
correlation function of our operators~\eqref{eq:24},~\eqref{eq:13}.
The key observation illustrated by these examples is the following:
For large operators, at any genus order, all skeleton graphs that
contribute are \emph{quadrangulations}, \ie all faces of these graphs are quadrangles.
This is because of what we call \textit{the octopus
principle}. It comes about because we have to distribute a large
number of propagators on the edges of the skeleton graphs. For
example, consider the $k$ propagators connecting operators $\op{O}_1$
and $\op{O}_2$ in~\figref{fig:octopus}. In this case, operators $\op{O}_1$ and
$\op{O}_2$ are connected by two bridges, and we have to sum over all
ways of distributing $k$ propagators on these two bridges. For large
$k$, the overwhelming number of terms will have $\order{k}$ propagators
on both bridges,
and the sum of these terms will produce a factor $k$. The sum of all
terms where any of the bridges is populated by only a finite number of
propagators is finite, and thus suppressed in the large-$k$ limit.
Hence we immediately see that all
propagator bundles want to be heavily populated, evoking the picture
of an octopus who wants to spread its tentacles over all possible
cycles of the Riemann surface. More generally, if there are $n$
edges connecting two operators, we have to sum over the number $k_i$ of
propagators on each edge $i$, with the constraint that
$\sum_ik_i=k$. This sum expands to
\begin{equation}
\sum_{\substack{k_1,\dots,k_n\\k_1+\dots+k_n=k}}
=\frac{k^{n-1}}{(n-1)!} + \order{k^{n-2}}
\,,
\label{eq:combiFactors}
\end{equation}
and the leading term only receives contributions from configurations
where all $k_i=\order{k}$. This has two consequences: At large $k$,
\emph{(i)} all edges of all skeleton graphs are occupied by $\order{k}$ propagators, and
\emph{(ii)} only graphs where the total number of edges between all operators is maximal
will contribute. All terms that violate any of these two conditions
will only contribute at subleading orders in large $k$. At every fixed
genus, we will call graphs whose total number of edges is maximal
\emph{maximal graphs}.
As will be seen below, the number of edges in a maximal graph of genus
$g$ is equal to $4g+4$. Hence the contribution at each genus $g$
comes with an additional $k^4$ enhancement compared to
the genus $g-1$ contribution. This explains the powers of $k$
in the series~\eqref{eq:polynomials}.%
\footnote{For a more detailed discussion of
this octopus principle (unbaptized until now) see the discussions around
equation (6) in~\cite{Bargheer:2017nne} or equation (6.10)
in~\cite{Bargheer:2018jvq}. This phenomenon was actually encountered
long before, in the times of the BMN explorations; see most
notably the discussion on pages~5 and~6 in~\cite{Kristjansen:2002bb}, where it
was already identified that the large-charge limit would project out skinny handles in
the genus expansion. As mentioned in the introduction, the key
difference compared to those earlier BMN works is that here several R-charge
directions are taken to be large, and that only now we can take advantage of
the great control over the 't~Hooft coupling, as fully captured by the
function $\oct$.}
This is also why the double-scaling limit $k\sim \sqrt{\Nc}$ is
precisely the regime that we probe when re-summing that expansion.

\begin{figure}[t]
\centering
\begin{tabular}{cc}
\resizebox{.25\totalheight}{!}{\includegraphics[width=\textwidth]{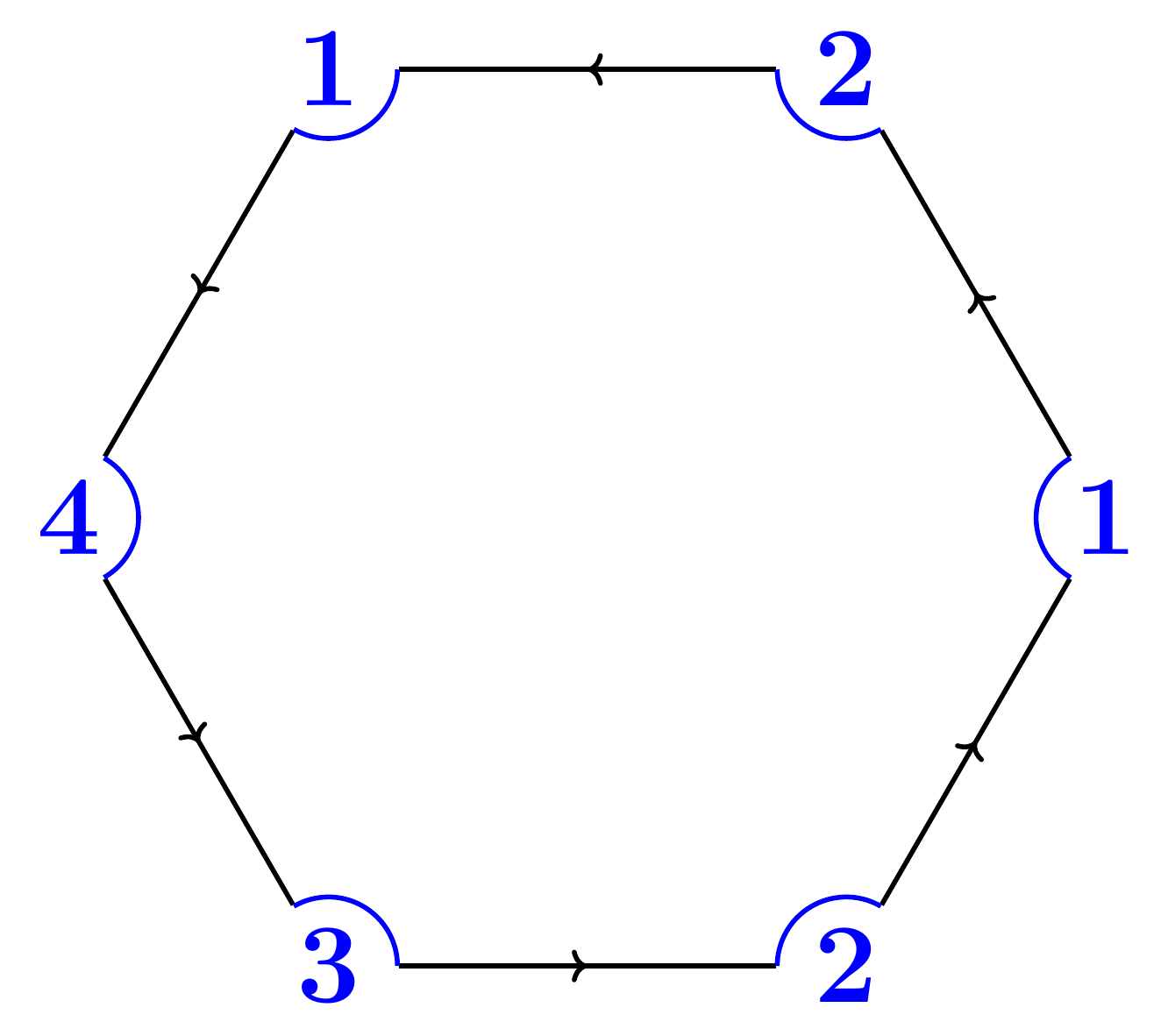}} &
\resizebox{.25\totalheight}{!}{\includegraphics[width=\textwidth]{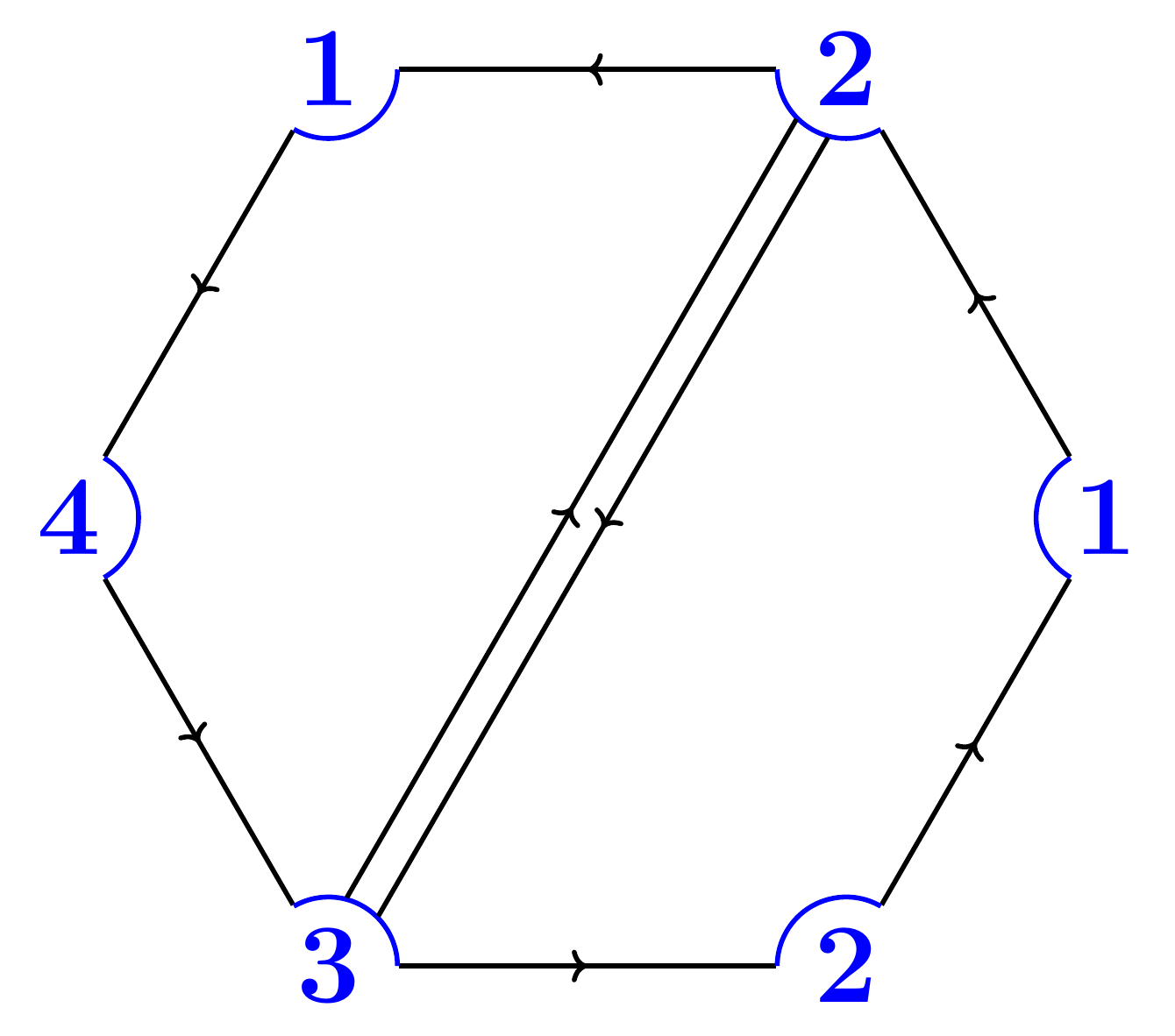}}\\
(a) & (b)
\end{tabular}
\caption{All bigger polygons, such as the hexagon shown in (a), can be
split into squares by adding further bridges, as shown in (b). Adding
such bridges does not increase the genus of the surrounding graph.}
\label{fig:breaking}
\end{figure}

We conclude that at large $k$, the dominating graphs are the so-called
maximal graphs, to which no extra propagator bundles can be added without
increasing the genus. In such graphs, all faces are bounded by as few edges as possible. For our
operator polarizations~\eqref{eq:24},~\eqref{eq:13}, the irreducible faces are
squares, and hence all maximal
graphs are quadrangulations. More precisely,
since each operator $\op{O}_i$ can only connect to operators
$\op{O}_{i\pm 1} $, all faces of all large-$k$ skeleton graphs
must be of one of the following three types:
\begin{align}
\oct&=
\texttt{square}[\op{O}_i {-} \op{O}_{i+1} {-} \op{O}_{i+2} {-} \op{O}_{i+3}]
=\texttt{square}[\op{O}_i {-} \op{O}_{i{-}1} {-} \op{O}_{i{-}2} {-} \op{O}_{i{-}3}]
\,,\nn \\
1=\oct\subrm{BPS}&=
\texttt{square}[\op{O}_i {-} \op{O}_{i+1} {-} \op{O}_{i} {-} \op{O}_{i+1}]
=\texttt{square}[\op{O}_i {-} \op{O}_{i{-}1} {-} \op{O}_{i} {-} \op{O}_{i{-}1}]
\,,\nn \\
1=\oct'\subrm{BPS}&=
\texttt{square}[\op{O}_i {-} \op{O}_{i+1} {-} \op{O}_{i+2} {-} \op{O}_{i+1}]
=\texttt{square}[\op{O}_i {-} \op{O}_{i{-}1} {-} \op{O}_{i{-}2} {-} \op{O}_{i{-}1}]
\,,
\label{eq:squaretypes}
\end{align}
as illustrated in~\figref{fig:squareTypes}.
All bigger polygons can always be split into squares by adding further
brigdes without increasing the genus, as
illustrated in~\figref{fig:breaking}.
Note that we cannot break the squares into triangles, since each operator $\op{O}_i$
can only connect to operators $\op{O}_{i\pm 1}$.
The squares in the last two lines of~\eqref{eq:squaretypes}
only contain at most three different BPS operators and are thus
protected by supersymmetry and simply give $1$. The square in the
first line is the non-trivial
function $\oct$ that appeared already at genus zero.

To summarize, all graphs that contribute at genus $g$ in the large-$k$ limit are quadrangulations
of a genus-$g$ surface, such that all faces
are squares of the types~\eqref{eq:squaretypes}. By Euler counting we
have
\begin{equation}
2-2g
=(V = 4)+(F = n)-(E=4n/2)
=4-n
\,,
\end{equation}
so that at genus $g$,
all graphs contain a total of $n=2g+2$ squares and twice as many
edges. Indeed, our single genus-zero
skeleton graph was simply given by two squares, as explained in the
previous section. In the three genus-one examples of~\figref{fig:threegraphs},
we have four squares.%
\footnote{As indicated by the colors, the
diagram in~\figref{fig:threegraphs}(a) contains only BPS squares,
the diagram in~\figref{fig:threegraphs}(b) contains two
copies of the non-BPS square $\oct$ (and two BPS squares), and
the diagram in~\figref{fig:threegraphs}(c) contains four copies of the
non-BPS square $\oct$.}
It is also simple to see that the number of
non-BPS squares $\oct$ in each graph ought to be even.\footnote{The
non-BPS square is bounded by four different types of edges, while the
perimeter of the other two types of squares is formed by even numbers
of edges of the same type, as can be seen
in~\figref{fig:squareTypes}. Since each square is glued to another
square along an
identical type of edge, the surface can only close if the number
of non-BPS squares is even.}
Therefore, we conclude that at each fixed genus $g$, all contributions
sum to a polynomial $P_{g+1}$ in $\oct^2$ of degree $g+1$, thus
leading to~\eqref{eq:polynomials}. Finding these polynomials is
tantamount to counting quadrangulations.

\begin{figure}
\centering
\resizebox{1.8\totalheight}{!}{\includegraphics[width=\textwidth]{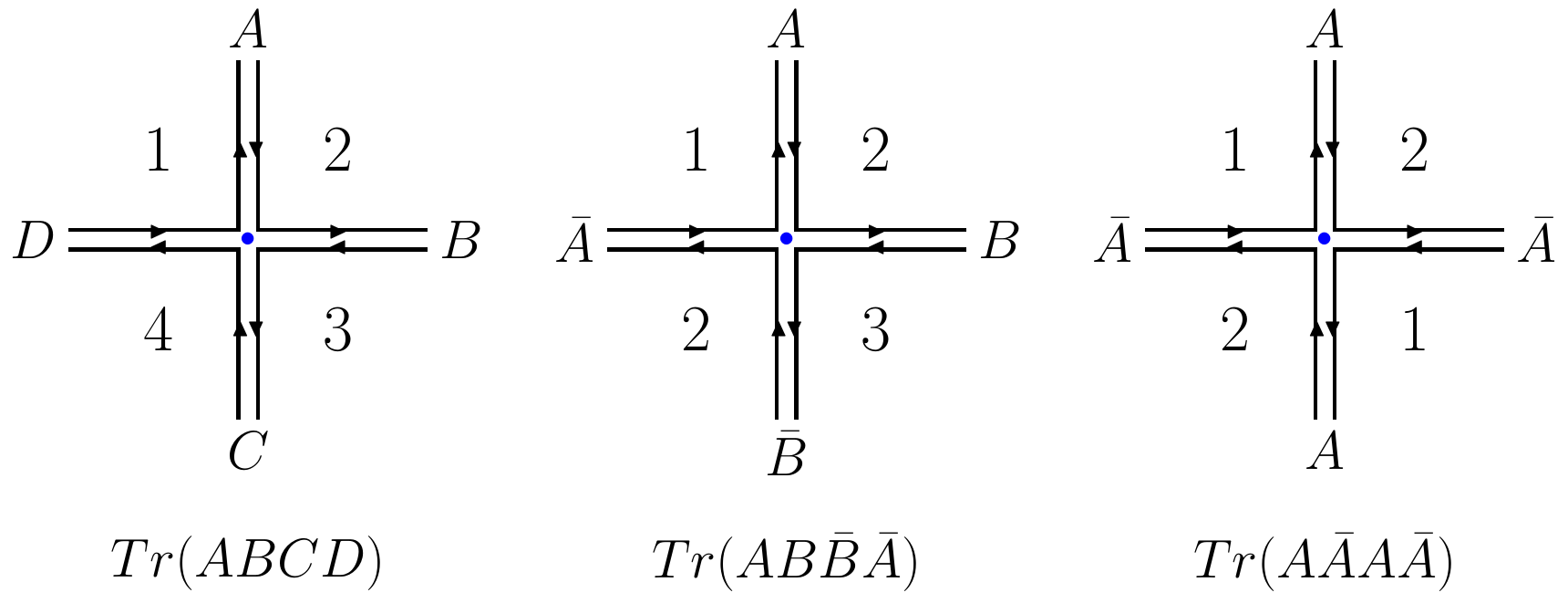}}
\caption{Matrix model vertices. Each is the dual counterpart of the
corresponding three square types represented in~\figref{fig:squareTypes}.}
\label{fig:vertices}
\end{figure}

\begin{figure}[tbh]
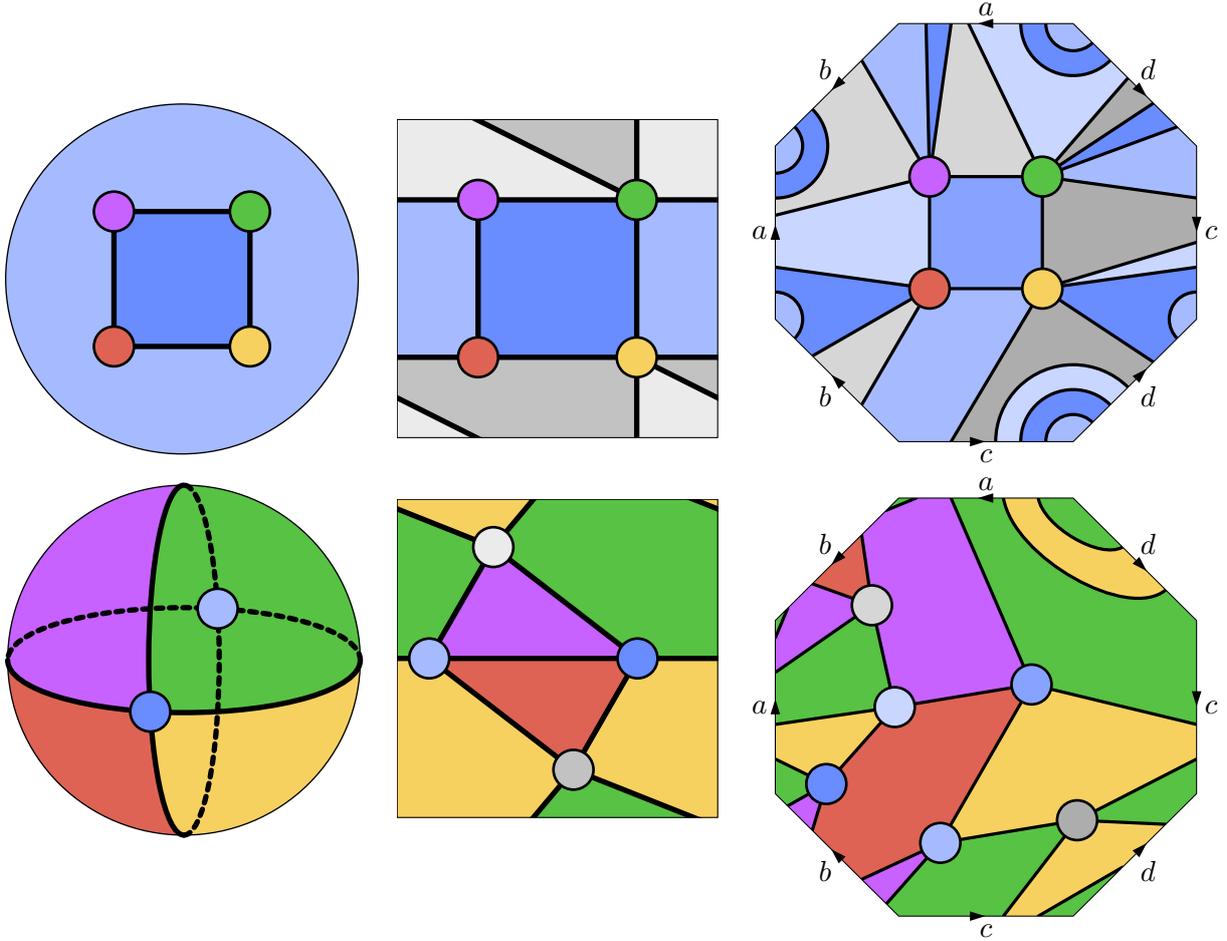

\centering
\begin{tabular}{@{}c@{}}
\includegraphics{FigExampleGraphGenusZero} \\[2mm]
\includegraphics{FigExampleGraphGenusZeroDual}
\end{tabular} \hfill
\begin{tabular}{@{}c@{}}
\includegraphics{FigGraphOct2Torus} \\[6.5mm]
\includegraphics{FigGraphOct2TorusDual}
\end{tabular} \hfill
\begin{tabular}{@{}c@{}}
\includegraphics{FigExampleGraphGenusTwo} \\
\includegraphics{FigExampleGraphGenusTwoDual}
\end{tabular}
\caption{Example graphs at genus $g=0,1,2$ (top) with their duals
(bottom). In the top row we have four vertices (the BPS operators) and
$2g+2$ faces (the squares). In the bottom row we have four faces (the
BPS operators) and $2g+2$ quartic vertices (the squares).
}
\label{fig:exampleDual}
\end{figure}

In order to count quadrangulations of surfaces of genus $g$
with $4$ vertices (punctures) and $2g+2$ squares, we introduce a
matrix model.%
\footnote{Two beautiful matrix model reviews
are~\cite{Kazakov:2000aq,Zvonkin:1997}.}
The matrix model naturally
describes the duals of the
skeleton graphs, where each of the $2g+2$ original square faces becomes
a quartic vertex, and where the original $4$ vertex operators
$\op{O}_1,\dots,\op{O}_4$ become four faces of the dual graph. Each bridge
connecting operators $\op{O}_i$ with $\op{O}_{i+1}$ is now pierced by a propagator
of the dual graph; since there are four types of bridges
$\op{O}_{i}{-}\op{O}_{i+1}$, we will have four complex matrices, one for each such
original bridge type. See~\figref{fig:vertices} for the vertices
and~\figref{fig:exampleDual} for example graphs with their duals.
There are $10$
different square faces in~\eqref{eq:squaretypes}, and so there will
be $10$ different
vertices in the matrix model. All in all, the partition function of
our matrix model is
\begin{equation}
Z\equiv
\int [\mathcal{D}A][\mathcal{D}B][\mathcal{D}C][\mathcal{D}D]
\exp\bigbrk{-S_\text{kin}[A,B,C,D]+S_\text{int}[A,B,C,D]}
\,,
\label{eq:MM}
\end{equation}
with the kinetic action term
\begin{equation}
S\subrm{kin}
=\tr\lrsbrk{\frac{A\bar A}{k_1}+\frac{B\bar B}{k_2}+\frac{C\bar C}{k_3}+\frac{D\bar D}{k_4}}
\end{equation}
and the interaction term
\begin{align}
S\subrm{int}
&=\oct\tr(ABCD)+\oct  \tr(\bar D\bar C\bar B\bar A)
\label{eq:Sint}\\
&+\tr\lrsbrk{
\frac{(A\bar A)^2+(B\bar B)^2+(C\bar C)^2+(D\bar D)^2}{2}
+AB\bar B \bar A+BC \bar C \bar B+C D \bar D \bar C+D A \bar A \bar D
}\,.\nn
\end{align}
The interaction part consists of two non-BPS vertices in the first
line (the duals of the
non-BPS squares,  which therefore come with a factor $\oct$),
and eight BPS vertices (which come with a factor of
$1$ since they are BPS) in the second line.

In the kinetic term, we have introduced parameters $k_i$,
$i=1,\dots,4$ as a means of counting the number of propagator bundles
connecting $\op{O}_i$ and $\op{O}_{i+1}$
in each skeleton graph by simply reading off the corresponding power of
$k_i$. This is quite important, because we have to dress each
quadrangulation by four factors of the type~\eqref{eq:combiFactors}, one for
each type of connection. Keeping track of the numbers of different
types of edges individually also allows us to calculate the correlator of a more general and considerably richer set of
operators (the sum over permutations for $\op{O}_1$ and $\op{O}_3$ is implicit)
\begin{equation}
\op{O}_{1} = \tr\brk{\bar{Z}^{k_4}\,{\color{red}\bar{X}}^{k_1}}
,\quad
\op{O}_{2}= \tr\brk{{\color{red}X}^{k_1+k_2}\,}
,\quad
\op{O}_{3} = \tr\brk{\bar{Z}^{k_3}\,{\color{red}\bar{X}}^{k_2}}
,\quad
\op{O}_{4}= \tr\brk{{Z}^{k_3+k_4}}
.
\label{ops}
\end{equation}
At genus zero, there is again a single graph contributing to the correlator, and it is
again a nice rectangle frame as in~\figref{fig:intro}(a). The difference
is that now there are $k_i$ propagators connecting $\op{O}_i$ and
$\op{O}_{i+1}$. The limit of large charges now amounts to taking all the
$k_i$ to be of order $\sqrt{\Nc}$. At genus zero, for example, we have
\begin{equation}
\vev{\op{O}_1\dots\op{O}_4}_{g=0}
=\frac{{\color{blue}\oct^2}}{(x_1-x_2)^{2k_1}(x_2-x_3)^{2k_2}(x_3-x_4)^{2k_3}(x_4-x_1)^{2k_4}}
\,.
\label{eq:genus0General}
\end{equation}
At higher genus, in the large charge limit and with $\zeta_i\equiv k_i/\sqrt{\Nc}$,
\begin{equation}
\frac{\vev{ \op{O}_1\dots \op{O}_4 }}{\vev{ \op{O}_1\dots \op{O}_4}_{\lambda=0,g=0}}
\xrightarrow{\;k_i\sim\sqrt{\Nc}\;}
\sum_{g=0}^{\infty} \frac{P_{4g|g+1}(k_1,k_2,k_3,k_4,\oct^2)}{\Nc^{2g}}
\equiv \mathcal{A}(\zeta_1,\zeta_2,\zeta_3,\zeta_4,\oct)
\,,
\label{eq:generalKs}
\end{equation}
where $P_{4g|g+1}$ are polynomials of degree $g+1$ in $\oct^2$ whose
coefficients are \textit{homogeneous} polynomials of degree $4g$ in
the four $k_i$. When all $k_i$ are equal, then
\begin{equation}
P_{4g|g+1}(k_i|\oct^2)
\xrightarrow{\,k_i\to k\,}
k^{4g} P_{g+1}(\oct^2)
\,,
\label{eq:Pki-to-Pk}
\end{equation}
and we get back to our previous correlator~\eqref{eq:polynomials}.

To obtain the full correlator~\eqref{eq:generalKs} at genus $g$ from
the matrix model~\eqref{eq:MM}, we bring down $2g+2$ vertices, pick the
$N^4$ coefficient%
\footnote{The matrix
model comes with its own number of colors $N$, which we use to
identify numbers of faces and genus, as illustrated in the example~\eqref{eq:Z1}. As usual
with such graph dualities, $N$
is not to be identified with the $\Nc$ of $\mathcal{N}=4$ SYM, see
\eg~\cite{Brown:2010af}. In fact,
we will soon explain that it is often convenient to introduce
rectangular matrices with sizes $N_i \times N_{i+1}$ in the matrix
model language, to better keep track of the different faces in the matrix
model, \ie the different operators in the original picture.}
(since we are after a four-point correlation function, which in
terms of the dual matrix model means that we are interested in graphs
with four faces), and focus on those contributions where
all $k_j$ appear. That last condition is due to R-charge conservation,
which implies that all types of bridges between operators $\op{O}_i$
and $\op{O}_{i+1}$ must appear. We thus discard any monomials such as
$k_1^2 k_2^2$ which do not contain all $k_i$. All in all, the term
that we are interested in is
\begin{equation}
Z=\dots +
{\color{blue}
N^4 k_1 k_2 k_3 k_4
\Bigbrk{\mathcal{Z}\equiv \sum_{g=0}^\infty \tilde P_{4g|g+1}(k_1,k_2,k_3,k_4|\oct^2)}}
+ \dots
\label{eq:Z1}
\end{equation}
These tilded polynomials $\tilde{P}_{4g|g+1}$ count our quadrangulations,
and are thus \textit{almost} the polynomials arising
in the correlator~\eqref{eq:generalKs}.
To get precisely those, however,
we also need to include the combinatorial factors~\eqref{eq:combiFactors}.
Since we strip out an overall $k_1,\dots,k_4$ factor in defining the
reduced partition function $\mathcal{Z}$, we finally conclude that
%
\begin{equation}
P_{4g|g+1}(k_1,k_2,k_3,k_4|\oct^2) =
\tilde P_{4g|g+1}(k_1,k_2,k_3,k_4|\oct^2)
\bigg|_{
\displaystyle k_1^{n_1} \dots k_4^{n_4}
\to
\frac{k_1^{n_1} \dots k_4^{n_4}}{n_1!\dots n_4!}
\,.}
\label{eq:replacement}
\end{equation}
or equivalently
\begin{equation}
\mathcal{A}(\zeta_1,\zeta_2,\zeta_3,\zeta_4,\oct)
=\mathcal{Z}\big|_{k_i^n \to {\zeta_i^n}/{n!}}
\label{eq:corFromZ}
\end{equation}
for the full correlator at any genus and any coupling.%
\footnote{Obviously, the replacement $k_i^n\to\zeta_i^n/n!$ should
only be made after expanding $\mathcal{Z}$.}
This is our main result.

As a trivial check, consider genus zero. We need to bring down $2g+2=2$
vertices, \ie we consider terms in the expansion of
$\exp\lrbrk{-S\subrm{int}}$ that are of degree $2$ in the
interaction vertices. If we bring down two vertices from the second line
in~\eqref{eq:Sint}, we see right away that we either get more than four
faces (from
$\langle \tr(AB\bar B \bar{A} )\tr(CD\bar D \bar{C})\rangle$ for
example) \textit{or} we generate terms which do not contain all
$k_i$'s (from $\langle \tr(A\bar A)^2\tr(B\bar B)^2\rangle$ for
example). Bringing down an odd number of vertices from the first line
in~\eqref{eq:Sint} gives a zero result by charge conservation. So we are
left with the possibility of bringing down two non-BPS vertices from
the first line. This leads to
\begin{equation}
\oct^2 \langle \tr(ABCD) \tr(\bar D\bar C\bar B\bar A) \rangle= N^4 k_1 k_2 k_3 k_4 {\color{blue} \oct^2}
\,,
\end{equation}
recognizing precisely the genus zero result in~\eqref{eq:genus0General}.

Bringing down further octagon vertices from $\exp\lrbrk{-S\subrm{int}}$, we
generate all the above-mentioned polynomials $\tilde P$
and thus their transformed partners $P$, which enter the four-point
correlation function~\eqref{eq:generalKs}.
We managed to compute the general polynomials $P_{4g|g+1}$ up to genus $g=4$. As we saw above,
the genus-zero polynomial is simply $P_{0|1}=\oct^2$. At genus one, we
find
\begin{align*}
P_{4|2}&=
k_1k_2k_3k_4 \lrsbrk{1
+\oct^2\frac{
\sum_{j=1}^4\sbrk{(k_j+k_{j+1})^4-k_j^4}
+12(k_1k_2+k_3k_4)(k_1k_4+k_2k_3)
}{24k_1k_2k_3k_4}
+\frac{\oct^4}{2}
}.
\nn
\end{align*}
The polynomials for $g=2,3,4$ are attached in the file \filename{polynomials.m}.
For equal charges, $k_i\equiv k$, the polynomials $P_{4g|g+1}$ reduce
to $k^{4g}$ times a polynomial $P_{g+1}$ in $\oct$ with rational
coefficients, see~\eqref{eq:Pki-to-Pk}. The resulting correlator
$\mathcal{A}$ is quoted in~\eqref{eq:fin2} below.
We have cross-checked the polynomials $P_{4g|g+1}$ obtained from the matrix model
against an explicit construction of all contributing skeleton graphs
up to genus three, see~\appref{sec:generating-maximal-graphs}.

Let us make three comments. The first one is that we are extracting the
term with~4 faces, proportional to $N^4$. This is actually the
\textit{smallest} power of $N$ arising in the perturbative expansion
if we keep only terms containing all four $k_i$, as we are instructed to do. (It
would be the next-to-smallest if we lift the latter restriction.) This
is in stark contrast with the usual large-$N$ expansion, where the
leading terms carry the largest powers of $N$. So the limit we are
interested in is a sort of $N\to 0$ limit of the matrix model.

In
vector models, the limit of small number of colors is a very
interesting one, related to polymers and other such
fascinating combinatorics, see
\eg~\cite{deGennes:1972339,deGennes:1980PhT,Cardy:1996book}.
Another interesting instance of such limits shows up in the study of
entanglement entropy and quenches disorder, where one often uses the replica
trick to study the~$n$ copies of a given system in the formal $n\to 0$
limit. This is done to generate logarithms (of a density matrix or partition
function) that were originally absent through the identity
$\lim_{n\to 0} (x^{n}-1)/n=\log(x)$. Sometimes the coupling between the
various copies is encoded in a matrix, see
\eg~\cite{Sherrington:1975zz,Almeida:1978stability,Morone:1409.2722}.
In those
cases, we are also interested in a formal limit where the matrix size goes
to zero. Such matrices, present most notably in spin-glass studies, have
extremely rich dynamics.
In matrix model theory,
dualities between $N\to\infty$ and $\tilde{N}\to0$ limits have in fact been
found before, in the context of the theory of intersection numbers of
moduli spaces of curves,
see~\cite{Brezin:2000characteristic,Brezin:2007aa,Brezin:2007wv,Brezin:2010ar}.%
\footnote{Another reference which mentions this limit and
dubs it as the anti-planar limit is~\cite{Bellucci:2004fh}.}
As we will see below, the $N\to0$ limit will show up again and again
in several simplified matrix model combinatorics in very amusing ways.
It would be
interesting to find a nice statistical mechanics application of this
zero-color limit.

A second small comment is that we can consider rectangular
matrices~\cite{DiFrancesco:2002mvz}, where matrix $A$ has dimensions
$N_1\times N_2$, matrix $B$ has dimensions
$N_2\times N_3$, and so on. Then $N^4$ would be replaced by
$N_1 N_2 N_3 N_4$, identifying precisely the four faces corresponding to
the four distinct operators. The terms containing this factor automatically
contain all $k_i$, so using rectangular matrices would allow us to
condense the instructions above into simply
\begin{equation}
\mathcal{Z}(k_i)
=\frac{1}{k_1 k_2 k_3 k_4}
\lim_{N_i \to 0}
\frac{\partial}{\partial N_1}
\frac{\partial}{\partial N_2}
\frac{\partial}{\partial N_3}
\frac{\partial}{\partial N_4}
Z(N_i,k_i) \,.
\label{eq:RectangularZ}
\end{equation}
This highlights once more the limit of small number of colors we are interested in.

The last comment pertains to the combinatorial replacement
in~\eqref{eq:replacement}. In effect, by introducing an extra $1/n!$ for
each coupling $k_i^n$, we are Borel transforming our matrix model
perturbative expansion. Interestingly, it renders the partition
function expansion finite, as we will see explicitly below. Namely,
the transformation removes the usual $(2g)!$ divergence that is due
to the proliferation of graphs at higher
genus~\cite{Shenker:1990uf,Polchinski:1994fq}, and thus leads to a fully
convergent expansion. While getting rid of divergences might be seen
as a feature, losing the D-brane physics they encode is bit of a bug.
This is presumably related to the fact that although we are resumming a
't~Hooft expansion, we are not generating arbitrarily complicated
multi-string intermediate states. The large-charge limit projects
onto folded strings, BMN strings and copies thereof, eliminating
non-perturbative effects arising from the more complicated
multi-string states which the D-branes source. It would be fascinating to
slowly decrease the size of our BPS operators to move away from our
fully convergent limit and carefully isolate these novel effects in a
controllable way.

\section{Matrix Model Simplification and Limits}
\label{sec:Simplifications}

Ideally, we would like to determine the full correlation function
$\mathcal{A}(\zeta_1,\zeta_2,\zeta_3,\zeta_4,\oct)$ by solving the
matrix model~\eqref{eq:MM}. That would be equivalent to
computing the polynomials~$P_{4g|g+1}$ to all genus, which we have
not succeeded thus far. What we \textit{did} manage to do is to
simplify this matrix model problem into an equivalent matrix model
problem where we have two hermitian matrices $\mathbb{M}_1$,
$\mathbb{M}_2$ and two complex matrices $\mathbb{X}$, $\mathbb{Y}$,
with a non-diagonal propagator between $\mathbb{X}$ and $\mathbb{Y}$
equal to the octagon function $\oct$, so that
\begin{equation}
\vev{F}\equiv \int
[\mathcal{D}\mathbb{M}_1]
[\mathcal{D}\mathbb{M}_2]
[\mathcal{D} \mathbb{X}]
[\mathcal{D} \mathbb{Y}]
F\exp\biggsbrk{
- \frac{1}{2}\tr \mathbb{M}_1^2
-\frac{1}{2}\tr \mathbb{M}_2^2
-\tr \begin{pmatrix} \mathbb{X} &\!\! \mathbb{Y} \end{pmatrix}
\!\begin{pmatrix} 1 &\!\! \oct \\ \oct &\!\! 1 \end{pmatrix}^{\!-1}
\!\!\begin{pmatrix} \bar{\mathbb{X}} \\ \bar{\mathbb{Y}} \end{pmatrix}
}.
\label{eq:MM2}
\end{equation}
Then we have the rather compact expression
\begin{equation}
\mathcal{Z} =
\frac{
\lrvev{
\tr\log\lrbrk{
\mathbb{I}-
\frac{k_2}{\mathbb{I}-k_2 \mathbb{M}_2} \bar{\mathbb{X}}
\frac{k_1}{\mathbb{I}-k_1 \mathbb{M}_1} \mathbb{X}
}
\tr\log\lrbrk{
\mathbb{I}-
\frac{k_3}{\mathbb{I}-k_3 \mathbb{M}_2}  \bar{\mathbb{Y}}
\frac{k_4}{\mathbb{I}-k_4 \mathbb{M}_1} \mathbb{Y}
}
}_{\text{two faces}}
}{k_1 k_2 k_3 k_4}
\label{eq:newRep}
\end{equation}
for the reduced partition function, from which we can readily extract
the correlator via~\eqref{eq:corFromZ}. When expanding the logarithms in
powers of $k$, we can drop all terms whose total power is not a multiple of
four, since the latter are the terms that correspond to an even number of
octagons as required (at genus $g$ we keep $4g+4$ powers of $k$). We
also extract the coefficient of $N^2$, which is the smallest power
of~$N$ on the right hand side. So again, in this alternative matrix
model formulation, we are after the $N\to 0$ limit. As a check,
we can expand to leading order in $k$ to get
\begin{equation}
\mathcal{Z} = \langle \tr (  \bar{\mathbb{X} }  \mathbb{X} ) \tr (\bar{\mathbb{Y}}   \mathbb{Y} ) \rangle_\text{two faces}
+ \order{k^4}
\,,
\end{equation}
which evaluates to $\oct^2$, since Wick contracting complex
fields of the same type would lead to four faces, and since each
off-diagonal propagator equals $\oct$.
This is exactly what we expect at genus zero.

The derivation of~\eqref{eq:newRep} follows the graphical manipulations
in~\appref{sec:integrating-in-and-out}. Technically, we open up all quartic vertices
in~\eqref{eq:Sint} into pairs of cubic vertices using auxiliary fields
as detailed in \appref{sec:all-quandrangulations}. If done
carefully, the resulting action is Gaussian in the original four
complex matrices. Integrating them out then leads to~\eqref{eq:newRep}.
In particular, the logarithms arise from the complex matrix identity
\begin{equation}
\int [\mathcal{D}\mathbb{A}] e^{-\tr(\mathbb{A}.\mathbb{Q}.\bar{\mathbb{A}})} = (\det \mathbb{Q})^{-N}
\,.
\end{equation}
It is particularly nice that in these integrations we explicitly
generate two such factors which automatically produce two factors of
$N$. That is why in~\eqref{eq:newRep} we extract two faces only rather
than four as in the original representation with four complex matrices.
Technically, this renders the representation~\eqref{eq:newRep} quite
powerful. Besides, there are less degrees of freedom as we went from
four complex matrices to two complex and two hermitian.

More physically, we started with a matrix model with four complex
matrices corresponding to the four types of consecutive propagators in
our large cyclic operators. The four-point function of these four
cyclic operators was mapped to a dual correlation function with four
faces in the dual matrix model with matrices $A,B,C,D$. The two-point
function with two faces in~\eqref{eq:newRep} is thus a hybrid
representation, where two of the four operators are represented as
vertices and the other two as faces, see~\figref{fig:TwoFaces}.
See~\cite{Maldacena:2004sn,Brown:2010af}, and also the very inspiring
talk~\cite{talk:Gopakumar:Johannesburg:4.2010} for very
similar (and often more general) dynamical graph dualities obtained by
integrating-in and -out matrix fields.

In practice, we compute~\eqref{eq:newRep} by expanding out the
expression to any desired monomial in the $k_i$ and then preforming the various
free Wick contractions. We found it particularly convenient to Wick
contract the complex matrices first and the hermitian matrices at the
end. Once $\mathbb{X}$ and $\mathbb{Y}$ are integrated out, because of
the alternating pattern in~\eqref{eq:newRep} it is easy to see that we
generate products of traces containing either
$U_{k_1}({\color{red}\mathbb{M}_1})$ and
$U_{k_4}({\color{red}\mathbb{M}_1})$ \textit{or}
$U_{k_2}({\color{blue}\mathbb{M}_2})$ and
$U_{k_3}({\color{blue}\mathbb{M}_2})$, but never both (here
$U_k(M) \equiv k/(\mathbb{I}-k M)$). Expanding further the $U_k$ in
terms of the fundamental Hermitian fields we conclude that our full
reduced partition function is given by a sum of factorized
one-matrix correlators,
\begin{equation}
\mathcal{Z} = \sum_{g=0}^{\infty} \sum_{n_1,\dots,n'_{M'}} \mathbb{C}^{(g)}_{n_1,\dots,n'_{M'}} \times \vev{\tr(\mathbb{M}_1^{n_1})\dots\tr(\mathbb{M}_1^{n_{M}})}_\text{one face}
\times
\vev{\tr(\mathbb{M}_2^{n'_1})\dots\tr(\mathbb{M}_2^{n'_{M'}})}_\text{one face} \label{eq:factorizedSum}
\,.
\end{equation}
In this sum, $n_1+\dots + n_M+n'_1+\dots+n'_{M'}\le 4g$, and the
combinatorial factors $\mathbb{C}^{(g)}$ arising from integrating out
the complex matrices $\mathbb{X}$ and $\mathbb{Y}$ are \emph{homogeneous} polynomials of degree
$4g$ in the $k_j$, with coefficients that are polynomials in~$\oct^2$ of maximally
degree~$g+1$.%
\footnote{Note in particular that we can have
$M=1$ and $n_1=0$, so that the first term in~\eqref{eq:factorizedSum}
would just give $\vev{\tr(\mathbb{I})=N}_\text{one face}=1$.}
Importantly, note that because of the
factorization in~\eqref{eq:factorizedSum}, each Hermitian correlator
now has to be restricted to a \textit{single} face, which is quite a bit simpler
than the previous two-face problem in~\eqref{eq:newRep}, which in turn
was a considerable simplification over the initial four-face problem
in~\eqref{eq:MM}.%
\footnote{Up to genus $1$, using the notation
$\langle\cdots\rangle_{\!{\color{red}1}}\equiv
\langle\cdots\rangle_{\!\text{one face}}$, we have
%
\bbn
\mathcal{Z} ={}&
\oct^{2}\vev{\tr(\mathbb{I})}^{2}_{\!\color{red}1}
+\oct^{4}\,k_1 k_2 k_3 k_4\,\vev{\tr(\mathbb{I})}^{2}_{\!\color{red}1}
+k_{1}k_{2}k_{3}k_{4}\vev{\tr(\mathbb{M}_{1})^{2}}_{\!\color{red}1}
 \vev{\tr(\mathbb{M}_{2})^{2}}_{\!\color{red}1}
+\oct^{2}\bigg(
\frac{(k_1{+}k_4)^4}{24}\vev{\tr(\mathbb{M}_{1}^{4})}_{\!\color{red}1}\vev{\tr(\mathbb{I})}_{\!\color{red}1}
\\ &
+\frac{k_{1}^{2}k_{2}^{2}}{4}\langle\tr(\mathbb{I})\rangle^{2}_{\!\color{red}1}
+\frac{k_{1}^{3}k_{2} +3 k_{1}^{2} k_{2} k_{4} +3 k_{1}k_{3}k_{4}^{2}+k_{3}k_{4}^{3}}{6}\,\langle \tr(\mathbb{M}_{1})^{2}\rangle_{\!\color{red}1}\langle \tr(\mathbb{I})\rangle_{\!\color{red}1}
+
{\scriptsize\begin{array}{l}k_{1}\leftrightarrow k_{3} \\ k_{2}\leftrightarrow k_{4} \\ \mathbb{M}_{1}\leftrightarrow\mathbb{M}_{2} \end{array} }
\bigg) +\dots\\
={}& \bigg(1+\frac{1}{2}\,\oct^{4}\bigg)\prod_{i=1}^{4}k_{i}\,+\,\oct^{2}\bigg(1+\sum_{i=1}^{4}k_{i}^{4}+k_{i}^{3}k_{i+1}+k_{i}k_{i+1}^{3}+k_{i}^{2}k_{i+1}^{2}+k_{i}k_{i+1}^{2}k_{i+2}\bigg)+\dots
\\
\Rightarrow\,\mathcal{A}&={} \bigg(1+\frac{1}{2}\,\oct^{4}\bigg)\prod_{i=1}^{4}\frac{\zeta_{i}}{\color{red}1!}\,+\,\oct^{2}\bigg(1+\sum_{i=1}^{4}\frac{\zeta_{i}^{4}}{\color{red}4!}+\frac{\zeta_{i}^{3}}{\color{red}3!}\frac{\zeta_{i+1}}{\color{red}1!}+\frac{\zeta_{i}}{\color{red}1!}\frac{k_{i+1}^{3}}{\color{red}3!}+\frac{\zeta_{i}^{2}}{\color{red}2!}\frac{\zeta_{i+1}^{2}}{\color{red}2!}+\frac{\zeta_{i}}{\color{red}1!}\frac{\zeta_{i+1}^{2}}{\color{red}2!}\frac{\zeta_{i+2}}{\color{red}1!}\bigg)+\dots \,.
\end{align*}}

\begin{figure}[t]
\begin{tabular}{cc}
\resizebox{1\totalheight}{!}{\includegraphics[width=\textwidth]{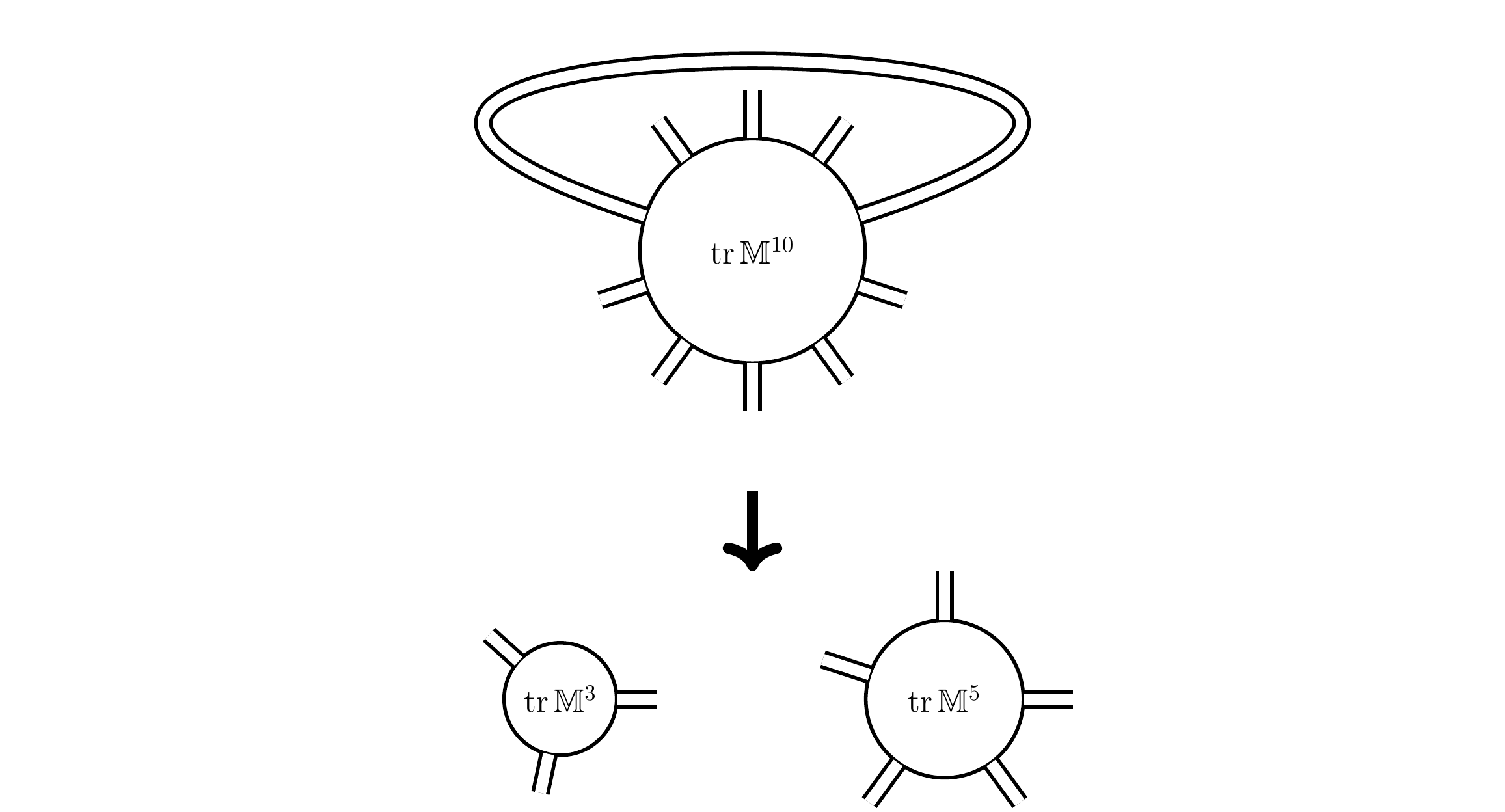}} &
\resizebox{.2\totalheight}{!}{\includegraphics[width=\textwidth]{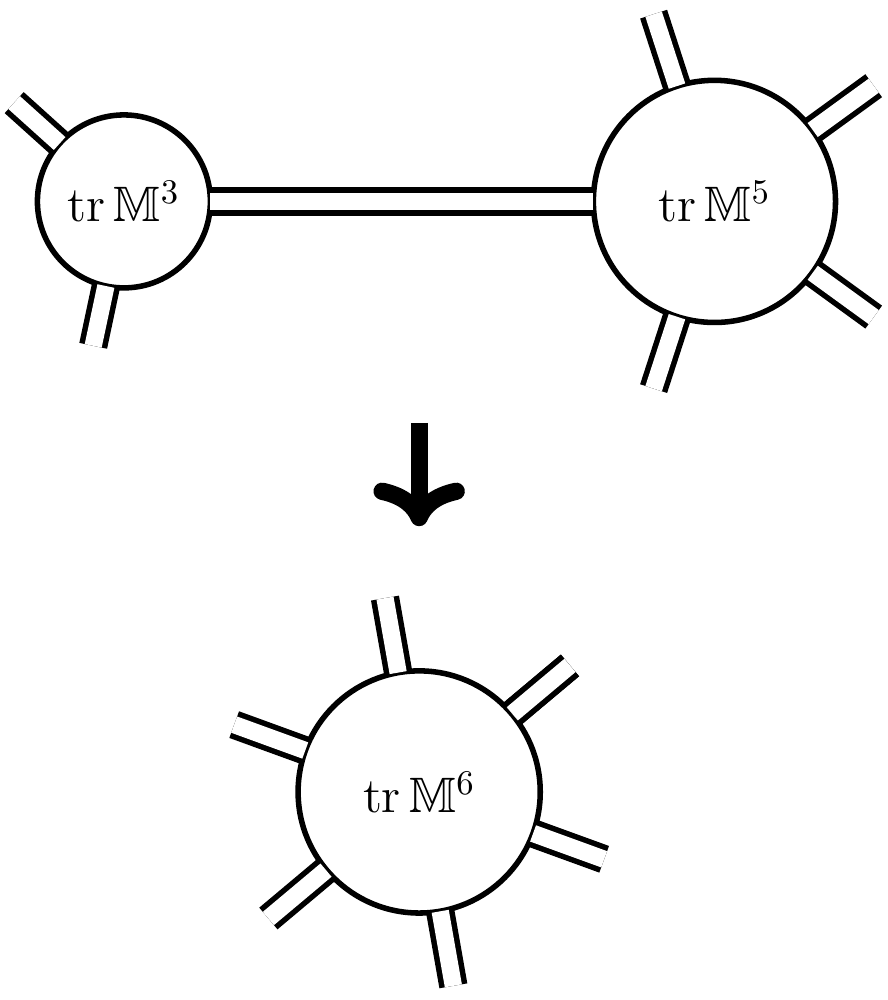}} \\
(a) & (b)
\end{tabular}
\caption{(a) Splitting and (b) joining of traces representend as
differential operators in~\protect\eqref{dif}.}
\label{fig:Split-Join}
\end{figure}

In fact, the one-face problem was solved already in the first days of
matrix models, see \eg the discussion below equation~(9) in~\cite{DiFrancesco:1992cn},
from which we readily get the generating function of all these
multi-trace Hermitian matrix model single-face expectation values as
\begin{multline}
\Bigvev{\exp{\sum\limits_{n=1}^\infty t_n \tr(\mathbb{M}^n)}}_{\!\text{one face}}
\\
=\sum_{j=1}^\infty \sum_{k=0}^{j-1}
\frac{(2k+1)!!(2j-2k-3)!! }{2j (-1)^k}
\Bigbrk{\frac{j-2k-1}{k+1/2}
\sum_{r=0}^{2k} \bar p_r p_{2j-r} - \bar p_{2k+1}p_{2j-2k-1}}
\,,
\label{oneFace}
\end{multline}
where $p_r$, $\bar
p_r$ are defined via the Schur polynomial identities
\begin{equation}
\sum\limits_{n=0}^\infty p_n z^n
\equiv \exp \biggbrk{ \sum\limits_{n=1}^\infty z^n t_n}
\quad \text{and} \quad
\sum\limits_{n=0}^\infty \bar p_n z^n
\equiv \exp \biggbrk{-\sum\limits_{n=1}^\infty z^n t_n}
\,.
\end{equation}
Another beautiful representation follows from the single-face limit
of~\cite{Mironov:2017och}, which gives
\begin{equation}
\Bigvev{\exp{\sum\limits_{n=1}^\infty t_n (\mathbb{M}^n) }}_\text{\!\text{one face}}
=\Bigbrk{\exp{\sum\limits_{a,b=0}^{\infty}
{
\overbrace{abt_{a}t_{b}\frac{\partial}{\partial t_{a+b-2}}}^{\text{splitting}}
+\overbrace{(a+b+2)t_{a+b+2}\frac{\partial}{\partial t_{a}}\frac{\partial}{\partial t_{b}}}^{\text{joining}}
}}}t_{0}
\,,
\label{dif}
\end{equation}
where the differential operator can be though of as implementing
fusion and fission of the various traces as one Wick-contracts all
these correlators, see~\figref{fig:Split-Join}. For general~$N$, we
would replace the $t_{0}$ at the end
by $e^{N t_0}$, and the $N^F$ coefficient in the final expansion would
compute the $F$--face result; it is quite a huge simplification to
simply linearize this exponent to get a single face in~\eqref{dif}, as
needed for our problem.%
\footnote{\label{footnoteNovel}We also found, experimentally, yet
another beautiful and even more compact representation for these
single-face correlators:
\begin{equation*}
\biggvev{\prod_{a=1}^{n}\tr\left(e^{t_a\mathbb{M}}-\mathbb{I}\right)}_{\!\text{one face}}
=\frac{1}{T^{2}} \prod_{a=1}^{n} 2\sinh\lrbrk{\frac{t_a T}{2}}
\,, \qquad
T\equiv \sum_{b=1}^n t_b\,,
\end{equation*}
see also~\appref{sec:all-extremal} for a multi-hermitian matrix
generalization. In fact, this representation has been found before in the
context of intersection theory on moduli spaces~\cite{Brezin:2007wv}.}
%

We could also try to integrate out the Hermitian matrices first.%
\footnote{In this regard, note that when we set all couplings
$k_j\equiv k$ equal, then we can rescale the matrices $\mathbb{X} \to \sqrt{1-k
\mathbb{M}_1} \mathbb{X} \sqrt{1-k \mathbb{M}_2}$, and similarly
for~$\mathbb{Y}$, to get rid of all Hermitian matrices in the
logarithms in~\eqref{eq:newRep}, and put them in new Yukawa-like
interactions upstairs, linear in both $\mathbb{M}_1$ and
$\mathbb{M}_2$ (because the square roots nicely combine when the
couplings $k_j$ are equal). Then we could use the tricks of~\cite{Kostov:1988fy} to
integrate out the Hermitian matrices and generate some new logarithmic
potentials for the remaining two complex matrices. The problem is that
by setting all couplings to be the same, we naively lose the
possibility of Borel resumming w.r.t.\ each individual coupling, as
needed to get our correlator in~\eqref{eq:corFromZ}. It would be very nice
to overcome this obstacle.}

This concludes the general description of our matrix model, and how we
dealt with it in practice to produce higher-genus
predictions. Next, we focus on interesting simplifying limits. There
are at least three obvious interesting limits where our correlator
should simplify:
\begin{equation}
{\bf (A)}\,\,\oct\to 1
\,,\qquad
{\bf (B)}\,\,\oct\to 0
\,,\qquad
{\bf (C)}\,\,\oct\to \infty
\,.
\label{eq:limits}
\end{equation}
The first corresponds to $\lambda=0$ or tree level (but all genus
orders). The other limits are more interesting \cite{TillPedroFrankVascoWorkInProgress}. The second shows up, for example, at strong coupling, and also
in interesting null limits at finite coupling. The last one would be
realized for instance in the so-called bulk-point
limit~\cite{Maldacena:2015iua}.

\subsection{Small Octagon Limit, \texorpdfstring{$\oct\to 0$}{O->0}}

In the small octagon limit where $\oct\to 0$, we are left with the
eight BPS square vertices in the second line of~\eqref{eq:Sint}, all with
the same weight $=1$. Now, there is another setup where we would have
encountered these, and only these, type of vertices, namely if we were
to study extremal correlators in the double-scaling limit of~\cite{\BMNworks}
using quadrangulations. There, we
know how to solve the original problem directly, using single-matrix
model technology, and the results found for these correlators are
basically given by products of
$\sinh( \omega_{ij} J_i J_j)$ factors times some simple rational function,
where $J_i$ are the charges of the involved operators, and the
frequencies $\omega_{ij}$ are pure numbers. We cannot directly apply
the very same techniques to our case, since we are now dealing with
several complex matrices. Instead, we guessed that the result ought to
take a similar form. We made an ansatz with a number of sinh factors and
fixed the various frequencies and prefactors by matching with the
first few terms of our matrix-model perturbative expansion. Then we computed
a few more orders to cross-check the validity of this guess up to
genus six. It all
works out beautifully, and the result turns out to be remarkably
simple:
\begin{equation}
\mathcal{A}(\zeta_{1},\zeta_{2},\zeta_{3},\zeta_{4}|\oct=0)
=\prod_{i=1}^{4} \,\frac{\sinh\left(\frac{1}{2}\,\zeta_{i}(\zeta_{i-1}+\zeta_{i}+\zeta_{i+1})\right)}{\frac{1}{2}\,(\zeta_{i-1}+\zeta_{i}+\zeta_{i+1})}
\,.
\label{eq:guessedBPS}
\end{equation}
In fact, we did a bit more than this: We considered a generalized
matrix model, where each of the eight BPS vertices is dressed with an
arbitrary coefficient, and guessed the general form of the resulting
``twisted'' correlator following the same strategy,
see~\appref{sec:all-quandrangulations}. We also studied these
generalized BPS quadrangulations for $n$-point extremal correlators
in~\appref{sec:all-extremal}, and for a setup where $n$ operators are
cyclically connected (the correlator discussed here is the $n=4$ case
of that) in~\appref{sec:bps-part-all-cyclic}.

It would be very interesting to find a first-principle honest
derivation of~\eqref{eq:guessedBPS}, starting from~\eqref{eq:MM}
or~\eqref{eq:MM2}. It might very well elucidate powerful tricks which we may
hope to use in the general case, where the octagon vertex is inserted
back.

It is important to stress that the result~\eqref{eq:guessedBPS} is
very non-perturbative, although it has no coupling dependence in it.
Note in particular that setting $\oct\to 0$ is not the same as setting
the coupling to zero, which is instead $\oct\to 1$. When $\oct\to 0$,
the various loop corrections must work hard to
completely cancel the tree-level result $\oct=1$. For example, at
genus zero and tree level, the correlator is equal to $1$, but it
becomes~$\oct^2$ at non-zero coupling. So when~$\oct\to0$, the
full genus zero result is washed out non-perturbatively.
In string terms, the
BPS vertices describe point-like string configurations, while~$\oct$
describes a large folded string. When the string tension is
very large, the BPS configurations with no area survive, while the big
extended strings are suppressed. Recall that the planar result
consists of just two copies of a big folded string glued together. Interestingly, this
contribution gets suppressed for $\oct\to0$.
At genus one, we start having configurations that are free of folded
strings and those should survive. Indeed, the
genus expansion of~\eqref{eq:guessedBPS}
starts at genus one:
\begin{align}
\frac{\mathcal{A}}{k_{1}k_{2}k_{3}k_{4}} &
= 0\times \Nc^{0}\,+\, \frac{1}{\Nc^{2}}+\frac{1}{\Nc^{4}}\,\frac{1}{24}\sum_{j=1}^{4} k_{j}^{2}\left(k_{j-1}+k_{j}+k_{j+1}\right)^{2}
\nonumber\\ &\qquad
+\frac{1}{\Nc^{6}}\frac{1}{1152}\,\sum_{i=1}^{4}\sum_{j=1}^{4}\left(1-\sfrac{2}{5}\delta_{i,j}\right)k_{i}^{2}k_{j}^{2}(k_{i-1}+k_{i}+k_{i+1})^{2}(k_{j-1}+k_{j}+k_{j+1})^{2}
\nonumber\\ & \qquad
+\mathcal{O}(1/\Nc^{8})
\label{ExpansionGuessedBPS}
\end{align}
It would be cute if we could understand the numbers in this expansion
directly from string theory by carefully counting these degenerate
string configurations. A good starting point could
be~\cite{Janik:2010gc,Klose:2011rm}, where degenerate point-like string
configurations for two- and three- point functions were analyzed, see
also~\cite{Minahan:2012fh,Bargheer:2013faa}. It would be interesting to
generalize them to our BPS squares, and to understand how those can be put
together purely in string terms.

\subsection{Large Octagon Limit, \texorpdfstring{$\oct\to \infty$}{O->oo}}

Another interesting limit is the regime where the octagon $\oct$ becomes very
large. In this case, only the maximal power of~$\oct$ survives at each
order in the genus expansion, which is~$\oct^{2g+2}$. In
other words, we only use the two non-BPS squares in our
quadrangulations, and set all BPS squares to zero. This dramatically simplifies
the representation~\eqref{eq:newRep} to
\begin{equation}
\mathcal{Z} =
\sum_{n=1}^\infty \frac{(k_1 k_2 k_3 k_4)^{n-1} \oct^{2n}}{n^2}
\vev{\tr((\mathbb{X}\mathbb{Y})^n)\tr((\bar{\mathbb{X}}\bar{\mathbb{Y}})^n)}_\text{two faces}
\,,\label{Znon}
\end{equation}
where $\mathbb{X}$ and $\mathbb{Y}$ here are complex matrices
with diagonal propagator normalized to~$1$, \ie with kinetic term
simply given by
$-\tr(\mathbb{X}\bar{\mathbb{X}})-\tr(\mathbb{Y}\bar{\mathbb{Y}})$. To
arrive at this expression, we notice that \emph{(i)} we can set to
zero all $\mathbb{M}_j$ matrices, since they describe BPS
quadrangulations, and \emph{(ii)} we only keep the off-diagonal Wick
contractions between the complex matrices in~\eqref{eq:newRep}, since
those generate octagons, while self-contractions do not. Since we are
only off-diagonally contracting $\mathbb{X}$ with $\bar{\mathbb{Y}}$ and
$\mathbb{Y}$ with $\bar{\mathbb{X}}$, we can swap $\bar{\mathbb{Y}}$ and
$\bar{\mathbb{X}}$ and replace the off-diagonal by a purely diagonal
propagator  equal to
$\oct$ for the matrices $\mathbb{X}$ and $\mathbb{Y}$.
Finally, we can rescale the propagator to $1$ by taking all factors
of~$\oct$ out of the correlator. In this way, upon expanding the
logarithms, we obtain~\eqref{Znon}.

The representation~\eqref{Znon} immediately leads to a very compact
expression for our correlator in the large octagon limit, since the
dependence on~$k$ is explicit, and thus the Borel-transform procedure can be
done straightforwardly, yielding
\begin{equation}
\mathcal{A} =
\frac{1}{\zeta_1 \zeta_2 \zeta_3 \zeta_4}
\sum_{n=1}^\infty
\frac{(\zeta_1 \zeta_2 \zeta_3 \zeta_4 \oct^2)^n}{n^2\,{\color{red} (n-1)!^4}}
\langle\tr((\mathbb{X}\mathbb{Y})^n)\tr((\bar{\mathbb{X}}\bar{\mathbb{Y}})^n)\rangle_\text{two faces}
\,.
\label{AnonBPS}
\end{equation}
The two-point function in this expression can be evaluated
analytically at any $N$, that is for any number of faces, starting from
its single-complex-matrix counterpart
\begin{equation}
\vev{\tr(\mathbb{X}^n)\tr(\bar{\mathbb{X}}^n)} =
\sum_{k=1}^n \prod _{i=1}^k (i+N-1)  \prod_{m=1}^{n-k} (N-m)
\,.
\label{hookSum}
\end{equation}
This expression is derived by
decomposing each trace into characters of hook representations
(generating one sum per trace) and then using character
two-point function orthogonality, thus killing one of the two sums.
See for instance formula~(B.2) in~\cite{Beisert:2002bb}.
The final sum over hooks is~\eqref{hookSum}. We want the same
expression with $\mathbb{X} \to \mathbb{X} \mathbb{Y}$. To find it, we
proceed as for the single-matrix case, except that we use so-called
fission relations (see \eg~\cite{Kostov:1997bs}) to open up characters
$\chi_\lambda(\mathbb{X} \mathbb{Y})$ into
$\chi_\lambda(\mathbb{X})\chi_\lambda(\mathbb{Y})$ upon doing the
relative matrix angle integral between the two matrices. Each
character thus splits into two, so the
representation~\eqref{hookSum} ends up being modified to
\begin{equation}
\lrvev{\tr((\mathbb{X}\mathbb{Y})^n)\tr((\bar{\mathbb{X}}\bar{\mathbb{Y}})^n)}
=\sum_{k=1}^n\lrbrk{\prod _{i=1}^k(i+N-1)\prod_{m=1}^{n-k}(N-m)}^{\color{red}\!2}
\,.
\label{hookSum2}
\end{equation}
We can now simply expand the summand at small $N$ to read off the
leading $N\to 0$ term, which is precisely the required two-face
contribution. Plugging that into~\eqref{AnonBPS}, we obtain our full
correlator
\begin{equation}
\mathcal{A} =
\frac{1}{\zeta_1 \zeta_2 \zeta_3 \zeta_4}
\sum_{g=0}^\infty
\frac{(\zeta_1 \zeta_2 \zeta_3 \zeta_4 \oct^2)^{g+1}}{(g+1)^2\,{g!^4}}
\sum_{m=0}^g m!^2 (g-m)!^2
\,.
\label{AnonBPSSum2}
\end{equation}
We can re-sum this expression into%
\footnote{The two terms in the integrand can be combined to the
simpler expression $2stI_0(2\sqrt{s\,t\,\omega})/(s+t-1)$. However,
the integration of the small-$\omega$ expansion of this expression is
badly defined, whereas after expanding the integrand
in~\eqref{integralRep}, the integration directly gives~\eqref{AnonBPSSum2}.}
\begin{equation}
\mathcal{A} =
\oct^2
\int\limits_0^1ds \int\limits_0^1dt \lrsbrk{
\frac{s\,t}{s+t-1}\,I_0\bigbrk{2\sqrt{s\,t\,\omega}}
-\frac{(1-s)(1-t)}{s+t-1}\,I_0\bigbrk{2\sqrt{(1-s)(1-t)\,\omega}}
}\,, \label{integralRep}
\end{equation}
where $\omega=\zeta_1 \zeta_2 \zeta_3 \zeta_4 \oct^2$, and where
$I_0$ is the modified Bessel function of the first kind. Note that
this expression is valid for $\oct$ large, $k_j$ large, $\Nc$
large, but $\omega = k_1 k_2 k_3 k_4 \oct^2/\Nc^2$ can be either
large or not, it depends on how these limits are taken. In particular
we find
\begin{equation}
\mathcal{A} \simeq
\frac{1}{\zeta_1 \zeta_2 \zeta_3 \zeta_4}
\times
\begin{cases}
\omega+O(\omega^2) & \omega \ll 1\,,
\\[3mm]
\displaystyle\frac{e^{2\sqrt{\omega}}}{\sqrt{\pi}\,\omega^{1/4}} & \omega \gg 1\,.
\end{cases}
\label{asymptotics}
\end{equation}
As mentioned above, in the two-point function
representation~\eqref{eq:newRep}, each of the two logarithms represents one
of the large cyclic operators, and the two faces encode the remaining
two operators. This is how this matrix model representation encodes our
original four-point correlator. There are other representations of
this fully non-BPS result, which
are instructive in their own right, such as the original matrix model
where the four operators are faces, and also two new representations
in~\appref{app:NonBPS}: A one-point function with
three faces, and a three-point function with one face,
see~\eqref{eq:summaryQ1}. In all these matrix model representations,
we are after the
leading term in the $N\to 0$ limit.

Finally, let us stress once more the very important effect of the
Borel $1/g!$ arising from the large operator combinatorics. It is the
four $1/g!$ factors in~\eqref{AnonBPSSum2} that are responsible for the
very nice convergence of this expression. Indeed,
\beq
\sum_{m=0}^g m!^2 (g-m)!^2 \simeq 2(g!)^2 \simeq \sqrt{4\pi g} \,4^{-g}\, (2g)! \qquad \text{for} \qquad g\gg 1
\,,
\eeq
exhibiting the usual large-genus behavior expected in such
string/matrix theories~\cite{Shenker:1990uf,Polchinski:1994fq}. This growth
would otherwise render the matrix model perturbative expansion asymptotic, with missing
non-perturbative effects hinting at the physics of D-branes, see
above. Because of the extra combinatorial factors in (\ref{AnonBPSSum2}) we obtain instead a perfectly convergent
expression~\eqref{integralRep} with asymptotic behavior~\eqref{asymptotics}.

\subsection{Free Octagon Limit, \texorpdfstring{$\oct\to 1$}{O->1}}

Having analyzed the very non-perturbative $\oct\to0$ and
$\oct\to \infty$ limits, we turn to what should naively be a much
simpler limit: The free octagon limit, where $\oct\to 1$. In this
case, the diagonal and off-diagonal propagators of the
complex matrices in~\eqref{eq:newRep} are identical. Hence the
matrices $\mathbb{X}$ and $\mathbb{Y}$
can be identified, thus leading to a simpler matrix model
representation with a single complex matrix $\mathbb{X}$ and two
Hermitian matrices $\mathbb{M}_1$ and $\mathbb{M}_2$, with partition
function
\begin{equation}
\mathcal{Z} =
\frac{
\lrvev{
\tr\log\lrbrk{
\mathbb{I}-
\frac{k_2}{\mathbb{I}-k_2 \mathbb{M}_2} \bar{\mathbb{X}}
\frac{k_1}{\mathbb{I}-k_1 \mathbb{M}_1} \mathbb{X}
}
\tr\log\lrbrk{
\mathbb{I}-
\frac{k_3}{\mathbb{I}-k_3 \mathbb{M}_2}  \bar{\mathbb{X}}
\frac{k_4}{\mathbb{I}-k_4 \mathbb{M}_1} \mathbb{X}
}
}_{\text{two faces, } k^{4m}}
}{k_1 k_2 k_3 k_4}
\,,
\label{eq:newRepBPS}
\end{equation}
where
\begin{equation}
\vev{F}\equiv \int
[\mathcal{D}\mathbb{M}_1]
[\mathcal{D}\mathbb{M}_2]
[\mathcal{D} \mathbb{X}]
F \exp\Bigsbrk{
 \frac{1}{2}\tr \mathbb{M}_1^2
+\frac{1}{2}\tr \mathbb{M}_2^2+\tr \mathbb{X}\bar{\mathbb{X}}
}
\,.
\label{eq:MM2BPS}
\end{equation}
Once Borel transformed, this matrix model partition function computes
the tree level correlator ($\lambda=0$) of the operators~\eqref{ops}
at any genus order in the double-scaling limit where $k_j/\sqrt{\Nc}$
is held fixed with $k_j$ and $\Nc$ both taken to infinity. As before,
it is easy to expand this correlator to very high genus order. However,
compared to the previous cases, we were not able to either derive or
guess the all-genus expansion. It would be very interesting to find the
proper matrix model technology allowing us to compute the expectation
value~\eqref{eq:newRepBPS} in this amusing $N\to 0$ limit where the
two-face contribution dominates.

Perhaps we could even expect more, and actually compute the
correlator for any $\Nc$ in this free theory limit. Note that the
space-time dependence at tree level is completely fixed by R-charge
conservation and thus factors out, as there must be exactly $k_i$ propagators connecting each two
consecutive operators. The free-theory all-genus correlator is thus given by a matrix
model of two complex matrices that are simply the complex
scalars $Z$ and $X$ in $\mathcal{N}=4$ SYM. Perhaps this model can be solved
using two-matrix model techniques following \eg~\cite{Kazakov:1987qg,Kazakov:1996ae,Kazakov:1995gm,Kazakov:1996zm}.

A related observation stemming from the absence of any non-trivial
space-time dependence at tree level, and from the fact that complex
fields cannot self-contract, is that the free-theory correlator can also be though
of as a two-point function of a \emph{holomorphic} double-trace operator
$\op{O}=\tr\brk{{X}^{k_1+k_2}\,} \tr\brk{{Z}^{k_3+k_4}}$ with an
\emph{anti-holomorphic} double trace operator
$\op{O}'= \tr\brk{\bar{Z}^{k_4}\,{\bar{X}}^{k_1}}
\tr\brk{\bar{Z}^{k_3}\,{\bar{X}}^{k_2}}$. If we could decompose these
operators into restricted Schur polynomials as in~\cite{Corley:2001zk,Brown:2007xh}, we
could exploit their orthogonality to evaluate the free
correlator at finite~$\Nc$ and~$k_i$.

Another final option would be to try to compute the free correlator for many more values of
$k_i$ and $\Nc$, observe a pattern and guess the full result.

\section{Conclusions}
\label{sec:Conclusions}

In this work, we considered the four-point function%
\footnote{A sum over permutations is implicit for the operators with
two scalars $\tr(\bar Z^k\bar X^k)$.}
\begin{equation}
\frac{\vev{
{\tr\brk{{\color{black}\mathbold{\bar Z}^{k}}{\color{red}\mathbold{\bar{X}}^{k}}}(0)}\,
{\tr\brk{{\color{red}\mathbold{X}^{2k}}}(z)}\,
{\tr\brk{{\color{red}\bar{\mathbold{X}}^{k}}{\color{black}\mathbold{\bar Z}^{k}}}(1)}\,
{\tr\brk{{\mathbold{Z}^{2k}}}(\infty)}
}}{\text{same at $\lambda=0$ and $\text{genus}=0$}}
\equiv
\mathcal{A}(\zeta|\oct)
\label{eq:corrconcl}
\end{equation}
in the double-scaling limit where $\Nc$ and $k$ are both very large with
\begin{equation}
\zeta=\frac{k}{\sqrt{\Nc}}
\label{DS}
\end{equation}
held fixed.%
\footnote{In the main text, we discussed a more general set of
correlators with four different $k_j$, but for this summarizing discussion we
stick to the simpler case of equal $k_j\equiv k$, as in the introduction.}
This correlator $\mathcal{A}(\zeta|\oct) $ is very rich, but
still simple enough that we can say a great deal about it, and often
even about its all-genus re-summation.

The reason for this is a nice decoupling of the large $\Nc$ expansion combinatorics -- which are
encoded in the dependence of the function $\mathcal{A}$ on the effective coupling $\zeta$ and
on the octagon function $\oct$ -- and the finite 't~Hooft
coupling dynamics and conformal field theory geometry -- which enter
through the octagon function alone as
$\oct=\oct(z,\bar z|\lambda)$. We deal with the very
interesting dynamics of~$\oct$
in~\cite{TillPedroFrankVascoWorkInProgress},
while here we attacked the combinatorial problem.  In fact, the
separation of combinatorics
and dynamics relies on nothing but a little bit of supersymmetry, on
the large $\Nc$ limit, and on having large $R$-charges to play with. We
should therefore be able to find octagons and perform very similar -- if not
identical -- re-summations in other gauge theories with
less supersymmetry.

We found for instance that as $\oct\to 0$, the
correlator~\eqref{eq:corrconcl} simplifies to
\begin{equation}
\mathcal{A}=
\lrbrk{\frac{\sinh(\frac{3}{2}\zeta^{2})}{\frac{3}{2}\zeta}}^4
\,,
\label{eq:fin1}
\end{equation}
while as $\oct\gg 1$, we obtain instead
\begin{align}
\mathcal{A}&=\oct^2
\int\limits_0^1dt\int\limits_0^1ds\, \left[
\frac{t s}{t+s-1}\, I_0\bigbrk{2\sqrt{t\,s}\,\zeta^2 \oct} - \frac{(1-t) (1-s)}{t+s-1}\, I_0\bigbrk{2\sqrt{(1-t)(1-s)}\,\zeta^2 \oct}  \right].
\label{eq:fin2}
\end{align}
For general finite $\oct$, we could compute the function $\mathcal{A}$
through genus four, \ie as a Taylor expansion in $\zeta$ as
\begin{align}
\mathcal{A} = \oct^{2}
&+\zeta^4\lrbrk{
1+\sfrac{9}{2}\oct^2+\sfrac{1}{2}\oct^4
}
\nn\\
&+\zeta^8\lrbrk{
\sfrac{3}{2}
+\sfrac{607}{80}\oct^2
+\sfrac{97}{36}\oct^4
+\sfrac{1}{16}\oct^6
}
\nn\\
&+\zeta^{12}\lrbrk{
\sfrac{81}{80}
+\sfrac{7321}{1120}\oct^2
+\sfrac{953}{216}\oct^4
+\sfrac{5689}{12960}\oct^6
+\sfrac{5}{1296}\oct^8
}
\nn\\
&+\zeta^{16}\lrbrk{
\sfrac{459}{1120}
+\sfrac{75553}{22400}\oct^2
+\sfrac{44971}{12600}\oct^4
+\sfrac{5587171}{7257600}\oct^6
+\sfrac{2903}{86400}\oct^8
+\sfrac{31}{207360}\oct^{10}
}
\nn\\
&+\order{\zeta^{20}}
\,.
\label{eq:fin3}
\end{align}
These are our main results. It would be formidable to find the full form
of $\mathcal{A}$, interpolating between~\eqref{eq:fin1} and~\eqref{eq:fin2},
and reproducing~\eqref{eq:fin3} in the 't~Hooft expansion. Obtaining one
more simplifying point of data, such as the free correlator
$\oct\to 1$, might provide us with some inspired guess.

We conclude with some further comments on generalizations and future directions.

The hexagonalization prescription
of~\cite{Fleury:2016ykk,Eden:2016xvg,Eden:2017ozn,Fleury:2017eph,Bargheer:2017nne,Bargheer:2018jvq},
or the octagonalization prescription for large operators described
here, splits the study of correlators into a problem of combinatorics
of skeleton graphs and the dynamics related to filling in the faces of
these graphs by integrability-computed objects (octagons in our case).
In the dual graph picture, we end up taming these skeleton graphs with
matrix models with a small fixed number of faces that correspond to the
vertices in our correlator, since graph duality interchanges
vertices and faces. So we end up with matrix models where we are
interested not in a large $N$ expansion of the matrix model -- where
the maximal number of faces dominate in the 't~Hooft expansion -- but
rather in the small $N\to 0$ limit, which projects onto the required
correlators with a small number of faces. We developed some
technology for dealing with this interesting $N\to 0$ limit whenever
needed in our examples.
Matrix model dualities between large/small matrix rank
limits have been studied before in the context of intersection number
theory~\cite{Brezin:2000characteristic,Brezin:2007aa,Brezin:2007wv,Brezin:2010ar}.
It would be
fascinating to further explore the mathematics of these dualities, as
well as their use in our context.

\begin{figure}[ht]
\centering
\begin{tabular}{c@{\quad\quad\quad}c}
\resizebox{.4\totalheight}{!}{\includegraphics[width=\textwidth]{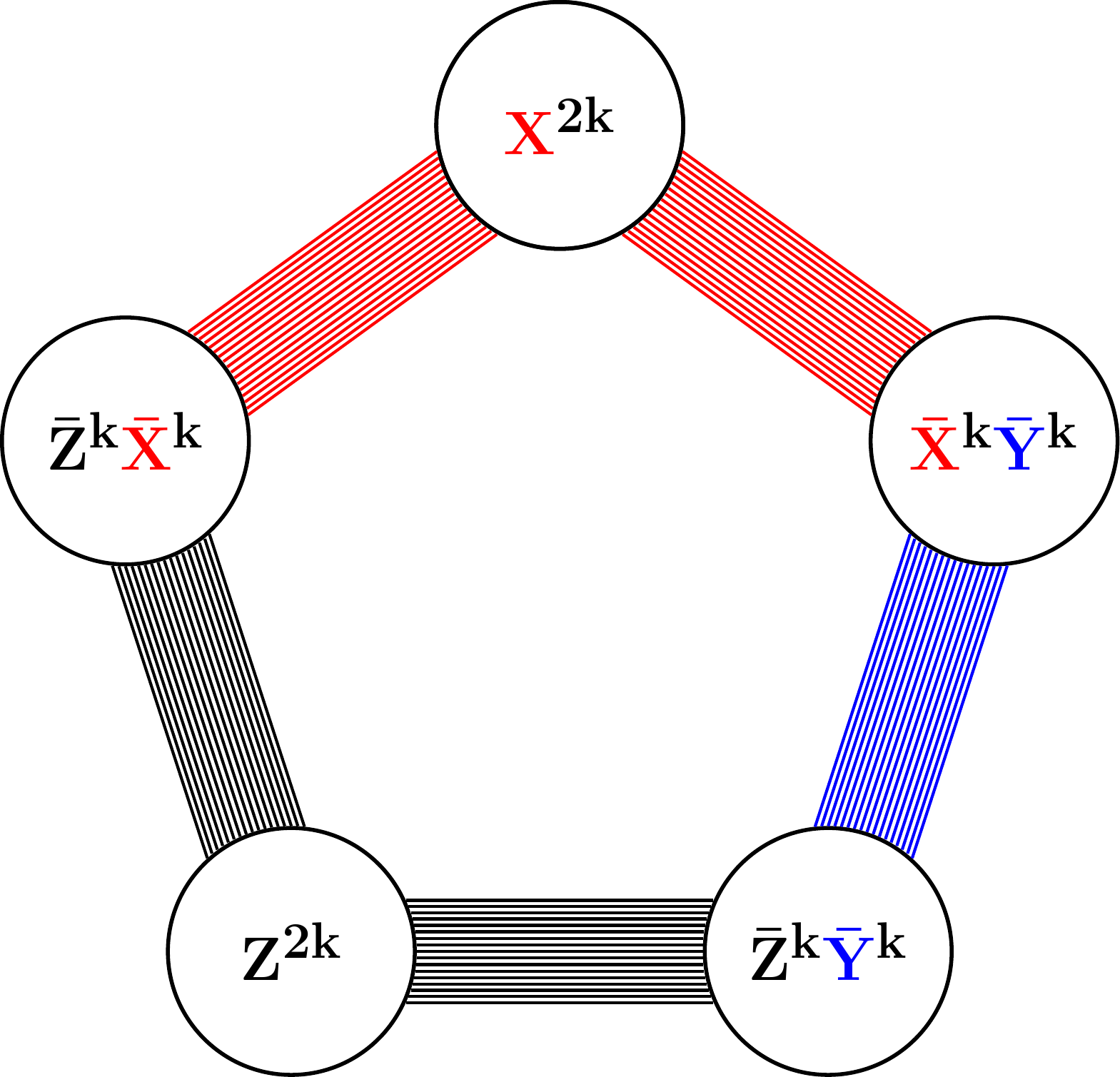}} &
\resizebox{.47\totalheight}{!}{\includegraphics[width=\textwidth]{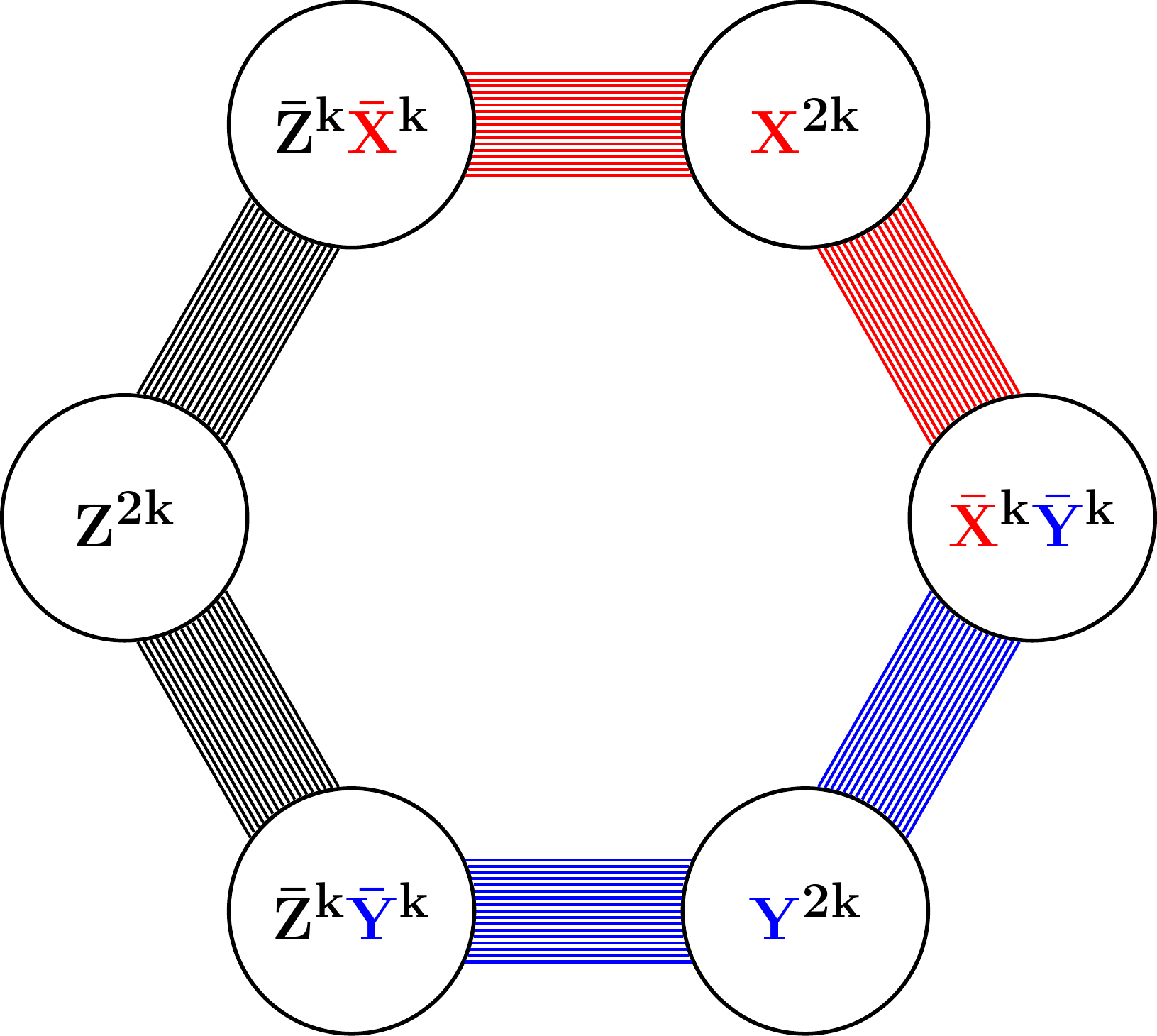}} \\
(a) & (b)
\end{tabular}
\caption{Polarized (a) five-point and (b) six-point functions}
\label{fig:Five-Six}
\end{figure}

We dubbed our correlator as cyclic, because the R-charge polarizations of the
operators are chosen such that operator $\op{O}_i$ is forced to connect
only to operators $\op{O}_{i \pm 1}$. Since we have six scalars in
$\superN=4$ SYM, we
can as well construct generalized cyclic configurations with five and with
six operators (but not more), see~\figref{fig:Five-Six}.
What happens to these higher cyclic correlators
in the double scaling limit~\eqref{DS}? Again, because of the large charges,
the contributing graphs will have to be maximal, which means that all
faces are bounded by a minimal number of edges.
However, contrary to the four-point case, the indivisible faces are not just
squares (\ie octagons). We now have non-BPS pentagons (for the five-point function)
and non-BPS hexagons (for the six-point function) besides the various
BPS squares (there are $10$/$12$ different BPS squares for the
five/six point function).
For a skeleton graph made of $n\subrm{s}$ squares, $n\subrm{p}$ pentagons, and
$n\subrm{h}$ hexagons, the number of edges is $n\subrm{e}=(4n\subrm{s}+5n\subrm{p}+6n\subrm{h})/2$.
By Euler's formula for~$n$ vertices at genus~$g$, we find for the
number~$n\subrm{e}$ of edges
\begin{equation}\label{eq:Euler5&6}
\frac{n\subrm{e}}{2}=2g-2+n-\frac{n\subrm{p}}{4}-\frac{n\subrm{h}}{2}
\,.
\end{equation}
Hence maximizing the number of edges requires to saturate the tiling of the
surface with squares only! That means that these correlators in the
double-scaling (DS)
limit actually have no coupling dependence, they are purely given by
BPS quadrangulations, which we found in~\appref{sec:bps-part-all-cyclic}.
Explicitly, we find
\begin{equation}
\frac{G_5^{\text{DS}}}{G^{\text{DS}}_{\text{5,free,planar}}}
=\lrbrk{\frac{\sinh(\frac{3}{2}\zeta^{2})}{\frac{3}{2}\zeta}}^5
\,,\qquad
\frac{G_6^{\text{DS}}}{G^{\text{DS}}_{\text{6,free,planar}}}
=\lrbrk{\frac{\sinh(\frac{3}{2}\zeta^{2})}{\frac{3}{2}\zeta}}^6
\,.
\end{equation}
Of course, we could look for subleading corrections to the double-scaling
limit where interesting coupling dependence would show up.
This would be particularly interesting for the six-point case, since we
expect the non-BPS hexagon -- being akin to a four-dimensional
six-point function -- to probe the genuine bulk-point singularity in
four dimensions~\cite{Maldacena:2015iua}.

We could also consider non-cyclic operators, and in particular
maximally connected configurations where all operators share
propagators with all others.
Here, the relevant maximal graphs will be BPS triangulations, and the
relevant scaling would be $k \sim \Nc^{1/3}$. It would be nice to
consider this limit and the corresponding matrix model.

In this last example, as well as in the previous five-point and six-point cyclic
function examples, we end up with tessellations where all building
blocks are BPS polygons with trivial expectation value $=1$. We expect
them to become interesting functions of the coupling as we twist the
theory, thus breaking supersymmetry.
An extreme and very interesting
example to analyze would be the fish-net
deformation~\cite{Gurdogan:2015csr,Caetano:2016ydc}, whose hexagons
have been introduced in~\cite{Basso:2017jwq,Basso:2018cvy}.

It is common to think of the sums over skeleton graphs in
hexagonalization and octagonalization prescriptions as some sort of
discretization of the string moduli
space~\cite{Fleury:2016ykk,Bargheer:2018jvq}. The picture could be slightly
different in a string-bit-like description, as recently put forward
in the context of $\grp{AdS}_5\times\grp{S}^5$ strings
in~\cite{Berkovits:2019ulm}, and in the fish-net theory
in~\cite{Gromov:2019aku}. In~\cite{Berkovits:2019ulm}, for instance, the
underlying theory is topological, and the sum over ways of connecting
the various string bits could perhaps be identified with the summation
over the various skeleton graphs. It would be nice to see the hexagons
and octagons more explicitly in the language of these works.

Everything so far was about very large operators.
Can we go beyond the large-operator limit and construct a dual matrix
model formulation of $\mathcal{N}=4$ SYM, with hexagons as vertex
building blocks describing any correlation function at any genus and
any coupling? In a way, it would be a
concrete gauge theory realization of the very inspiring proposal~\cite{Kazakov:2000ar}.
\textit{All our dreams can come true, if we have the
courage to pursue them}, said Walt Disney. So we should try.

\section*{Acknowledgements}

We thank
Nathan Berkovits,
Freddy Cachazo,
Jo{\~a}o Caetano,
Simon Caron-Huot,
Kevin Costello,
Davide Gaiotto,
Vasco Gon\c{c}alves,
Andrea Guerrieri,
Thiago Fleury,
Ivan Kostov,
Juan Maldacena,
Robert de Mello Koch,
Jo{\~a}o Penedones,
Amit Sever,
Paul Wiegmann,
Karen Yeats,
Sasha Zhiboedov,
and especially Vladimir Kazakov and Shota Komatsu
for numerous enlightening discussions and
suggestions. Research at the Perimeter Institute is supported in part
by the Government of Canada through NSERC and by the Province of
Ontario through MRI.
This work was additionally supported by a grant from the Simons
Foundation \#488661.

\appendix

\protect\addtocontents{toc}{\setcounter{tocdepth}{1}}

\section{Constructing Graphs Explicitly}
\label{sec:generating-maximal-graphs}

In the following, we want to explicitly construct all skeleton graphs
up to a given genus. Listing the explicit graphs will allow us to
compute the polynomials $P_{4g|g+1}(k_1,k_2,k_3,k_4)$ entering the
correlator~\eqref{eq:generalKs}, and hence will
provide an important cross-check of the results obtained with the help
of matrix model techniques in~\secref{sec:matrix-model-large}
and~\secref{sec:Simplifications}. Moreover,
constructing all contributing graphs explicitly is of more general
interest: In the present paper, we consider the case where all bridges
(propagator bundles) contain a large number of propagators, such that
all faces are isolated from each other (the sum over mirror
excitations\,/\,open string states reduces to the vacuum\,/\,ground
state). Hence for the purpose of this paper, it is sufficient to know
the \emph{number of graphs} that can be formed from a given set of
faces; how exactly these faces are arranged in each individual graph is irrelevant.
However, the more general hexagonalization
prescription~\cite{Fleury:2016ykk,Fleury:2017eph,Bargheer:2017nne,Bargheer:2018jvq}
for finite-charge operators requires to include non-trivial states
that propagate between faces, hence one \emph{does} have to know the
local structure of each graph explicitly. Hence it is important to
have techniques to construct the relevant graphs.

Mathematically, the graphs that we need to construct are \emph{ribbon
graphs} (also called fat graphs). In short, a ribbon graph is an
ordinary graph equipped with a cyclic ordering of the edges incident
to each vertex. More precisely, an ordinary graph consists of a set
$V$ of vertices, a set $H$ of half-edges, a map $s:H\to V$ that maps
each half-edge to the vertex that it is incident on, and a map
$i:H\to H$ (involutive, without fixed points) that maps each half-edge
to its other half. A ribbon graph is an ordinary graph $(V,H,s,i)$
together with a bijection $\sigma:H\to H$ whose cycles correspond to
the sets $s^{-1}(v)$ of half-edges incident on vertices $v\in V$. The
ordering of each cycle prescribes the ordering of the incident
half-edges at the vertex $v$. Topologically, each vertex of a ribbon
graph can be thought of as a disk, and each edge as a narrow rectangle
(or ``ribbon'', hence the name ``ribbon graph'') attached to two of
the vertex disks. The boundaries of these ribbons together with
segments of the vertex disks naturally form the \emph{faces} of a
ribbon graph (each face bounded by $n$ ribbons is a cycle of
$(i\circ\sigma)^{\circ n}$). Inserting an open disk into each of these
faces completes every ribbon graph to a compact oriented surface with
a definite genus, which we call the genus of the graph.

For our purposes, each vertex represents a single-trace operator, and
we think of it as a disk whose perimeter is formed by the ordered
fields within the trace. Each edge of a ribbon graph represents a
bundle of parallel propagators that connect a number of adjacent
fields within the single traces of the two operators connected by the
edge. We will alternatively call the edges ``propagator bundles'' or
``bridges'', depending on context. Because they are propagators, and
because our operators are local, we exclude edges that connect an
operator to itself. Also, we exclude ``parallel'' edges that connect
to identical operators next to each other (in a planar fashion): Since
the edges represent bundles of parallel propagators, such parallel
edges could be merged into a single edge (in the above language, such
parallel edges would form cycles of $i\circ\sigma\circ i\circ\sigma$).
Hence we only consider graphs where all faces are bounded by at least
three edges.

To summarize, we want to consider ribbon graphs with $n$ vertices
(punctures) of a given genus, ruling out edges that connect any vertex
to itself, and demanding that all faces are bounded by at least three
edges, \ie all faces are triangles or bigger polygons. In the
following, we call ribbon graphs with these properties just
``graphs''. We are particularly interested in graphs that are complete
in the sense that no further edge can be added to them (without
increasing the genus). We call such graphs \emph{maximal graphs}.
Obviously, every graph can be promoted to a maximal graph by adding
bridges. Hence conversely, every graph can be obtained from some
maximal graph by deleting edges. For this reason, we shall focus on
constructing the complete set of maximal graphs at a given genus.

It is easy to see that all faces of maximal graphs are either
triangles or squares. All bigger polygons can be split into smaller
polygons by inserting further bridges. But squares whose diagonally
opposite vertices are identical cannot be split, because we exclude
edges that connect any vertex to itself. Hence every face in a maximal
graph is either a triangle touching three different vertices, or a
square whose diagonally opposite vertices are identical. By imagining
fictitious edges that also split all squares into triangles, we can
think of every maximal graph of genus $g$ as a triangulation of a
genus-$g$ surface.

A given triangulation of a Riemann surface can be transformed into a
different triangulation by flipping some of its edges, where flipping
an edge means the transformation
\begin{equation}
\text{I}:
\qquad
\includegraphics[align=c]{FigFlipruleIleft.mps}
\quad\longrightarrow\quad
\includegraphics[align=c]{FigFlipruleIright.mps}
\,.
\label{eq:flip}
\end{equation}
Here, the circles are the vertices, and we
have labeled them arbitrarily.
Now, it is a mathematical theorem that the space of
triangulations of a surface of fixed genus and with a given
number of punctures is connected under the action of flipping edges
(see \eg~\cite{Hatcher:1991triangulations}).
In other words, any two triangulations are related by a sequence of
edge flips.%
\footnote{We thank Davide Gaiotto for discussions on this point.}
This means
that, starting with \emph{any single} triangulation, one can obtain
\emph{all} other triangulations by iteratively flipping edges. Since
we can associate a triangulation to every maximal graph, we can also
obtain all maximal graphs from a single maximal graph by flipping
edges. This requires flipping real edges as well as fictitious edges
that we added in order to split all squares into triangles. However,
we can shortcut the introduction of fictitious edges by supplementing
the flip operation~\eqref{eq:flip} with further transformations that
operate on squares. Namely, when an edge separates a triangle and a
square, we have to consider the following transformation:
\begin{equation}
\text{II}:
\qquad
\includegraphics[align=c]{FigFlipruleIIleft.mps}
\quad\longrightarrow\quad
\includegraphics[align=c]{FigFlipruleIIright.mps}
\,.
\label{eq:flipts}
\end{equation}
Here, the labels are again arbitrary, but their distribution is unique.
An edge may also separate two squares. Such edges can be transformed
in two inequivalent ways, and we have to include both of them:
\begin{equation}
\text{III}:
\qquad
\includegraphics[align=c]{FigFlipruleIIIleft.mps}
\quad\longrightarrow\quad
\left\{\,
\includegraphics[align=c]{FigFlipruleIIIright1.mps}
\,,\;
\includegraphics[align=c]{FigFlipruleIIIright2.mps}
\,\right\}
\,.
\label{eq:flipss}
\end{equation}
The flip move~\eqref{eq:flipts} cannot be undone by iterations of move I
without introducing self-contractions. In order to exhaust the space
of maximal graphs, we thus also need to include the inverse of~\eqref{eq:flipts}:
\begin{equation}
\text{IV}:
\qquad
\includegraphics[align=c]{FigFlipruleIIright.mps}
\quad\longrightarrow\quad
\includegraphics[align=c]{FigFlipruleIIleft.mps}
\,.
\label{eq:fliptsinv}
\end{equation}
Again, the labels are arbitrary, but their distribution matters and is unique.
Each of the transformations II--IV is the result of a sequence of
simple flip moves~\eqref{eq:flip} acting on real as well as fictitious
bridges (that split the squares). By the above considerations, it is
clear that the complete set of maximal graphs at a given genus can be
obtained by starting with an arbitrary maximal graph and iterating the
operations~\eqref{eq:flip}--\eqref{eq:fliptsinv} in all possible ways.

The result of the above discussion is the following iterative algorithm that
constructs all maximal graphs at a given genus:
\begin{enumerate}
\item Start with any maximal graph of the desired genus. This can for
example be constructed by iteratively adding random edges to the
empty graph until the target genus is reached, and then splitting all
faces of the resulting graph with as many further edges as possible.
\item For each edge of each graph in the list constructed in
the previous iteration step, generate a new graph by applying one of
the transformations~\eqref{eq:flip}--\eqref{eq:flipss} to that edge
(transformation~III generates two new graphs). For each pair of
adjacent edges with vertex structure as in~\eqref{eq:fliptsinv},
generate a new graph by applying transformation~IV.
\item The list of graphs constructed in the previous step may contain
graphs that are identical to graphs constructed in earlier iteration
steps. It can also contain several copies of identical graphs. Drop
all graphs that are identical to graphs already constructed earlier,
and drop all duplicates. The resulting list contains the new graphs.
\item Iterate steps 2--3 until the list of new graphs is empty, \ie
until all edge transformations only generate copies of graphs already
found earlier.
\end{enumerate}
We can implement this algorithm on a computer, and construct the
space of maximal graphs for various genera
and numbers of insertions (vertices). In order to reduce overcounting,
we treat all vertices as identical, \ie we use unlabeled vertices.%
\footnote{The algorithm works equally for labeled and unlabeled vertices.}
The size of the space of graphs grows
rapidly, see~\tabref{tab:numgraphs}.
\begin{table}
\centering
\begin{tabular}{c@{\;\;}rrrrrr}
&&\multicolumn{5}{c}{\quad\quad\; genus} \\[0.2ex]
&      & $0$ & $1$ & $2$ & $3$ & $4$ \\
\midrule
\multirow{9}*{\rotatebox{90}{number of insertions}}
&  $2$ : & $1$     & $1$      & $4$      & $82$     & $7325$ \\
&  $3$ : & $1$     & $3$      & $38$     & $661$    & \\
&  $4$ : & $2$     & $16$     & $760$    & $122307$ & \\
&  $5$ : & $4$     & $132$    & $18993$  & & \\
&  $6$ : & $14$    & $1571$   & $487293$ & & \\
&  $7$ : & $66$    & $20465$  & & & \\
&  $8$ : & $409$   & $278905$ & & & \\
&  $9$ : & $3078$  & & & & \\
& $10$ : & $26044$ & & & &
\end{tabular}
\caption{Numbers of maximal graphs for various genera and numbers
of insertions. Here, the vertices of the graphs are unlabeled, \ie all
vertices are treated as identical.}
\label{tab:numgraphs}
\end{table}
We note the following properties of maximal graphs of genus $g$ with
$n$ vertices:
\begin{itemize}
\item The planar two-point graph has one edge.
\item For $g=0$ and $n\geq3$, all maximal graphs consist of $2n-4$
triangles and no squares, they have $3n-6$ edges.
\item For $g\geq1$ and $n=2$, all maximal graphs consist of $2g$
squares and no triangles, they have $4g$ edges.
\item For $g\geq1$ and $n\geq3$, maximal graphs may consist of
triangles and squares, their maximum edge number is $6g+3n-6$ (no
squares, $2n+4g-4$ triangles), and their minimum edge number is
$4g+3n-7$ ($2g+1$ squares, $2n-6$ triangles).
\end{itemize}

A note on the implementation:
We found it convenient to represent ribbon graphs as
lists of vertices, where each vertex is an ordered list of incident
edges. For example, the graphs on the left in~\figref{fig:exampleDual}
can be represented in \mathematica as
\begin{lstlisting}[basicstyle={\small\ttfamily}]
graph[v[1,2], v[1,3], v[2,4], v[3,4]]
graph[v[1,2,3], v[1,4,2,5,6], v[3,7,8], v[4,8,5,6,7]]
graph[v[1,2,3,4,5,6], v[1,7,8,3,9,10,5], v[2,4,11,6,12], v[7,11,8,9,12,10]]
\end{lstlisting}
Here, the edges have been given arbitrary integer labels. The
bijection $\sigma$ is explicit in this representation, whereas the
incidence and half-edge identification maps $s$ and $i$ are implicit.
Of course, graphs in this representation are separately invariant
under \emph{(i)} permuting
the vertices \lstinline!v[..]! within \lstinline!graph[..]!, \emph{(ii)} rotating
the edge labels within individual vertices \lstinline!v[..]!, and \emph{(iii)}
relabeling the edges. When checking for equality of two graphs, these
invariances have to be taken into account. A brute-force way of
canonicalizing the representation is to tabulate over all permutations
of the vertices \lstinline!v[..]! as well as over all possible
rotations of the edge labels within each vertex, enumerating the edges
in order of appearance in each of the resulting representations, and to
select the lexicographically smallest representative of the set.

Now that we have obtained all maximal graphs at a given genus, it is
easy to construct \emph{all} graphs of that genus by iteratively
removing bridges in all possible ways, taking care to drop duplicate
graphs at each step. In particular, it is straightforward to obtain
all graphs that contribute to the four-point
correlator~\eqref{eq:generalKs}. Namely, the contributing graphs still
have a maximal number of edges, but now under the constraint that
$\op{O}_i$ only connects to $\op{O}_{i\pm1}$, but not to
$\op{O}_{i+2}$ (mod $4$). In other words, the four vertices of the
graph have to split into two pairs, where the members of each pair are
not connected by any edge. We call such graphs \emph{maximal cyclic
graphs}. To find them, we can take our list of maximal four-point
graphs, group the four vertices into pairs in all (three) possible
ways, and delete all edges connecting the members of each pair. Some
of the resulting graphs will not be maximal,%
\footnote{For example, if one of the deleted edges was adjacent to a
square, the resulting graph will have a non-minimal face and hence
cannot be a maximal cyclic graph.}
those have to be dropped (in practice, this can be done by keeping
only graphs with $4g+4$ edges). Following this procedure,
we find $6$, $215$, and $26779$ maximal cyclic graphs at genus
$1$, $2$, and $3$, respectively, which we attach in the file
\filename{maxcycgraphs.m}

Armed with these lists of maximal cyclic graphs, we can now construct
the polynomials $P_{4g|g+1}$.
Since we have treated all vertices as
identical (unlabeled) thus far, we first have to sum over all inequivalent
vertex labelings for each unlabeled graph. In addition, each labeled
graph comes with combinatorial factors from summing over all ways of
distributing the propagators on all edges (bridges) of the graph.
According to~\eqref{eq:combiFactors}, summing over the distribution of
$k_i$ propagators on $b_i$ bridges results in a factor
$k_i^{b_i-1}/(b_i-1)!$. Hence each labeled graph comes with a
combinatorial factor
\begin{equation}
\prod_{i=1}^4
\frac{k_i^{b_i-1}}{(b_i-1)!}
\,,
\label{eq:bridgefactors}
\end{equation}
where $b_i$ is the number of edges (bridges) connecting vertices
(operators) $\op{O}_i$ and $\op{O}_{i+1}$ (mod $4$) in the given graph.

There is one more point that we need to take into account: When we organize the
sum over all Wick contractions into a sum over skeleton graphs and a
sum over distributions of propagators on the edges of those skeleton
graphs, it may happen that two or more seemingly \emph{different}
distributions of propagators on the same skeleton graph may actually
represent \emph{identical} Wick contractions. The reason for this is
that we implicitly treat all edges as distinguishable (\ie labeled)
when we perform the sum over distributions of propagators. In
particular, this assumption is implicit in the
counting~\eqref{eq:combiFactors} leading to~\eqref{eq:bridgefactors},
therefore resulting in an overcounting that we have to compensate. At
the level of skeleton graphs, this overcounting manifests itself in
terms of non-trivial \emph{ribbon graph automorphisms}. Such
automorphisms are defined as follows: In a given ribbon graph (with
unlabeled vertices and edges), temporarily pick
unique labels for all vertices and edges, and mark a fixed point on the
perimeter of each of the vertices, in between any two adjacent
incident edges. There are many different
possible positions for these marked points. A non-trivial automorphism
is a combination of edge and vertex relabelings that transform the
graph with any other choice of marked points to the same graph with
the previously fixed chosen positions of marked points.%
\footnote{In gauge-theory terms, automorphisms map different choices
of ``beginnings/end'' of all operator traces to each other by
relabeling the operators and edges. See the final part of Section~2.2
in~\cite{Bargheer:2018jvq} for a more detailed definition with
examples.}
The set of automorphisms for a given ribbon graph $\Gamma$ form the
automorphism group $\aut\Gamma$. This group does not depend on the
initially chosen positions of marked points. In order to compensate
the overcounting explained above, one has to divide the
propagator-distribution factor~\eqref{eq:bridgefactors} by the size
$\abs{\aut\Gamma}$ of the automorphism group.%
\footnote{In order to find the automorphism group in practice for a
graph \lstinline!graph[..]! as represented above, we tabulate over all
cyclic rotations of the edge labels within individual vertices
\lstinline!v[..]!, and over all permutations of the vertices
\lstinline!v[..]! within \lstinline!graph[..]!. For each element of
the resulting list, we label the edges canonically (for example by
enumeration in order of appearance). We then collect identical
elements in the canonicalized list. The size of each group of
identical elements (all groups have the same size) is the size
$\abs{\aut\Gamma}$ of the automorphism group. The attached file
\filename{maxcycgraphs.m} also contains the explicit automorphism
factors.}
We find $(1,3,24)$ graphs with $\abs{\aut\Gamma}=2$ at genus
$(1,2,3)$, two graphs with $\abs{\aut\Gamma}=3$ at genus two,
and three graphs with $\abs{\aut\Gamma}=4$ at genus three. All other
graphs up to genus three have trivial automorphism group.

Now all that remains is to count within each graph $\Gamma$ the number
$p(\Gamma)$ of faces that touch all four vertices. Each of these faces
will be home to one octagon function~$\oct$,
see~\eqref{eq:squaretypes}. To construct the desired polynomials
$P_{4g|g+1}$, we have to sum over the set $\mathbold{\Gamma}_g$ of all
maximal cyclic ribbon graphs of genus $g$ with four vertices, and, for
each graph $\Gamma\in\mathbold{\Gamma}_g$, over all inequivalent ways
$\ell$ of assigning the operators $\op{O}_i$, $i=1,\dots,4$ to the
four vertices. The polynomials then are%
\footnote{In addition to the number of faces $\oct$, we can also count
the numbers of all other types of vertices~\eqref{eq:squaretypes} and
thus obtain a polynomial in all~$9$ types of faces. Doing so, we find
a complete match with the result
of~\secref{sec:all-quandrangulations}, again up to genus three.}
\begin{equation}
P_{4g|g+1}
=\sum_{\Gamma\in\mathbold{\Gamma}_g}
\frac{1}{\abs{\aut\Gamma}}
\sum_{\ell}
\prod_{i=1}^4
\frac{k_i^{b_i(\Gamma_\ell)-1}}{(b_i(\Gamma_\ell)-1)!}
\,\oct^{p(\Gamma)}
\,.
\end{equation}
Here, $b_i(\Gamma_\ell)$ is the number of edges connecting vertices
(operators) $\op{O}_i$ and $\op{O}_{i+1}$ in the labeled
graph~$\Gamma_\ell$. This concludes the construction of the
polynomials $P_{4g|g+1}$ from explicit graphs. We computed these
polynomials up to $g=3$ in this way, and found a perfect match with
the polynomials computed with matrix-model techniques as explained
in~\secref{sec:matrix-model-large} and~\secref{sec:Simplifications}.

\section{From Minimal to Maximal Graphs}

In this appendix we present a complementary approach to that of
\appref{sec:generating-maximal-graphs} on the construction of
skeleton graphs. We propose to start by finding the \textit{minimal}
graphs which are graphs with a single face or \textit{minimum} number
of edges for given fixed genus $g$ and number of vertices $n$. Using
these as a seed we can find all other graphs by adding new edges
recursively such that we do not change the genus of the original
graphs. This procedure stops when we saturate the graphs, such that
any additional edge would change the genus. This final stage
corresponds to the maximal graphs described in the previous appendix.

\begin{figure}
\centering
\resizebox{.5\totalheight}{!}{\includegraphics[width=\textwidth]{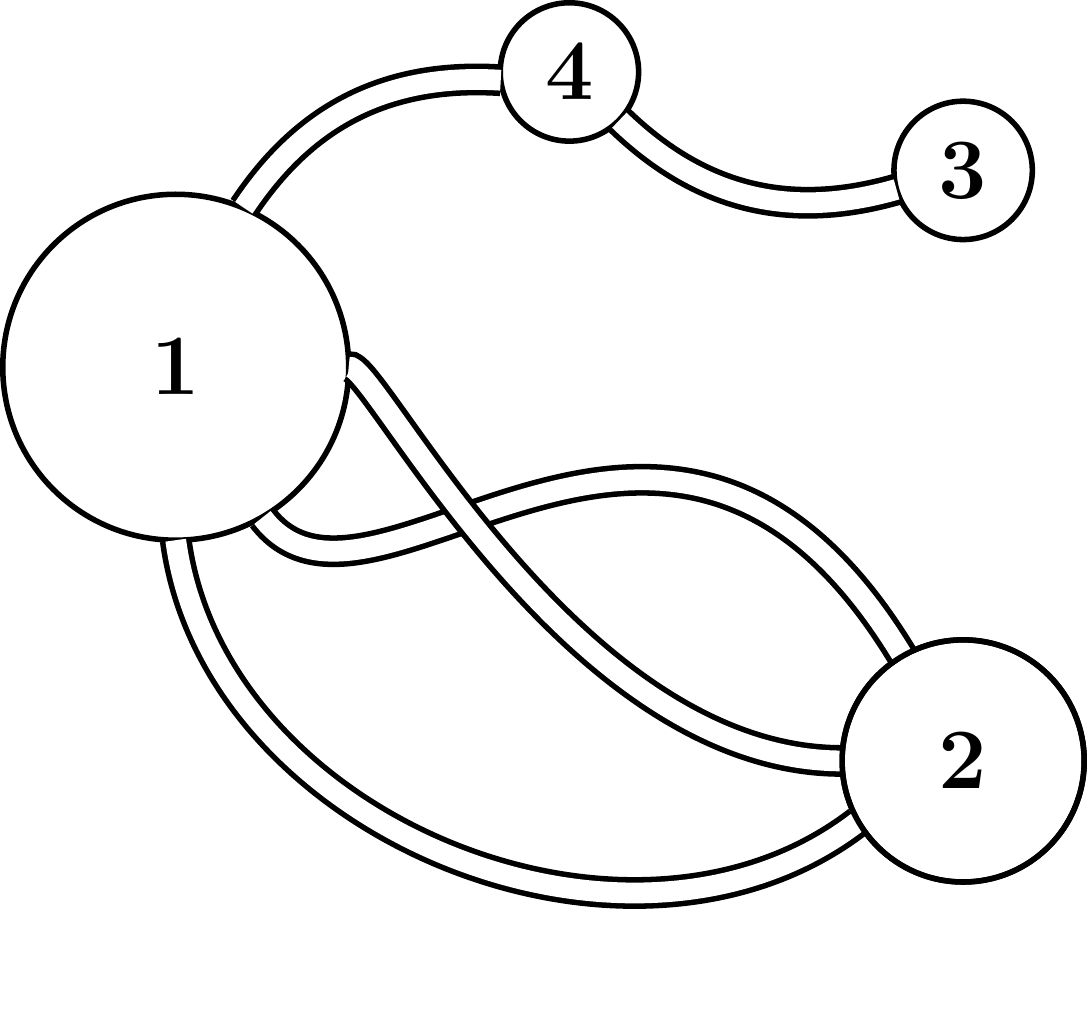}}
\caption{Minimal graph example with $n=4$ and $g=1$.}
\label{fig:exampleMin}
\end{figure}

\begin{figure}[ht]
\centering
\begin{tabular}{cc}
\resizebox{.35\totalheight}{!}{\includegraphics[width=\textwidth]{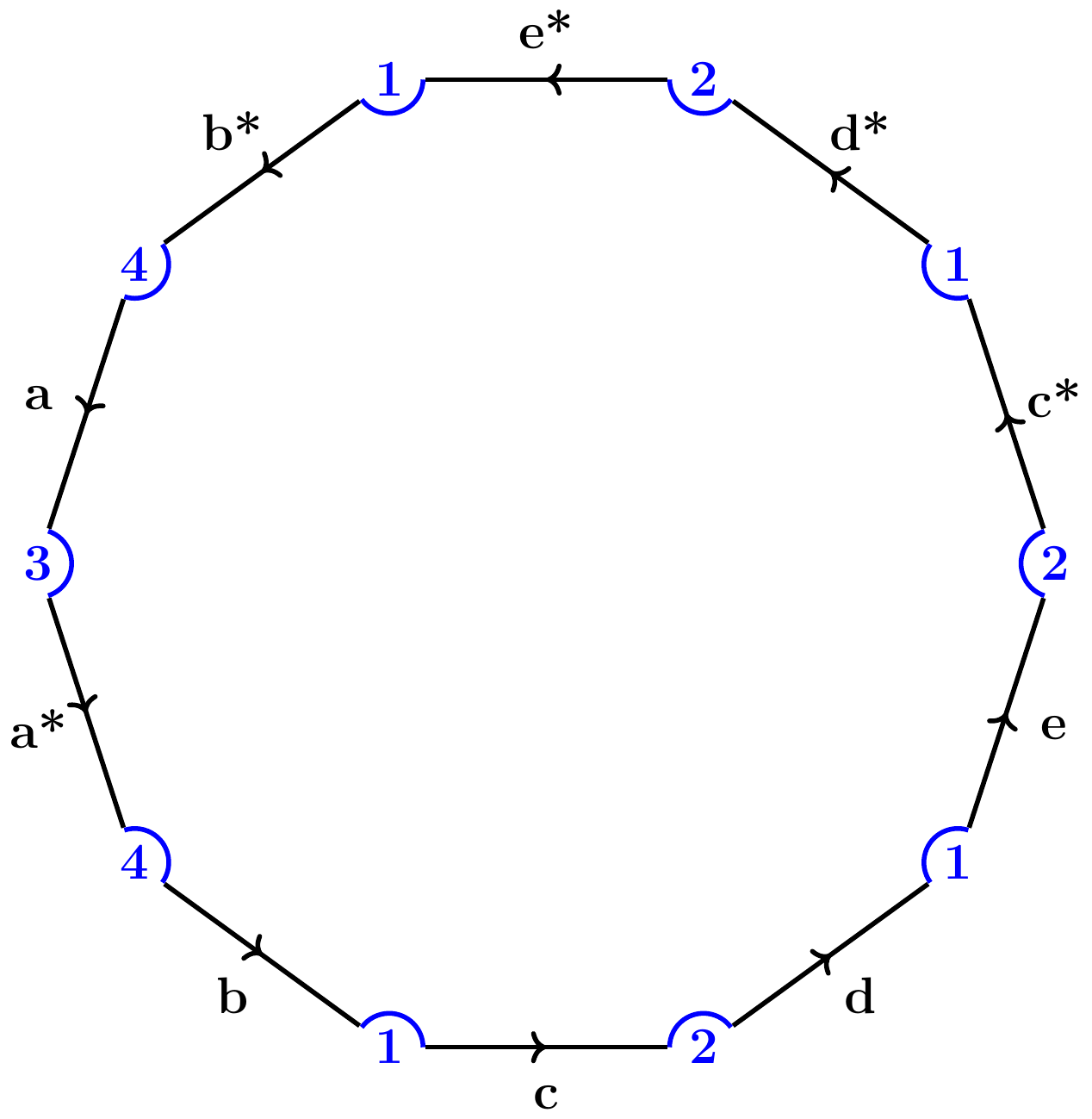}} &\quad\quad
\resizebox{.35\totalheight}{!}{\includegraphics[trim=100 0 200 150,clip,width=\textwidth]{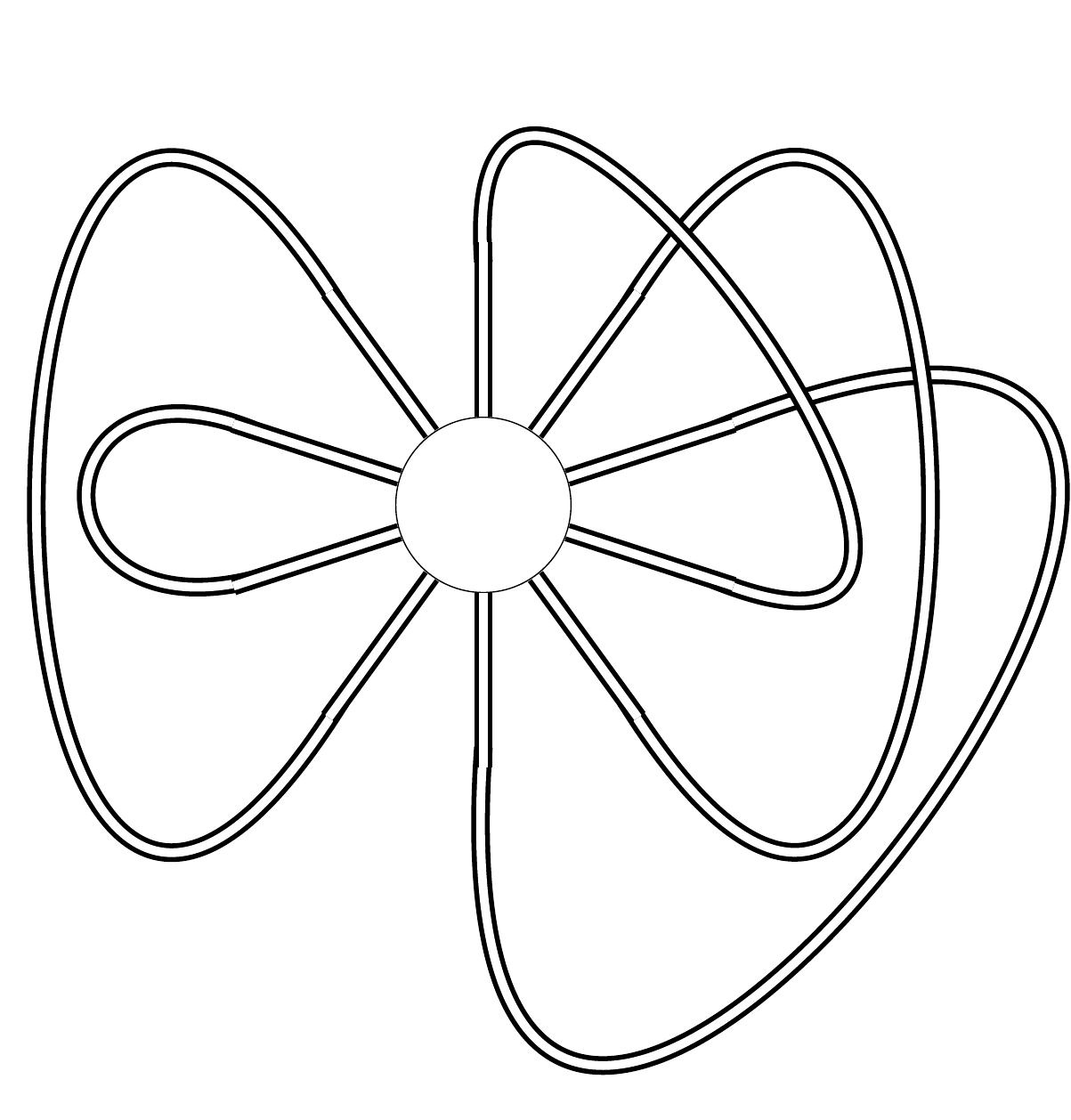}} \\
(a) & (b)
\end{tabular}
\caption{The genus-one minimal graph of~\figref{fig:exampleMin}
represented as the polygon of its single face (a), and its dual represented by Wick
contractions in a one-point matrix correlator (b).}
\label{fig:OneFaceandDual}
\end{figure}

A graph with a fixed number of vertices $n$ and genus $g$ is minimal
when it has a single face. From the Euler formula it follows that it
also has the minimum number of edges
\begin{equation}
2-2g=(F\subrm{min}=1)+(V=n)-E\subrm{min} \quad\to\quad E\subrm{min} = n+2g-1
\end{equation}
An instance of a minimal graph with four punctures and genus one
is presented in~\figref{fig:exampleMin}.
Another useful way of representing a minimal graph is given in
\figref{fig:OneFaceandDual} (a), where we present the single face of the graph as a
polygon whose sides represent the $2E\subrm{min}$
half-edges (these are of the
form $1\rightarrow 2$ and $2 \leftarrow 1$ which reconstruct an edge
$1 \rightleftharpoons 2$ in the fat graph),\footnote{Notice here we use a different notion of half-edge
compared to~\appref{sec:generating-maximal-graphs}} and its vertices given by
partitions of the original punctures of the graph.
This latter representation allows us to recognize that a minimal graph
can be found by starting with a $(2E\subrm{min})$-gon and identifying its
edges in a pairwise fashion such that we encapsulate $n$ vertices or
punctures. In more detail we follow these steps to construct the
minimal graphs:
\begin{itemize}
\item We start with a polygon with $2E\subrm{min}$ sides and some orientation.
\item We label the vertices with numbers from $1$ to $n$ in all
possible ways, allowing for repetitions in order to cover all vertices,
but we do not allow for neighboring vertices with the same label as
this would represent a self-contraction that we must dismiss. If we
consider special polarizations as in the main text, then we should also
dismiss the polygons with pairs of neighboring vertices labeled by operators
that cannot connect.
\item For each of the labeled polygons generated in the previous
step, we identify pairs of sides (half-edges) of the form
$1\rightarrow 2$ and $1\leftarrow2$ to reconstruct the edges
of the graph $1 \rightleftharpoons 2$. By doing so, all
the vertices of the polygon with the same label also get together to
reconstruct a puncture with that given label in the graph. We obtain a
consistent graph when we get a total of $n$ punctures with labels from
$1$ to $n$, with no repetition.
\end{itemize}
An alternative to this procedure can be found in the space of dual
graphs, where we trade faces by vertices. The dual of a minimal graph
has a single vertex, $E\subrm{min}$ edges, and $n$ faces. The advantage is
that all these dual $n$-faced dual graphs can be found from Wick
contractions in the Gaussian one-point function of a Hermitian matrix
$\langle \tr(\mathbb{M}^{2E\subrm{min}})\rangle$. For
instance see \figref{fig:OneFaceandDual} (b), each Wick contraction
there tells us how to identify the sides of the polygon in
\figref{fig:OneFaceandDual} (a). This dual point of view also
facilitates the counting of the minimal graphs as nicely explained
in~\cite{Zvonkin:1997} and derived in~\cite{HarerZagier:1986}. However
the counting in those
references has to be adapted to include labels in order to apply to
the counting of our minimal graphs. We do not pursue this here, as our
aim is only to provide a way to construct the minimal graphs.

Skeleton graphs with a higher number of edges can be simply
constructed by adding edges to the minimal graphs.
We would like to maintain the genus, so each additional edge must
increase the number of faces by one to satisfy the Euler formula
\beq
2 - 2g = (1+F\subrm{add})-n+(E\subrm{min}+E\subrm{add})
\quad \Rightarrow \quad
F\subrm{add}=E\subrm{add}
\,.
\eeq
To achieve this we simply start with the polygon representing a
minimal graph, and consider the additional edges as non-intersecting
diagonals of the polygon. These divide the polygon into sub-polygons
which represent the faces of the new non-minimal graphs.

Furthermore, we should only allow for diagonals that connect vertices
of the polygon with different labels, otherwise we would be including
self-connections. In the case of special polarizations, as considered in
the main body of this paper, we should also disallow diagonals representing
prohibited connections.

\begin{figure}
\centering
\begin{tabular}{c@{\qquad\quad}c}
\resizebox{.35\totalheight}{!}{\includegraphics[width=\textwidth]{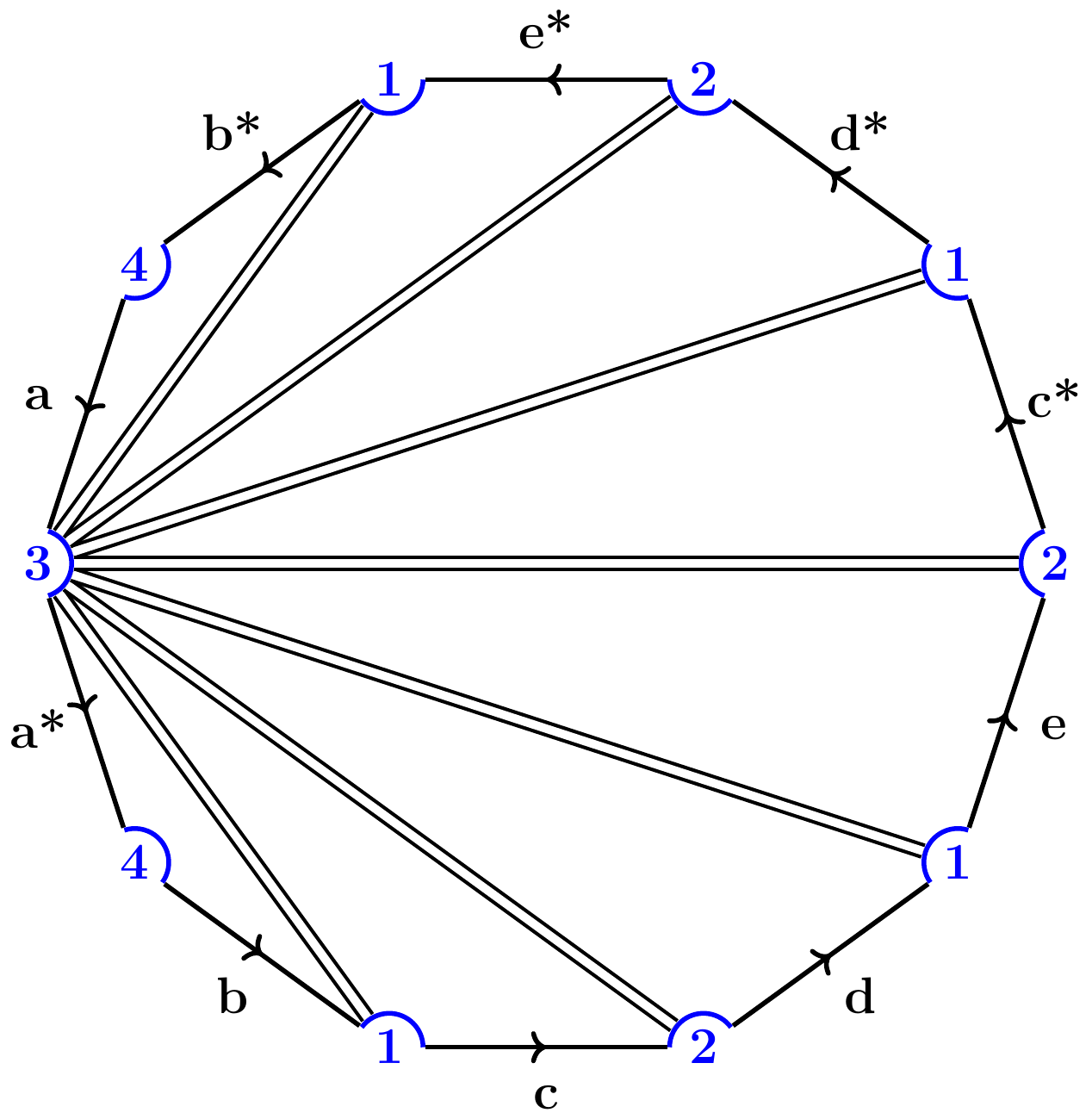}} &\resizebox{.35\totalheight}{!}{\includegraphics[width=\textwidth]{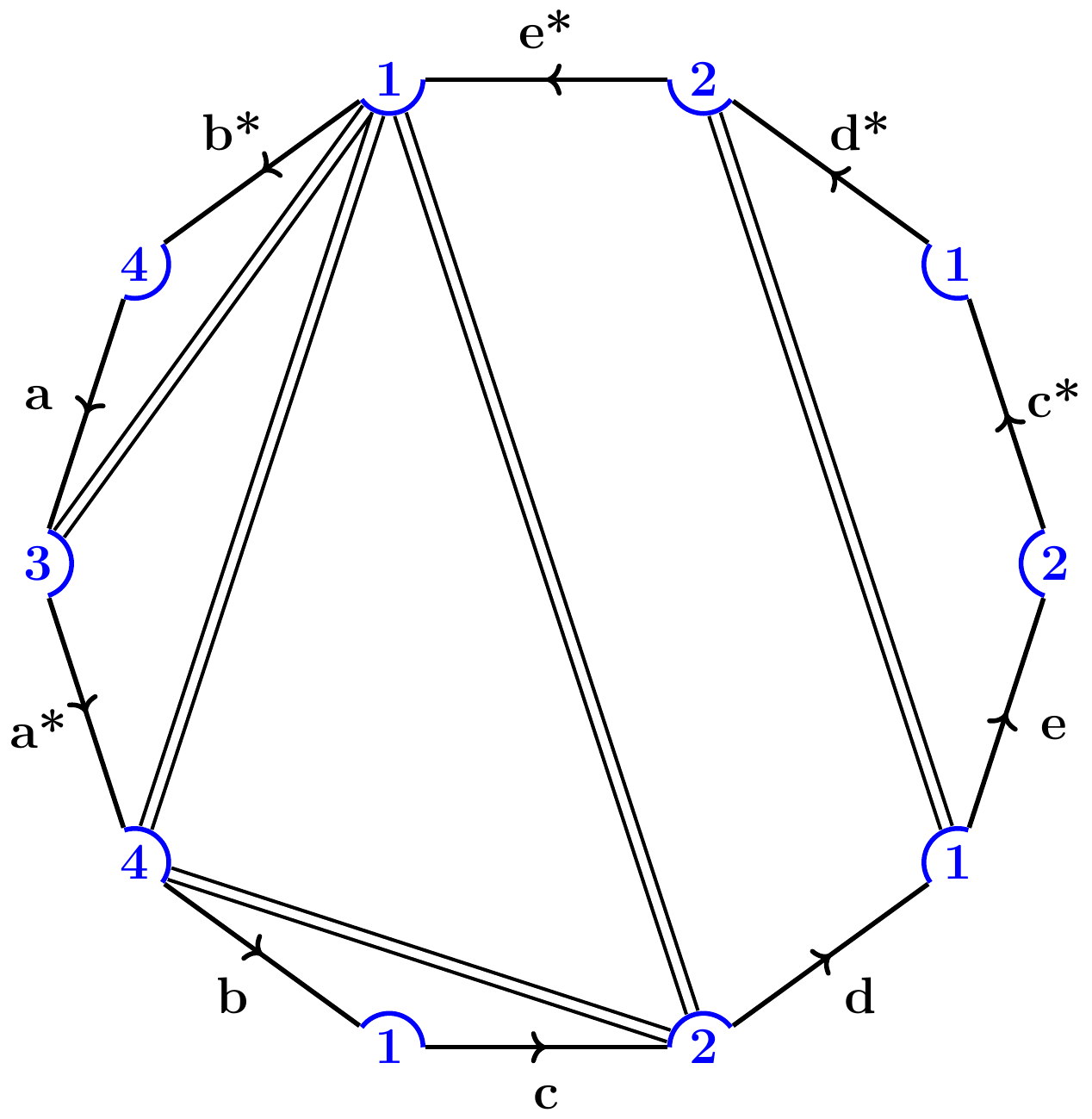}}
\\
(a) & (b)
\end{tabular}
\caption{Genus One: A maximal skeleton graph with only triangular
faces (a), and an edge-saturated skeleton graph with triangular and square
faces (b).}
\label{fig:MaximalExamples}
\end{figure}

Adding non-intersecting diagonals one by one to each polygon of a
minimal graph, we generate all skeleton graphs.
In general the saturation of the number of edges happens when we turn
on all possible non-intersecting diagonals forming a triangulation of
the polygon of a minimal graph, see \figref{fig:MaximalExamples} (a).
From this consideration it follows that the maximum number of edges
and faces a maximal graph can have are
\begin{align}
\label{eq:EmaxFmax}
E\subrm{max} &= (E\subrm{min}=n+2g-1) + (E\subrm{add}=2E\subrm{min}-3) = 3(E\subrm{min}-1)\nonumber\\
&= 3\,(n+2g-2)\,,\\
F\subrm{max} &= 2E\subrm{min}-2=2\,(n+2g-2)\,,
\end{align}
where the additional number of edges $E\subrm{add}$ simply corresponds to
the maximal number of non-intersecting diagonals in the
$(2E\subrm{min})$-gon and the maximum number of faces $F\subrm{max}$ is the
number of triangles.

Due to the restriction of no self-connections, the saturation of edges can also
happen before we reach the maximum value of edges~\eqref{eq:EmaxFmax}.
This is the case for the maximal graph in~\figref{fig:MaximalExamples}
(b) represented by a tessellation containing both triangular and
square faces.

In order to find all maximal graphs, we need to find all ways of
triangulating the polygons of the minimal graphs. This can be achieved
by following a recursive procedure of bifurcation of polygons.
Performing this procedure, we generate the list of all maximal graphs
starting with the minimal graphs as a seed. We obtained results up to
genus $3$ which confirm the maximal graph generating
algorithm of~\appref{sec:generating-maximal-graphs}. The disadvantage is that
the final list of graphs is redundant, since some originally
different minimal graphs get identified after adding new edges. In
practice we noticed that we only need to consider triangulations of
relatively few minimal graphs to obtain the full list of
maximal graphs. It would be nice to better
understand how to single out a minimal subset of all
minimal graphs that generates all maximal graphs.

\section{Counting Quadrangulations Including Couplings}

\subsection{Introduction}

For the correlator studied in this paper we have specific
polarizations that restrict the connections to be only between
neighbors $1-2-3-4-1$. This condition dismisses triangles, so we only
need to consider squares to find the corresponding maximal graphs that dominate in the
double scaling limit (DSL) considered in the main text.

\begin{figure}[ht]
\centering
\begin{tabular}{cc}
\resizebox{.35\totalheight}{!}{\includegraphics[width=\textwidth]{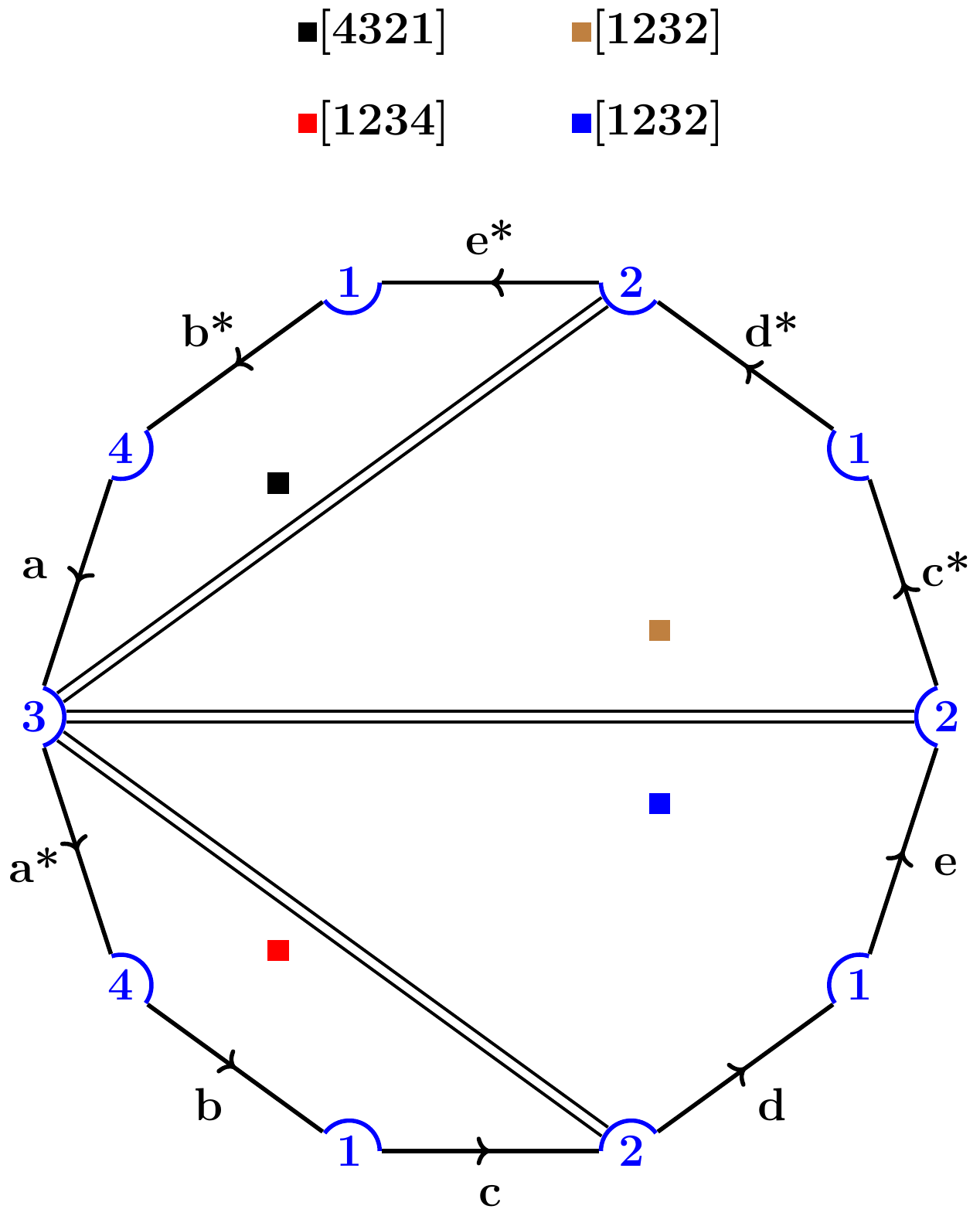}} &\quad\quad
\resizebox{.35\totalheight}{!}{\includegraphics[trim=100 0 200 100,clip,width=\textwidth]{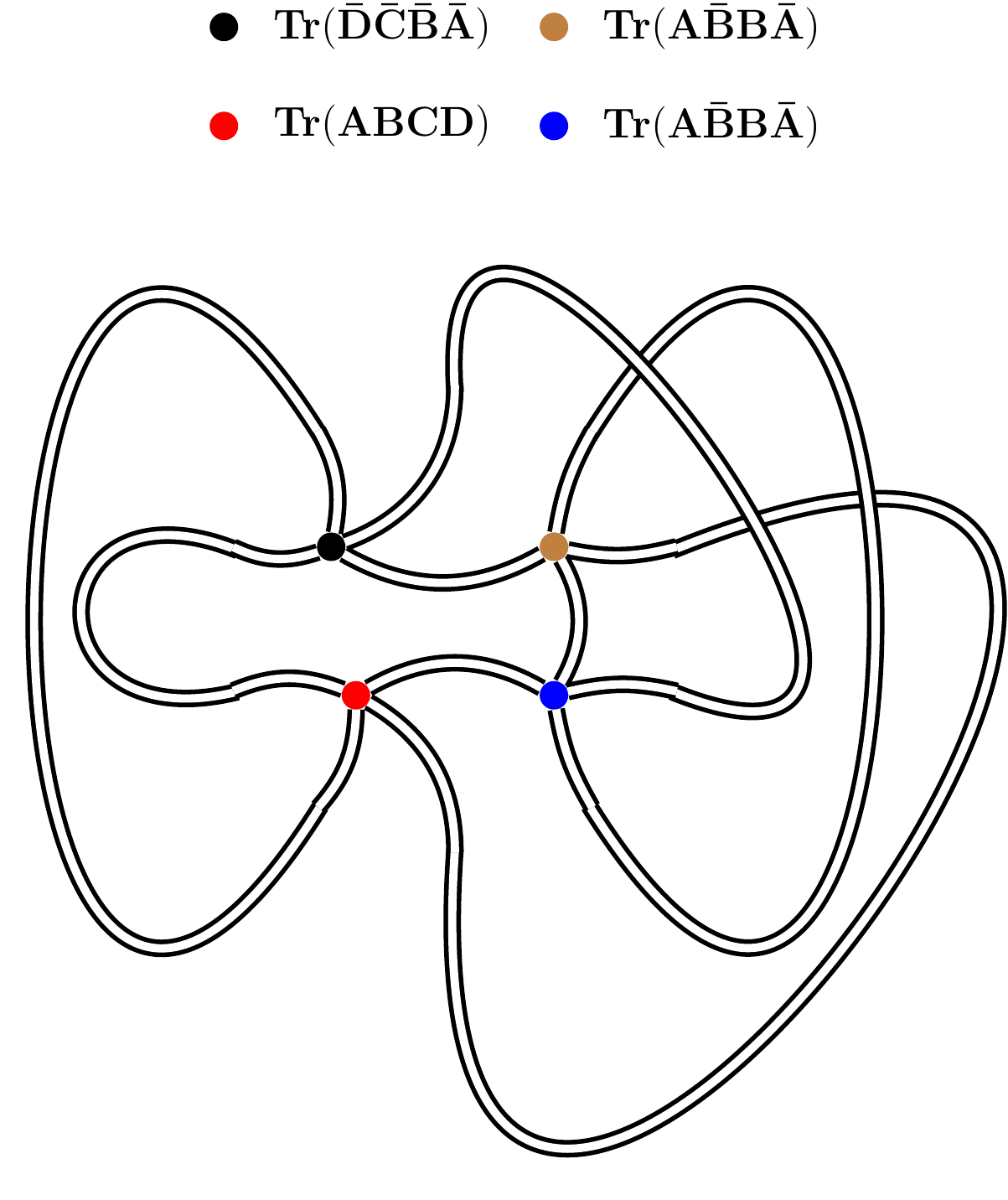}} \\
(a) & (b)
\end{tabular}
\caption{A genus-one quadrangulation and its graph dual as Wick contractions of four-valent vertices.}
\label{fig:QuadrangulationAndDual}
\end{figure}

In \figref{fig:QuadrangulationAndDual} (a) we present a genus-one
quandrangulation obtained from the minimal graph of \figref{fig:OneFaceandDual} (a) by adding
non-intersecting charged-allowed diagonals only, or from the (truly)
maximal graph in \figref{fig:MaximalExamples} (a) by erasing the
charge-disallowed connections $1-3$ and $2-4$.

\begin{figure}[ht]
\centering
\resizebox{2.5\totalheight}{!}{\includegraphics[width=\textwidth]{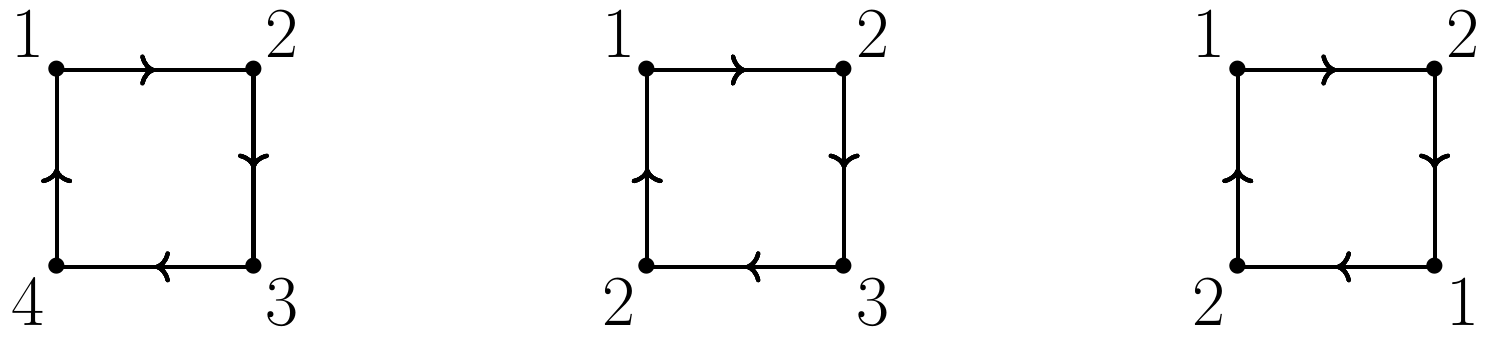}}
\caption{The three types of squares that enter a quadrangulation. The
vertices denote the punctures.}
\label{fig:3squares}
\end{figure}

The squares entering a quadrangulation of our correlator can be
classified according to the labels (operators 1,2,3 or 4) at its
vertices. We have three types of squares as presented
in~\eqref{eq:squaretypes} and in~\figref{fig:3squares}.
The first type includes the non-BPS squares
$[1234]$ and $[4321]$ that evaluate to
$1$ in the free theory, and to the octagon function $\oct$ when the
coupling is turned on. The other two types are BPS squares of them
form $[abcb]$ and the other four of form $[abab]$. These latter
squares still
evaluate to $1$ when turning on the coupling.

\begin{figure}[ht]
\centering
\resizebox{1.5\totalheight}{!}{\includegraphics[width=\textwidth]{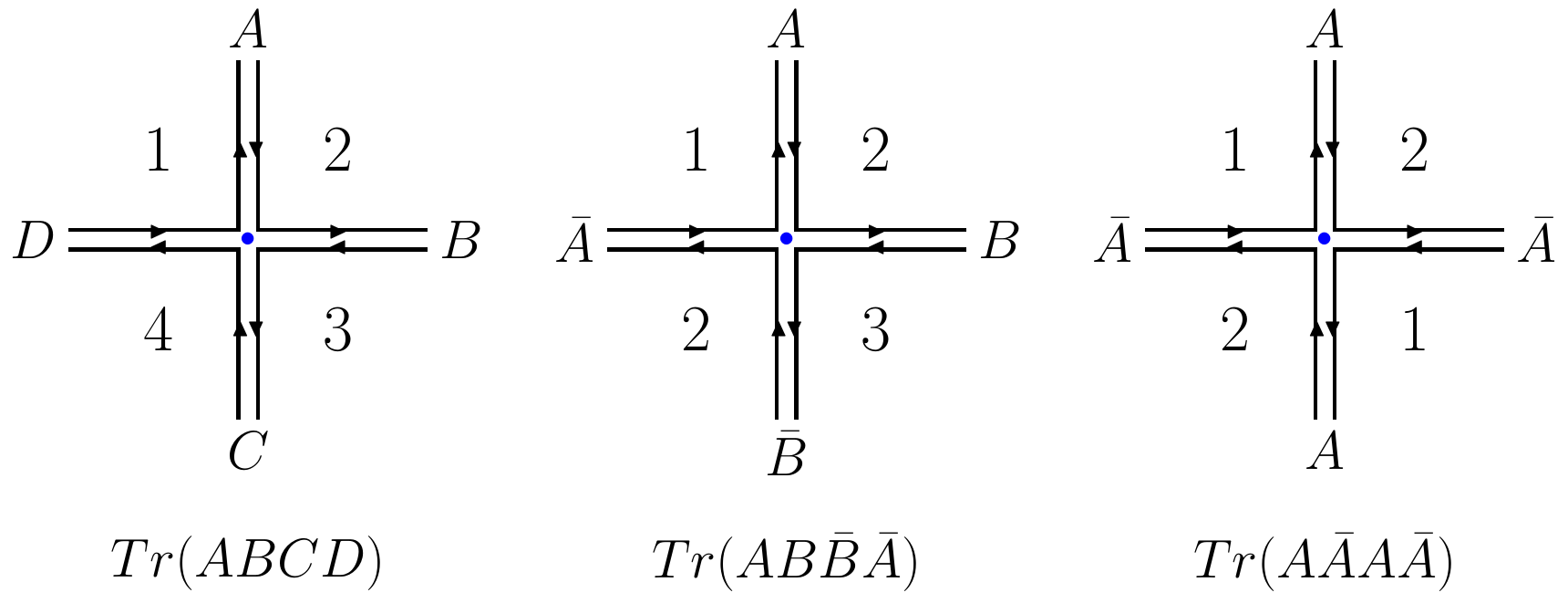}}
\caption{The three types of squares that enter a quadrangulation. The
original vertices $1,2,3,4$ are now faces tha lie between the new
edges.}
\label{fig:Vertices4}
\end{figure}

In order to find the graph's form by gluing these squares, we prefer to
work in the dual space, where these faces are traded by four-valent
vertices, see \figref{fig:Vertices4}. In this dual space, we have a total of 10 vertices,
which define a matrix model with action
\begin{align}
\label{eq:SquaresCouplings}
S &= \,-\tr A\bar{A}-\,\tr B\bar{B}-\,\tr C\bar{C}-\,\tr D\bar{D} +\oct\,\left(\tr(ABCD) +\,\tr(\bar{D}\bar{C}\bar{B}\bar{A})\right)\nonumber\\
&\qquad + \frac{\alpha_{1}}{2}\,\tr(A\bar{A}A\bar{A})+ \frac{\alpha_{2}}{2}\tr(B\bar{B}B\bar{B})+ \frac{\alpha_{3}}{2}\tr(C\bar{C}C\bar{C})+ \frac{\alpha_{4}}{2}\tr(D\bar{D}D\bar{D})\nonumber\\
&\qquad
+\beta_{1}\,\tr(DA\bar{A}\bar{D})
+\beta_{2}\,\tr(AB\bar{B}\bar{A})
+\beta_{3}\,\tr(BC\bar{C}\bar{B})
+\beta_{4}\,\tr(CD\bar{D}\bar{C})
\end{align}
Unlike the action in \eqref{eq:Sint}, which only includes the coupling
$\oct$ for non-BPS squares, here we include couplings
associated to each four-valent vertex to distinguish each type of
square. This gives the advantage of keeping track of the specific
squares that form a quadrangulation, which can help recognizing
symmetries or patterns essential for a genus resummation.

In this dual space, a quadrangulation is given by Wick contractions in
a Gaussian correlator as represented in
\figref{fig:QuadrangulationAndDual} (b). The correlator of this
particular set of (dual) vertices can be
explicitly computed as:
\beq
\frac{1}{2!}\left\langle \tr(ABCD) \tr(\bar{D}\bar{C}\bar{B}\bar{A})\tr(AB\bar{B}\bar{A})^{2}\right\rangle = N^{4} + 12 N^{6} + \frac{9}{2} N^{8} +\frac{1}{2} N^{10}
\eeq
On the left hand side we add a symmetry factor due to the two
identical vertices in the correlator. The result on the right hand
side is given as a polynomial in $N$, the rank of the complex
matrices, and from the exponents we can read off the number of faces
of the graphs constructed by Wick contractions. We are only interested
in the four-faced graphs, as they give the original four operators when
dualized back. Furthermore in order to guarantee the dualized four
faces give four different operators, we must have all $A,B,C,D$
matrices present in our correlators of four-valent vertices.

The relevant 4-faced partition function, extracted from the matrix
model with action~\eqref{eq:SquaresCouplings}, is explicitly given by
\beq\label{eq:partitionZ}
\mathbf{Z}(\oct,\alpha_{i},\beta_{i}) = \sum_{g=0}^{\infty}\sum_{T_{g}=\{t_{1},\cdots t_{2g+2}\}\in V_{4} } \frac{1}{\text{\text{sym}($T_{g}$)}}\left\langle \prod_{m=1}^{2g+2}\,t_{m}  \right\rangle_{4\text{ faces}}
\eeq
where we use the notation $\langle \cdots\rangle_{4\text{ faces}}$ to
indicate we extract the coefficient of $N^{4}$ only. The subset
$T_{g}=\{t_{1},\cdots t_{2g+2}\}$ is a list of $2g+2$ vertices,
which are picked from the list of ten four-valent vertices with
couplings
$V_{4}=\{\oct\,\tr(ABCD),\cdots
,\beta_{4}\,\tr(CD\bar{D}\bar{C})\}$ announced
in~\eqref{eq:SquaresCouplings}, with the extra condition of containing
all matrices $A,B,C,D$. The symmetry factor $\text{sym}(T_{g})$
contains a factor of $2$ for each vertex of the form
$\tr(X\bar{X}X\bar{X})$ and a factor $n!$ when we have $n$ identical
vertices $t_{m}$.

The partition functions $\mathbf{Z}$ of~\eqref{eq:partitionZ}
and $\mathcal{Z}$ of~\eqref{eq:Z1} or~\eqref{eq:RectangularZ} are
identical up to a simple replacement of couplings:
\beq\label{eq:repZ}
k_{1}k_{2}k_{3}k_{4}\,\mathcal{Z}(\oct,k_{i})= \mathbf{Z}(\oct,\alpha_{i},\beta_{i})\big{|}_{\alpha_{i}\to k_{i}^{2},\,\beta_{i}\to k_{i-1}k_{i},\,\oct^{2}\to \oct^{2}k_{1}k_{2}k_{3}k_{4}}
\eeq
As explained in~\secref{sec:matrix-model-large}, the partition function $\mathcal{Z}$ requires a
Borel-type transformation to give the cyclic correlator $\mathcal{A}$,
see \eqref{eq:corFromZ} . The analog transformation for $\mathbf{Z}$
defines the partition function
\beq\label{eq:Apartition}
\mathbf{A} = \sum_{g=0}^{\infty}\sum_{T_{g}=\{t_{1},\cdots t_{2g+2}\}\in V_{4} } \frac{1}{\text{\text{sym}($T_{g}$)}}\frac{1}{\text{weight}(T_{g})} \left\langle \prod_{i=1}^{2g+2}\,t_{i}  \right\rangle_{4\text{ faces}}
\eeq
where the sole difference with \eqref{eq:partitionZ} is the inclusion of the factorials:
\beq
\text{weight}(T_{g})\,=\, (n_{A}-1)! (n_{B}-1)!(n_{C}-1)! (n_{D}-1)!
\eeq
with $n_{X}$ counting the number of appereances of $X$ in the subset
$T_{g}$ of $2g+2$ vertices. Notice by construction we always demand
$n_{X}\geq 1$.

The partition function $\mathbf{A}$ of~\eqref{eq:Apartition} is
identified with the cyclic correlator $\mathcal{A}$ under the replacement
\beq\label{eq:repA}
\zeta_{1}\zeta_{2}\zeta_{3}\zeta_{4}\,\mathcal{A}(\oct,\zeta_{i}) =\mathbf{A}(\oct,\alpha_{i},\beta_{i})\big{|}_{\alpha_{i}\to \zeta_{i}^{2},\,\beta_{i}\to \zeta_{i-1}\zeta_{i},\,\oct^{2}\to \oct^{2}\zeta_{1}\zeta_{2}\zeta_{3}\zeta_{4}}
\eeq
with $\zeta_{i}={k_{i}}/{\sqrt{\Nc}}$.

By a direct computation of the correlators
$\langle\cdots\rangle_{4\text{ faces}}$ in~\eqref{eq:Apartition} we
obtain, up to genus one:
\begin{multline}
\label{eq:PartitionAg1}
\mathbf{A}(\oct,\alpha_{i},\beta_{i})
=\oct^{2}
+\frac{1}{2}\oct^{4}
\\
+\oct^{2}\left(\sum_{i=1}^{4}\left(\frac{1}{24}\alpha_{i}^{2}+\frac{1}{6}\alpha_{i}(\beta_{i}+\beta_{i+1})+\frac{1}{4}\beta_{i}^{2}\right) + \frac{1}{2}(\beta_{1}+\beta_{3})(\beta_{2}+\beta_{4})\right)
+\beta_{1}\beta_{2}\beta_{3}\beta_{4}
+\cdots
\end{multline}
where the dots indicate contributions from genus two and higher. This
latter expression can be compared with \eqref{eq:fin3} under the
replacement \eqref{eq:repA} and after setting $\zeta_{i}=\zeta$.

At higher genus, the correlators $\langle \cdots\rangle_{4\text{
faces}}$ become computationally more demanding, so in order to simplify
them we use integrating-in and -out operations that we describe in the
following section.

\subsection{Graph Operations}
\label{sec:integrating-in-and-out}

In order to simplify the correlators $\langle\cdots \rangle_{4\text{
faces}}$ of four-valent vertices, we now
introduce operations that reduce them to correlators with less number
of faces. We will present these operations at the level of graphs,
nevertheless they have an obvious translation into matrix theory
language as integrating-in and -out matrix fields.

\subsubsection{Integrating-In: Adding Edges}

We use this operation to split a four-valent vertex into two
three-valent vertices. This can be useful to restructure a graph and
set it up for the application of other simplifying operations.

\begin{figure}[ht]
\centering
\resizebox{1.5\totalheight}{!}{\includegraphics[width=\textwidth]{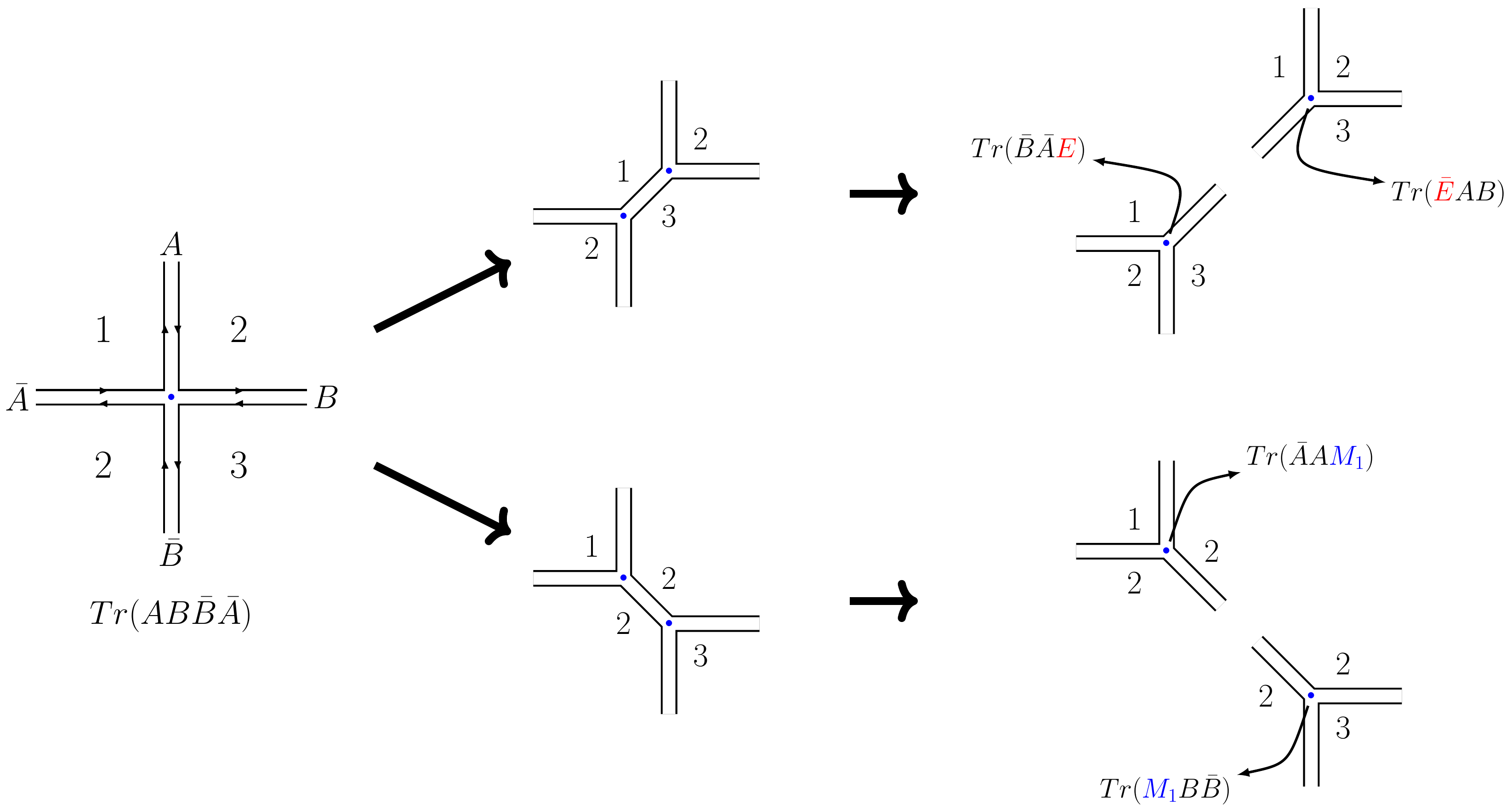}}
\caption{Adding an extra intermediate edge}
\label{fig:NewEdge}
\end{figure}

This operation is performed in two steps as shown in
\figref{fig:NewEdge}. In the first step we introduce a new edge and
increase the number of vertices by one, such that the genus of the
graph is maintained. As shown in the middle column of
\figref{fig:NewEdge}, there are two possibilities to add this
intermediate edge. In this specific example the two different options
require two different types of edges. The top type needs an edge with
different faces on its sides and can be represented by a complex
matrix in the matrix language. The bottom type needs an edge with the
same face on its sides and can be represented by a Hermitian matrix.
Finally in the second step we split this new edge resulting in two new
three-valent vertices.

Ultimately, we want to connect back the intermediate edge to
reconstruct the graph. But typically, we will first perform other
simplifications, such as integrating-out, before restoring the
intermediate edge, such that the final result will be simpler than the
original graph.

\subsubsection{Integrating-Out: Removing A Face}

We use this operation to decrease the number of faces, vertices and
edges all at the same time, such that the genus of the graph does not
change. In the matrix language, this corresponds to integrating out one
or more matrix fields.

\begin{figure}[ht]
\centering
\resizebox{1.5\totalheight}{!}{\includegraphics[width=\textwidth]{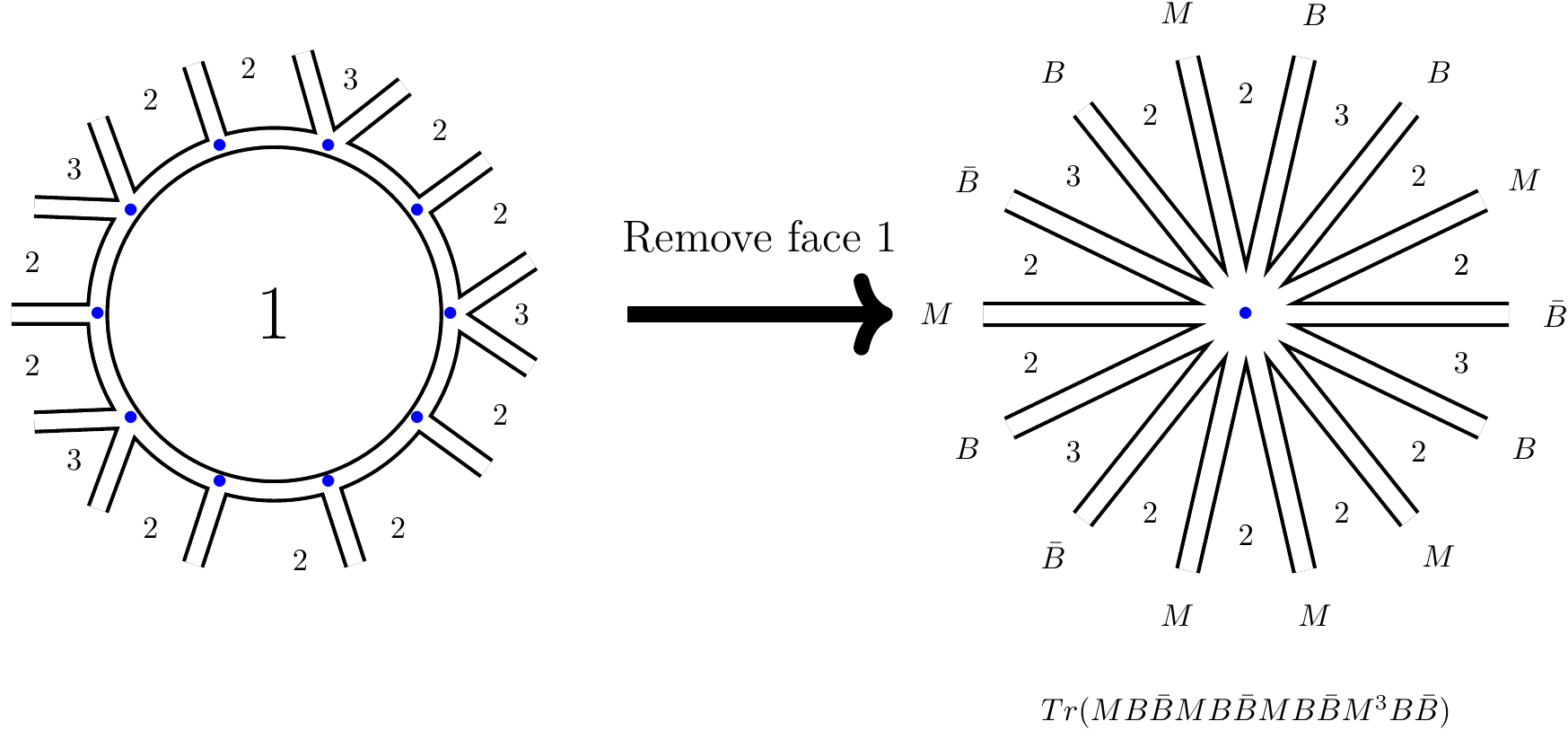}}
\caption{Vertices organize around a reference face}
\label{fig:OneFace}
\end{figure}

To perform this operation, we first choose a reference face, labeled
by $1$ for instance. Then we organize all vertices that have a $1$
appearing between their edges around the reference face
as shown on the left panel of \figref{fig:OneFace}. The next step is
to remove the reference face, such that all vertices on its perimeter
get contracted to a single effective vertex that inherits all the
outer edges, as shown on the right panel of \figref{fig:OneFace}.

In some cases it is possible to choose more than one reference face,
such that all vertices participate in an integrating-out operation. On
the other hand, in some cases it is not possible to pick a reference
face at first, for instance we can not place four-valent vertices with
faces between edges $[1212]$ around a reference face $1$ or $2$. In
these cases, making an integrating-in operation first can
allow to organize the resulting vertices of lower valence around a reference
face. In the following sections, we will perform combinations of these
graph operations to simplify the counting of quadrangulations.

\subsection{Non-BPS Quadrangulations}
\label{app:NonBPS}

As a warm-up, we consider quadrangulations formed by squares $[1234]$
and $[4321]$ only. As described in the main text, this addresses the
limit of large coupling $\oct$. In the dual space, the
relevant matrix model has action
\beq
S_{\text{large }\oct} =  \,-\tr A\bar{A}-\,\tr B\bar{B}-\,\tr C\bar{C}-\,\tr D\bar{D} +\oct\,\left(\tr(ABCD) +\,\tr(\bar{D}\bar{C}\bar{B}\bar{A})\right)
\,.
\eeq
The simplicity of this problem allows for the application of different
graph operations, which lead to different simplified outcomes. In what
follows, we list some of these results, summarized in \secref{sec:summaryNonBPS}.

\subsubsection{As a 1-vertex and 3-faces problem}

\begin{figure}
\centering
\resizebox{1.8\totalheight}{!}{\includegraphics[width=\textwidth]{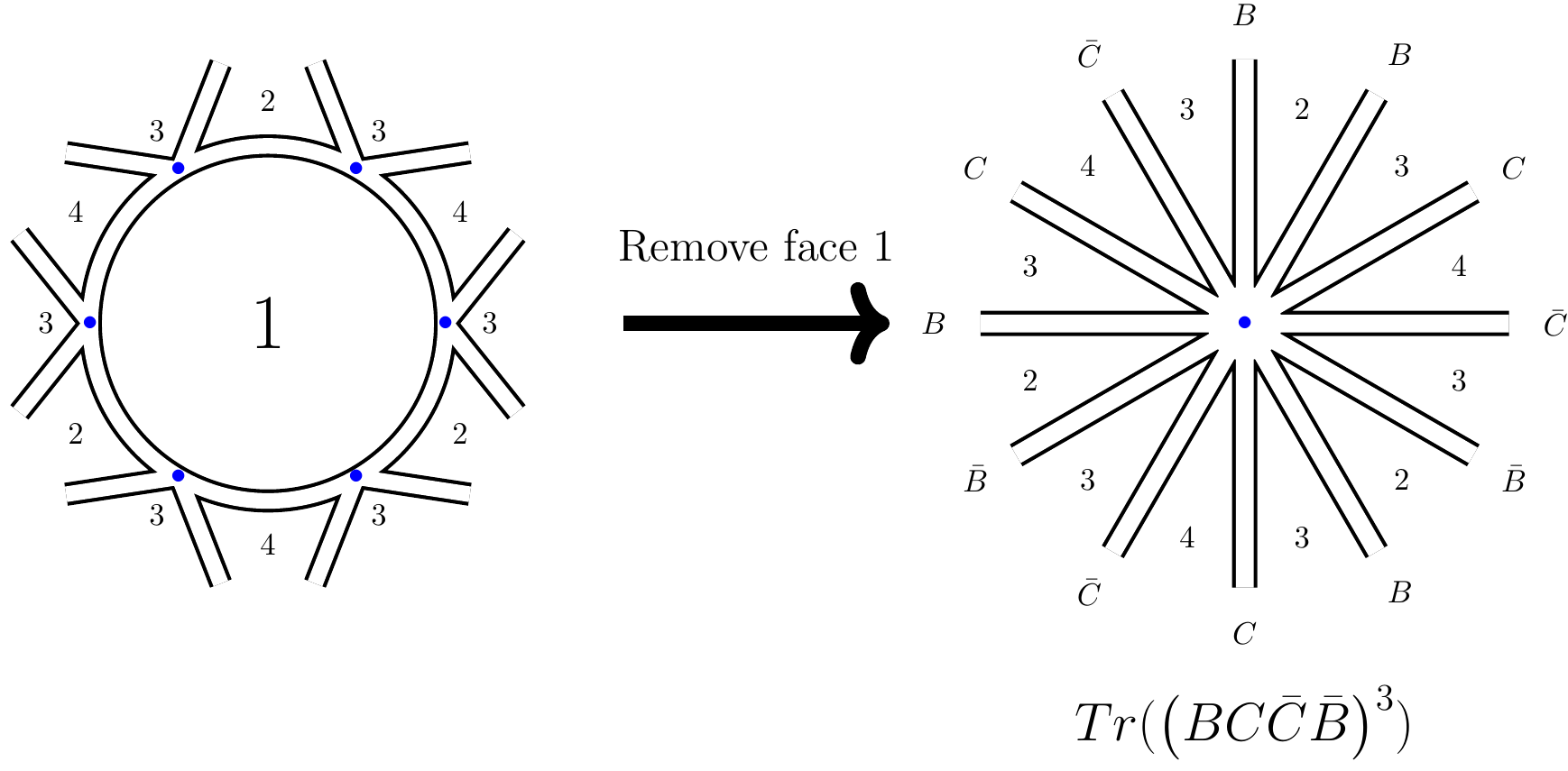}}
\caption{Genus-two example: The unique way of organizing the
four-valent vertices around a reference face $1$, which leads to a single
effective vertex, once we remove the reference face.}
\label{fig:NonBpsOnePoint}
\end{figure}

Having only vertices $\tr(ABCD)$ and
$\tr(\bar{D}\bar{C}\bar{B}\bar{A})$, we can easily apply the
integrating-out technique by picking as reference the face $1$ (or any
of the other three). Then, as shown in \figref{fig:NonBpsOnePoint},
there is a unique way of organizing the vertices on the perimeter of
this reference face, that is alternating the two types of vertices.
After removing the reference face $1$, the result is an effective
vertex with fields $B$ and $C$ (and conjugates) only, the fields $A$
and $D$ are integrated out.

The non-BPS quadrangulations counted by the correlator of four-valent
vertices $\langle\cdots\rangle_{4\text{ faces}}$ can now be counted by
a one-point correlator $\langle \tr(\cdots)\rangle_{3\text{ faces}}$,
see equation \eqref{eq:summaryQ3}.

\subsubsection{As a 2-vertices and 2-faces problem}

\begin{figure}[ht]
\centering
\includegraphics[width=\textwidth]{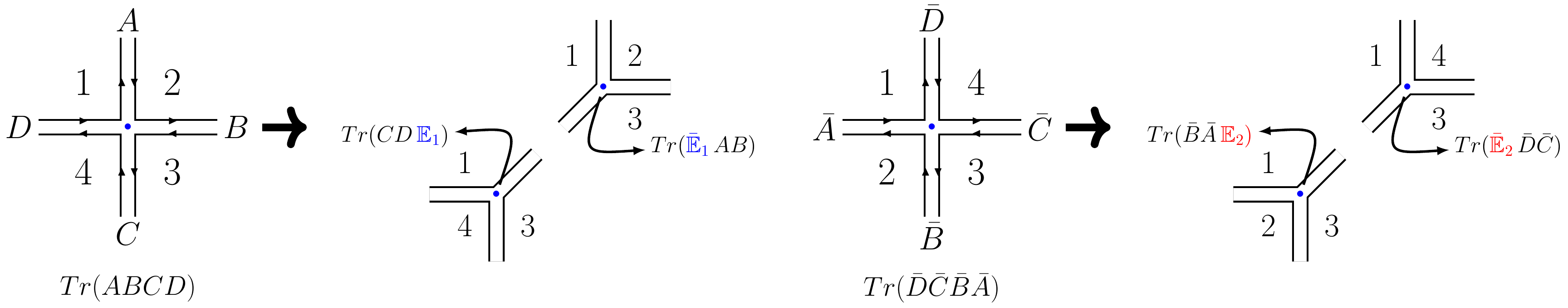}
\caption{Splitting non-BPS vertices.}
\label{fig:NonBpsSplit}
\end{figure}

\begin{figure}[ht]
\centering
\resizebox{1.2\totalheight}{!}{\includegraphics[width=\textwidth]{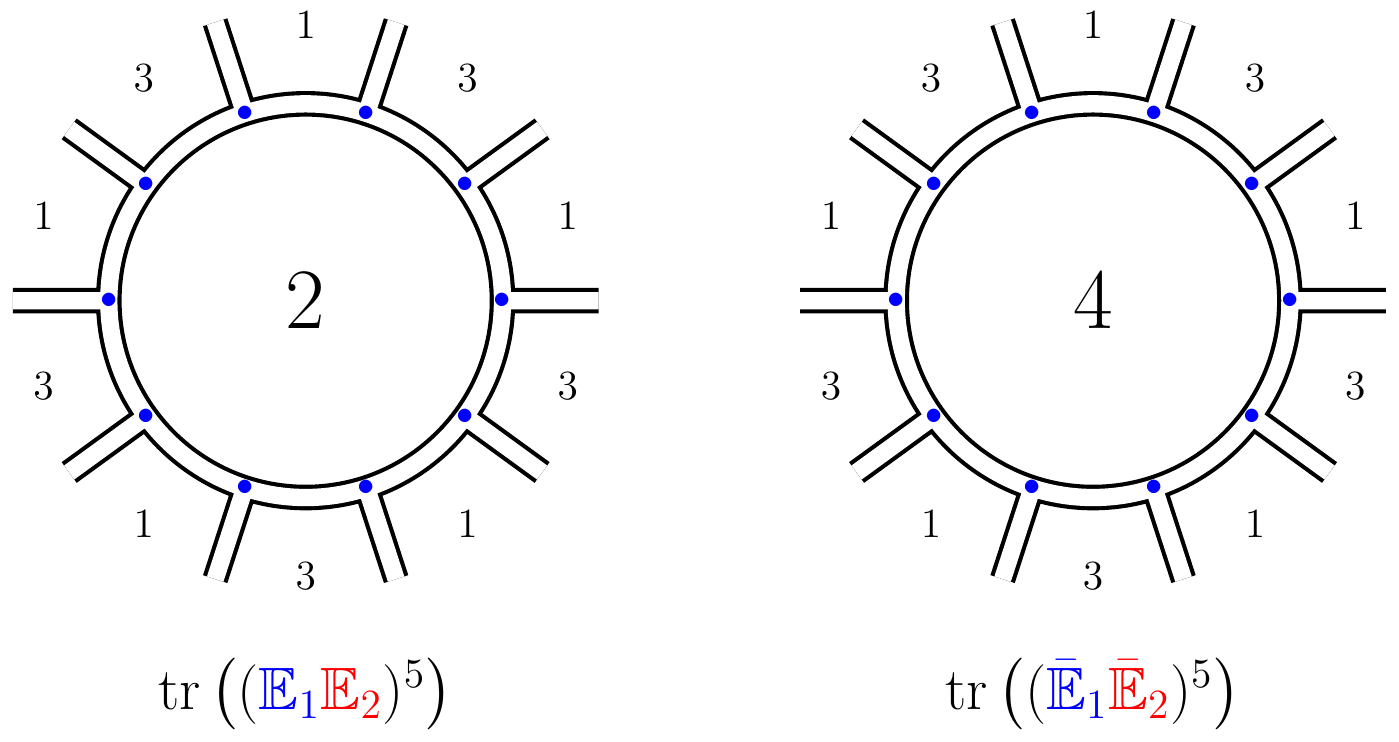}}
\caption{Genus four non-BPS example: Integrating out faces $2$ and $4$
gives a single two-point function of traces with two complex matrices.}
\label{fig:TwoNonBpsFaces}
\end{figure}

Another simplification of the non-BPS counting can be achieved by first
integrating-in complex fields in the non-BPS vertices as shown in
figure \figref{fig:NonBpsSplit}.
After splitting the four-valent vertices, we can arrange all
three-valent vertices around two reference faces ($2$ and $4$) as
shown in \figref{fig:TwoNonBpsFaces}.

Then we remove these faces $2$ and $4$, which effectively
integrates out all $A,B,C,D$, and obtain two effective vertices that
contain only the fields $\mathbb{E}_{1}$ and $\mathbb{E}_{2}$ (and
conjugates). Now the quadrangulation counting can be found by
computing a two-point function $\langle \tr(\cdots)
\tr(\cdots)\rangle_{2\text{ faces}}$, see equation \eqref{eq:summaryQ2}.

\subsubsection{As a 3-vertices and 1-face problem}

Naturally, dualizing the 1-vertex and 3-face problem, we obtain
a 3-vertices and 1-face problem. We have not worked out in detail
the necessary graph operations to get this latter outcome starting
with the four-valent vertices. Nevertheless, by working out a few examples,
we figured out what the correlator is, and present it in equation~\eqref{eq:summaryQ1}

\subsubsection{Summary for Non-BPS squares}\label{sec:summaryNonBPS}

We use $\mathcal{N}_{g}$ to denote the number of non-BPS
quadrangulations weighted by their corresponding symmetry factors
(automorphisms). From genus $g=0$ to $g=5$ these numbers, not
necessarily integers, are:
\begin{equation}
\{\mathcal{N}_{g}\}=\{1,\frac{1}{2},1,5,\frac{248}{5},840,\cdots\}
\end{equation}
they appear in the large $\oct$ limit of $\mathcal{A}$, see \eqref{Znon}, as:
\beq
\mathcal{A} =\frac{1}{\zeta_1 \zeta_2 \zeta_3 \zeta_4} \sum_{g=0}^\infty \frac{(\zeta_1 \zeta_2 \zeta_3 \zeta_4 \oct^2)^{g+1}}{g!^4}\,\mathcal{N}_{g}
\eeq
and can be computed in four different and equivalent ways
\bba
\mathcal{N}_{g}&=\frac{\left\langle\,\left(\tr(ABCD)\tr(\bar{D}\bar{C}\bar{B}\bar{A})\right)^{n}\,\right\rangle_{\text{4 faces}}}{n!^{2}}\label{eq:summaryQ4}\\
&=\frac{\left\langle\,\tr\left((BC\bar{C}\bar{B})^{n}\right)\,\right\rangle_{\text{3 faces}}}{n}\label{eq:summaryQ3}\\
&= \frac{\left\langle\,\tr\left((BC)^{n}\right)\tr\left((\bar{B}\bar{C})^{n}\right) \,\right\rangle_{\text{2 faces}}}{n^{2}}\label{eq:summaryQ2}\\
&=\frac{\left\langle\,\tr\left(B^{n}\right)\tr\left(C^{n}\right)\tr\left((\bar{B}\bar
C)^{n}\right)\,\right\rangle_{\text{1 face}}}{n^{3}}
\,,
\label{eq:summaryQ1}
\end{align}
where  $n=g+1$.

Notice that the weights in our correlators, denominators
in~\eqref{eq:summaryQ4,eq:summaryQ3,eq:summaryQ2,eq:summaryQ1},
correspond to symmetry factors. We have the
factor $n!$ for $n$ identical vertices, and the factor $n$ for traces
of the form $\tr(X^{n})$.

\subsection{All Quandrangulations}
\label{sec:all-quandrangulations}

We now address the full problem of counting quadrangulations including all ten
vertices. Out of the three possibilities we presented for the non-BPS
squares ($[1234]$ and $[4321]$) sector in~\secref{app:NonBPS}, we
found only the two-face simplification can be deformed to include the
BPS squares ($[abcb]$ and $[abab]$) and count all quadrangulations.

\begin{figure}[ht]
\centering
\includegraphics[width=\textwidth]{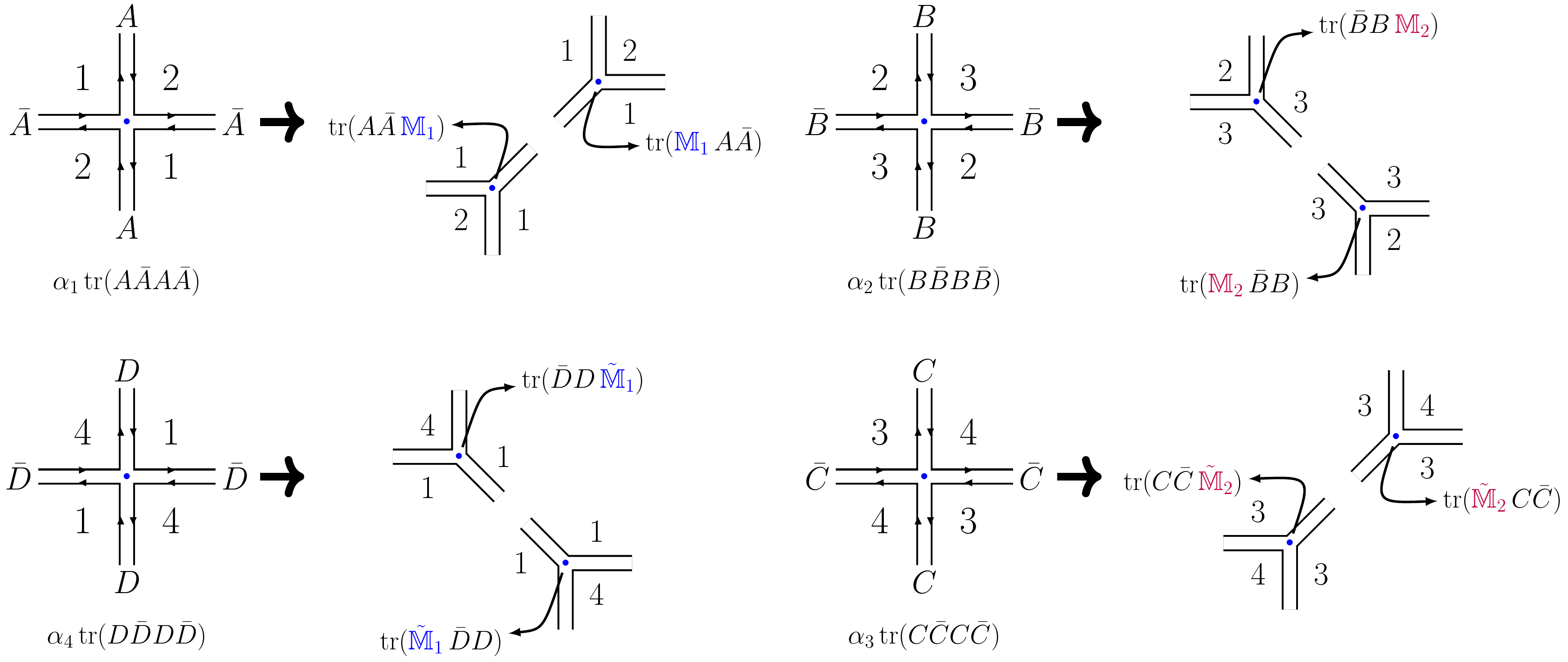}
\caption{Splitting vertices with couplings $\alpha_{i}$}
\label{fig:BPSa}
\end{figure}

\begin{figure}[ht]
\centering
\includegraphics[width=\textwidth]{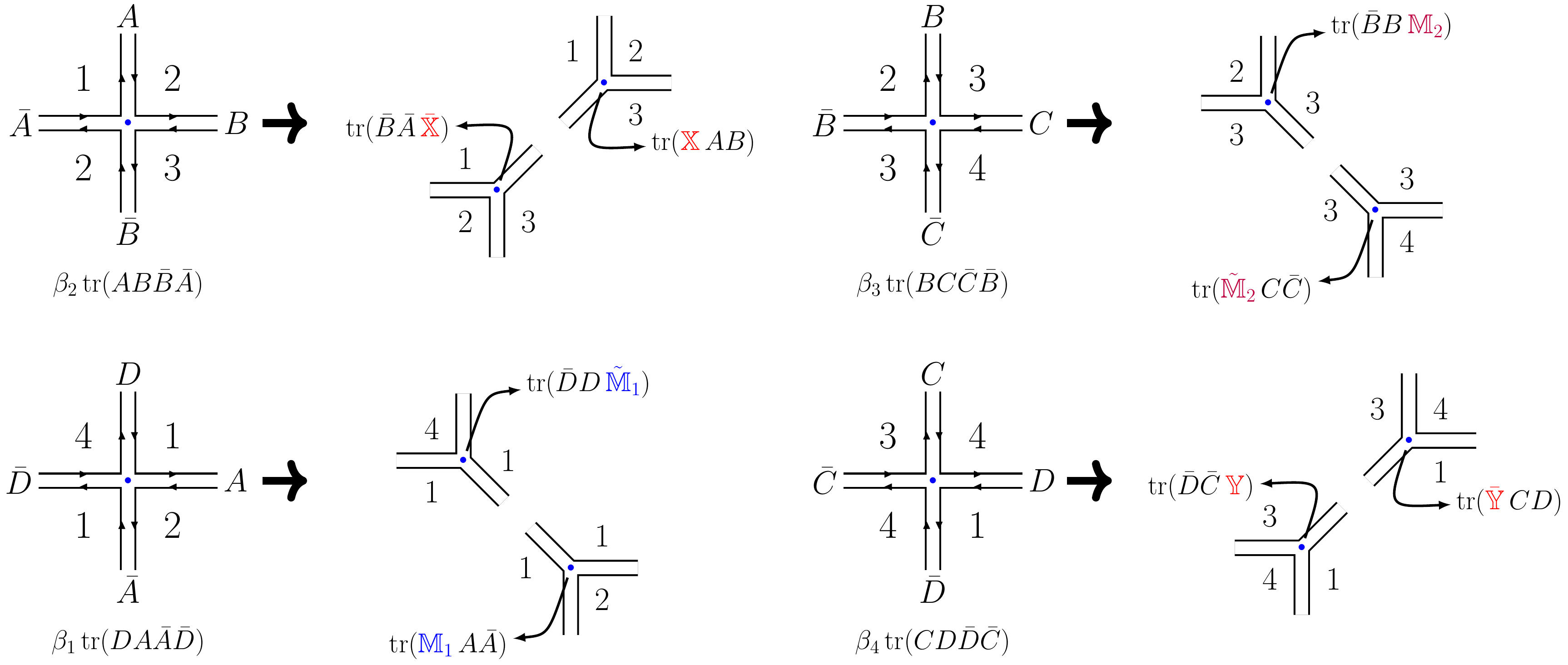}
\caption{Splitting vertices with couplings $\beta_{i}$}
\label{fig:BPSab}
\end{figure}

\begin{figure}[ht]
\centering
\includegraphics[width=\textwidth]{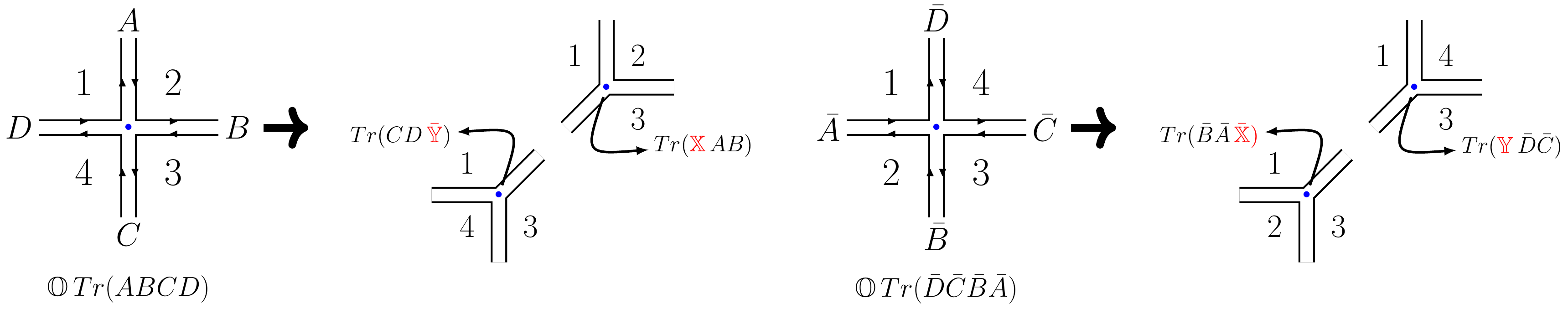}
\caption{Splitting vertices with coupling $\oct$ using complex matrices.}
\label{fig:NonBPSsplitXY}
\end{figure}

To get to this two-face simplification, we first integrate-in auxiliary
Hermitian matrices
$\mathbb{M}_{1},\mathbb{M}_{2},\tilde{\mathbb{M}}_{1},\tilde{\mathbb{M}}_{2}$
and complex matrices $\mathbb{X},\mathbb{Y}$ to split the BPS vertices
as shown in \figref{fig:BPSa} and \figref{fig:BPSab}. In addition to
that, to be consistent with this new auxiliary matrices, we relabel
the complex matrices in the splitting of the non-BPS vertices as shown
in \figref{fig:NonBPSsplitXY}.

\begin{figure}[ht]
\centering
\resizebox{1.5\totalheight}{!}{\includegraphics[width=\textwidth]{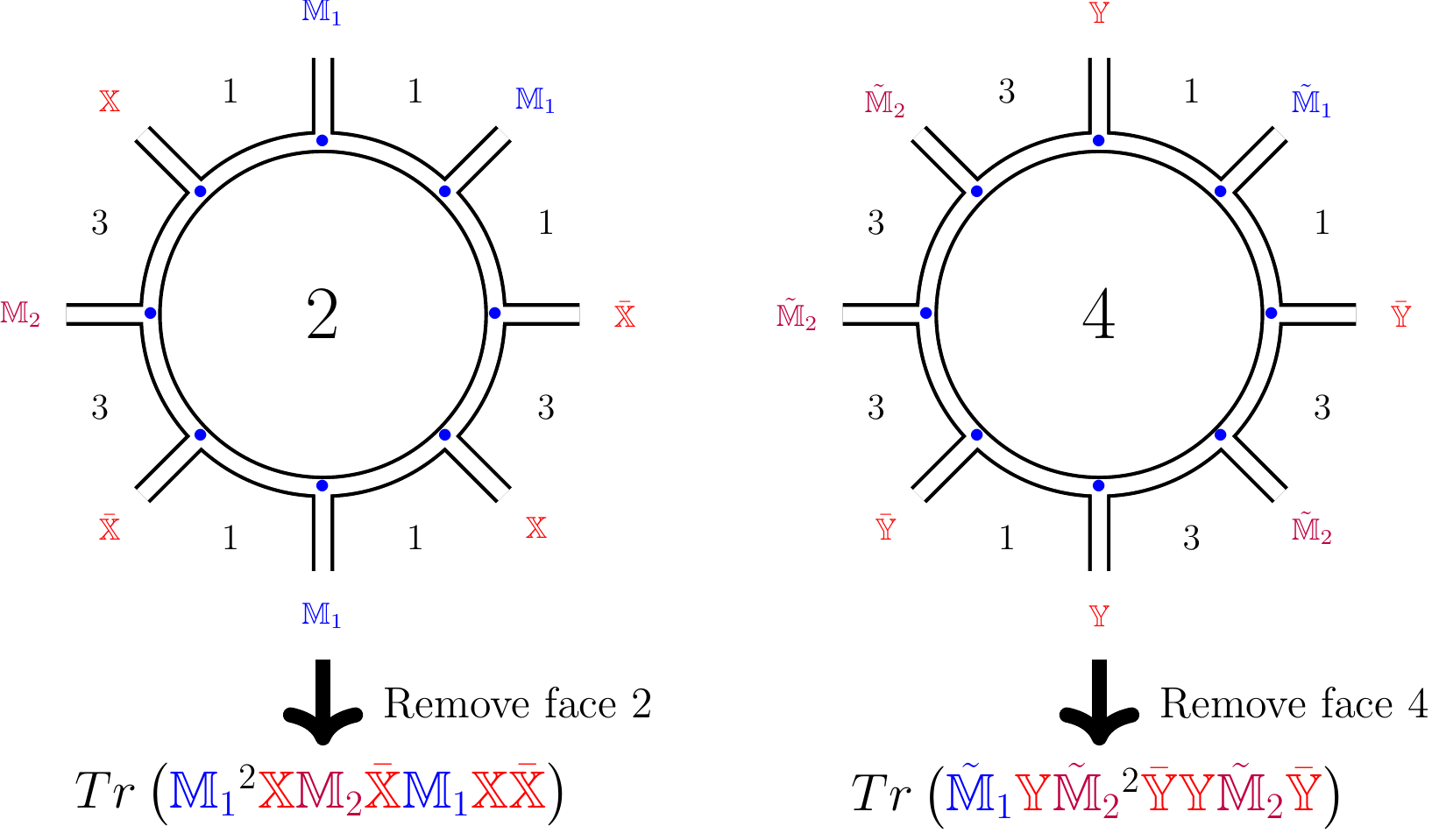}}
\caption{Integrating-out faces $2$ and $4$ gives two effective vertices, which under Wick contractions should encapsulate two faces: $1$ and $3$.}
\label{fig:TwoFaces}
\end{figure}

In order to reconstruct the couplings $\alpha_{i},\beta_{i}$ we impose
the auxiliary matrices satisfy:
\bba\label{eq:WickBPS}
\langle \mathbb{M}_{1}\,\mathbb{M}_{1}\rangle = \alpha_{1} &\qquad \langle \mathbb{X}\,\bar{\mathbb{X}}\rangle = \beta_{2} \nonumber\\
\langle \mathbb{M}_{2}\,\mathbb{M}_{2}\rangle = \alpha_{2} &\qquad \langle \mathbb{Y}\,\bar{\mathbb{Y}}\rangle = \beta_{4} \nonumber\\
\langle \tilde{\mathbb{M}}_{2}\,\tilde{\mathbb{M}}_{2}\rangle = \alpha_{3} &\qquad \langle \mathbb{M}_{2}\,\tilde{\mathbb{M}}_{2}\rangle = \beta_{3} \nonumber\\
\langle \tilde{\mathbb{M}}_{1}\,\tilde{\mathbb{M}}_{1}\rangle = \alpha_{4} &\qquad \langle \mathbb{M}_{1}\,\tilde{\mathbb{M}}_{1}\rangle = \beta_{1}
\end{align}
and similary for the non-BPS coupling:
\beq\label{eq:WickO}
\langle \mathbb{X}\,\bar{\mathbb{Y}} \rangle \,=\, \langle \bar{\mathbb{X}}\,\mathbb{Y} \rangle \,=\, \oct
\eeq
We can now arrange the three-valent vertices around reference faces
$2$ and $4$ as shown in \figref{fig:TwoFaces}. After removing
these faces we obtain effective vertices of the form:
\beq
T_{\{L,\{m_{i}\},\{n_{i}\}\}} \,=\, \tr\left(\prod_{i=1}^{L}\left(\mathbb{M}_{1}^{m_{i}}\,\mathbb{X}\,\mathbb{M}_{2}^{n_{i}} \bar{\mathbb{X}}\right)\right)
\quad \text{and} \quad
\tilde{T}_{\{\tilde{L},\{\tilde{m}_{i}\},\{\tilde{n}_{i}\}\}} \,=\, \tr\left(\prod_{i=1}^{\tilde{L}}\left(\tilde{\mathbb{M}}_{1}^{\tilde{m}_{i}}\,\mathbb{Y}\,\tilde{\mathbb{M}}_{2}^{\tilde{n}_{i}} \bar{\mathbb{Y}}\right)\right)
\eeq

The counting of quadrangulations now follows from computing the
two-point functions of these effective vertices with Wick contractions
dictated by~\eqref{eq:WickBPS} and~\eqref{eq:WickO} and extracting the
two-face coefficient $\langle T\,\tilde{T} \rangle_{2\text{ faces}}$.
For instance, at genus two we have a contribution from:
\begin{align}
\langle \tr\big{(}\mathbb{M}_{1}^{2}\,\mathbb{X}\,\mathbb{M}_{2}\,\bar{\mathbb{X}}\,\mathbb{M}_{1} \mathbb{X} \,\bar{\mathbb{X}}\big{)}\,&\tr\big{(}\tilde{\mathbb{M}}_{1}\,\mathbb{Y}\,\tilde{\mathbb{M}}_{2}^{2}\,\bar{\mathbb{Y}}\,\mathbb{Y}\,\tilde{\mathbb{M}}_{2}\,\bar{\mathbb{Y}}\,\big{)}\rangle_{2\text{ faces}} = \nonumber\\
&4\,\oct^{4}\alpha_{1}\alpha_{2}\alpha_{3}\beta_{1} \,+\, 16\,\oct^{4}\alpha_{1}\beta_{1} \beta_{3}^{2} \,+\, 16\,\oct^{2} \alpha_{1}\alpha_{2}\alpha_{3}\beta_{1}\beta_{2}\beta_{4}\nonumber\\
&+55\,\oct^{2} \alpha_{1}\beta_{1}\beta_{2}\beta_{3}^{2}\beta_{4} +4\,\alpha_{1}\alpha_{2}\alpha_{3}\beta_{1}\beta_{2}^{2}\beta_{4}^{2}+12\alpha_{1}\beta_{1}\beta_{2}^{2}\beta_{3}^{2}\beta_{4}^{2}
\,.
\end{align}
In order to compute the two-face partition function, we must consider
all possible effective vertices and their corresponding symmetry
factors:
\beq\label{eq:ZtwoPoint}
\mathbf{Z}(\oct,\alpha_{i},\beta_{i}) = \sum_{g=0}^{\infty}\,\sum_{\{\mathcal{N},\tilde{\mathcal{N}}\}\in \mathcal{M}_{g}}\,\frac{1}{s_{\mathcal{N}}s_{\tilde{\mathcal{N}}}}\,\langle T_{\mathcal{N}}\,\tilde{T}_{\tilde{\mathcal{N}}}\rangle_{2\text{-faces}}
\,,
\eeq
where we group the collections of indices as
\beq
\mathcal{N}\equiv\{L,\{m_{i}\},\{n_{i}\}\}\quad\tilde{\mathcal{N}}\equiv\{\tilde{L},\{\tilde{m}_{i}\},\{\tilde{n}_{i}\}\}
\,.
\eeq
Allowing for different orderings of $\{m_{i}\}$ and $\{n_{i}\}$
($\{1,2\}\neq \{2,1\}$) in the inner sum of
\eqref{eq:ZtwoPoint},\footnote{We could alternatively mod out orderings
$\mathcal{N}\equiv\{L,\{m_{i}\},\{n_{i}\}\}$ that are cyclically
equivalent within the trace. In that case we would have to modify the
symmetry factor $s_{\mathcal{N}}=\frac{L}{\#\,\text{equivalent
orderings}}$} then the symmetry factors are
simply giving by:
\beq
s_{\mathcal{N}}\,=\,L \qquad \text{and} \qquad s_{\tilde{\mathcal{N}}} \,=\,\tilde{L}
\,.
\eeq
Furthermore the inner sum is restricted to run over the group
$\mathcal{M}_{g}$, whose elements are all
$\{\mathcal{N},\tilde{\mathcal{N}}\}$ that satisfy:
\beq\label{eq:SquareRestriction}
2(L+\tilde{L})\,+\,\sum_{i=1}^{L}(n_{i}+m_{i}+\tilde{n}_{i}+\tilde{m}_{i})\,=\, 4g+4
\,,
\eeq
such that for each genus $g$ the number of fields is $4g+4$, leading
two $2g+2$ squares after Wick contractions. Furthermore, from
\eqref{eq:WickBPS} it follows that only when
$\sum_{i}m_{i}+\tilde{m}_{i}$ and $\sum_{i}n_{i}+\tilde{n}_{i}$ are
even numbers are the two-point correlators non-vanishing.

The expression~\eqref{eq:ZtwoPoint} admits a resummation that leads to the compact formula:
\beq\label{eq:TwoPointLogCouplings}
\mathbf{Z}(\oct,\alpha_{i},\beta_{i}) \,=\,
\lrvev{
\tr\log\lrbrk{
\mathbb{I}-
\frac{1}{\mathbb{I}- \mathbb{M}_2} \bar{\mathbb{X}}
\frac{1}{\mathbb{I}- \mathbb{M}_1} \mathbb{X}
}
\tr\log\lrbrk{
\mathbb{I}-
\frac{1}{\mathbb{I}- \tilde{\mathbb{M}}_2}  \bar{\mathbb{Y}}
\frac{1}{\mathbb{I}- \tilde{\mathbb{M}}_1} \mathbb{Y}
}
}_{\!2\text{ faces}}
\eeq
This is the analog of \eqref{eq:newRep}, but now with couplings
$\alpha_{i}$ and $\beta_{i}$ for the BPS squares. Notice that the
expansion of the $\log$s in \eqref{eq:TwoPointLogCouplings} leads to
terms which do not satisfy the restriction
\eqref{eq:SquareRestriction}. However, these unwanted terms do not have
a two-face contribution, $\langle \text{unwanted} \rangle_{\!2\text{
faces}} =0$.%
\footnote{This follows from the Euler formula. We have two vertices and
demand two faces, so the number of edges must be: \beq 2-2g = (F=2) +
(V=2) - E \quad \Rightarrow \quad E= 2g+2 \,, \eeq which means that only two-point
correlators with a total number of composite matrices given by a
multiple of four, condition \eqref{eq:SquareRestriction}, contribute
to $\langle\cdots\rangle_{2\text{ faces}}$. }
So effectively,~\eqref{eq:ZtwoPoint} and~\eqref{eq:TwoPointLogCouplings} are identical.

The partition function $\mathbf{A}$ that makes direct contact with the
cyclic correlator studied in this paper is now given by:
\beq\label{eq:AtwoPoint}
\mathbf{A}(\oct,\alpha_{i},\beta_{i}) = \sum_{g=0}^{\infty}\,\sum_{\{\mathcal{N},\tilde{\mathcal{N}}\}\in \mathcal{M}_{g}}\,\frac{1}{s_{\mathcal{N}}s_{\tilde{\mathcal{N}}}}\,\frac{1}{w_{\mathcal{N}}w_{\tilde{\mathcal{N}}}}\,\langle T_{\mathcal{N}}\,\tilde{T}_{\tilde{\mathcal{N}}}\rangle_{2\text{-faces}}
\eeq
with weights:
\beq
w_{\mathcal{N}}\,\equiv\,w_{\{L,\{m_{i}\},\{n_{i}\}\}}
\,=\,
\Bigbrk{-1+\sum_{i=1}^{L}(m_{i}+1)}!
\Bigbrk{-1+\,\sum_{i=1}^{L}(n_{i}+1)}!
\,,
\eeq
for which we have not found a compact re-summed formula as \eqref{eq:TwoPointLogCouplings}.

\subsubsection*{Perturbative genus computation}

The two-face formulations \eqref{eq:ZtwoPoint} and \eqref{eq:AtwoPoint} allows us to efficiently compute the
partitions $\mathbf{Z}$ or $\mathbf{A}$ up to genus 4. Then under the replacement~\eqref{eq:repA} we obtain the
perturbative result in~\eqref{eq:fin3}.

In the BPS limit $\oct\to 0$, we carry on up to genus 6 by
integrating out the complex matrices, and then evaluating one-face
correlators of Hermitian matrices as explained in
\secref{sec:Simplifications}. This larger amount of data, and the
inspiration we get from the extremal correlator in
\appref{sec:all-extremal} allows us to guess the re-summed series as:
\beq\label{eq:BPSlimitA}
\mathbf{A}(\oct=0,\alpha_{i},\beta_{i})= \prod_{i=1}^{4} \beta_{i}\frac{\sinh\frac{1}{2}\left(\beta_{i}+\alpha_{i}+\beta_{i+1}\right)}{\frac{1}{2}\left(\beta_{i}+\alpha_{i}+\beta_{i+1}\right)}
\,.
\eeq
Under the replacement \eqref{eq:repA}, we obtain the result in \eqref{eq:guessedBPS}.

In fact we further found that formula \eqref{eq:BPSlimitA} can be extended to find the BPS
part of a larger class of cyclic correlators, as we present in \appref{sec:bps-part-all-cyclic}.

\section{Other Results On Quadrangulations }
\subsection{\texorpdfstring{$(n+1)$}{(n+1)}-Point Extremal Correlators in DSL}
\label{sec:all-extremal}

In this appendix, we review the computation of protected extremal
correlators of the form\footnote{Here, the normalization is
such that the genus expansion goes as
$\Nc^{0}(\cdots)+\Nc^{-2}(\cdots)+\cdots$}
\begin{equation}\label{eq:Extremaln}
E_{n}=\langle \tr(Z^{J_{1}})\tr(Z^{J_{2}})\cdots \tr(Z^{J_{n}})\tr(\bar{Z}^{J_{R}})\rangle
\end{equation}
with $J_{R}=J_{1}+\cdots + J_{n}$,
in the double scaling limit $J_i\to \infty$, $\Nc\to \infty$ and
$J_i/\sqrt{\Nc}$ fixed. In fact the result is known at finite $\Nc$
and $J_{i}$ from \cite{Kristjansen:2002bb,Beisert:2002bb}. In the DSL
it is given by:
\beq\label{eq:ExtDSL}
E_{n}(J_{i},\Nc) \,\overset{\text{DSL}}{=} \,J_{R}^{-2}(2 \Nc)^{n}\, \prod_{i=1}^{n}\sinh\left(\frac{J_{i}J_{R}}{2 \Nc}\right)
\,.
\eeq
Here, we would like to present how to reproduce this result by counting
quadrangulations and using integrating-in and -out graph operations.

In the DSL, the correlator $E_{n}$ can be reconstructed in a similar
fashion as the cyclic correlator of this article. The skeleton graphs
that dominate are also given by quadrangulations, and to obtain $E_{n}$
we must count them and dress them with the lengths $J_{i}$ by
performing a Borel-type transformation.

All squares involved include the reservoir $R$ twice, and the
corresponding dual four-valent vertices are given in the matrix
action:
\beq\label{eq:Eaction}
S^{(\text{E})}_{n}=-\sum_{a=1}^{n}\tr(A_{a}\bar{A}_{a})\,+\,\frac{1}{2}\sum_{a=1}^{n}\sum_{b=1}^{n}\,\alpha_{a,b}\,\tr(A_{a}A_{b}\bar{A}_{b}\bar{A}_{a})
\eeq
with couplings $\alpha_{a,b}=\alpha_{b,a}$ associated to squares of the form [RaRb].

The relevant Borel-transformed partition function (the analog
of~\eqref{eq:Apartition}) coming from the matrix
model~\eqref{eq:Eaction} is
\beq\label{eq:ExtApartition}
\mathbf{A}^{(\text{E})}_{n}(\alpha_{a,b}) = \sum_{g=0}^{\infty}\,\sum_{T_{g}=\{t_{1},\cdots t_{n+2g-1}\}\in V^{(\text{E})}_{4} } \frac{1}{\text{\text{sym}($T_{g}$)}}\frac{1}{\text{\text{weight}($T_{g}$)}}\left\langle \prod_{m=1}^{n+2g-1}\,t_{m}  \right\rangle_{(n+1)\text{ faces}}
\,,
\eeq
where now $T_{g}$ is a subset of $(n+2g-1)$ four-valence vertices with
couplings chosen from $V^{(\text{E})}_{4} =
\{\alpha_{a,b}\tr(A_{a}A_{b}\bar{A}_{b}\bar{A}_{a})\}_{a,b
=1,\cdots,n}$ with at least one occurrence of $A_{a}$ for
$a=1,\cdots,n$.

The symmetry factor $\text{sym}(T_{g})$ has a factor of 2 for each
vertex of the form $\tr(A_{a}A_{a}\bar{A}_{a}\bar{A}_{a})$, and a $k!$
when $T_{g}$ contains a vertex repeated $k$ times. The weight factor
is given by:
\beq
\text{weight}(T_{g})\,=\, \prod_{i=1}^{n}(m_{A_{i}}-1)!
\eeq
where $m_{A_{i}}$ counts the number of occurrences of $A_{i}$ in the subset $T_{g}$.

This partition function $\mathbf{A}$ can be identified with the DSL of the extremal
correlator under the replacement:
\beq\label{eq:ExtReplaceA}
E_{n}(J_{i},\Nc) \,\overset{\mathrm{DSL}}{=} \Nc^{n-1}\,\mathbf{A}^{(\text{E})}_{n}(\alpha_{a,b})\big{|}_{\alpha_{a,b}\to {J_{a}J_{b}}/{\Nc}}
\eeq

\begin{figure}[ht]
\centering
\resizebox{2\totalheight}{!}{\includegraphics[width=\textwidth]{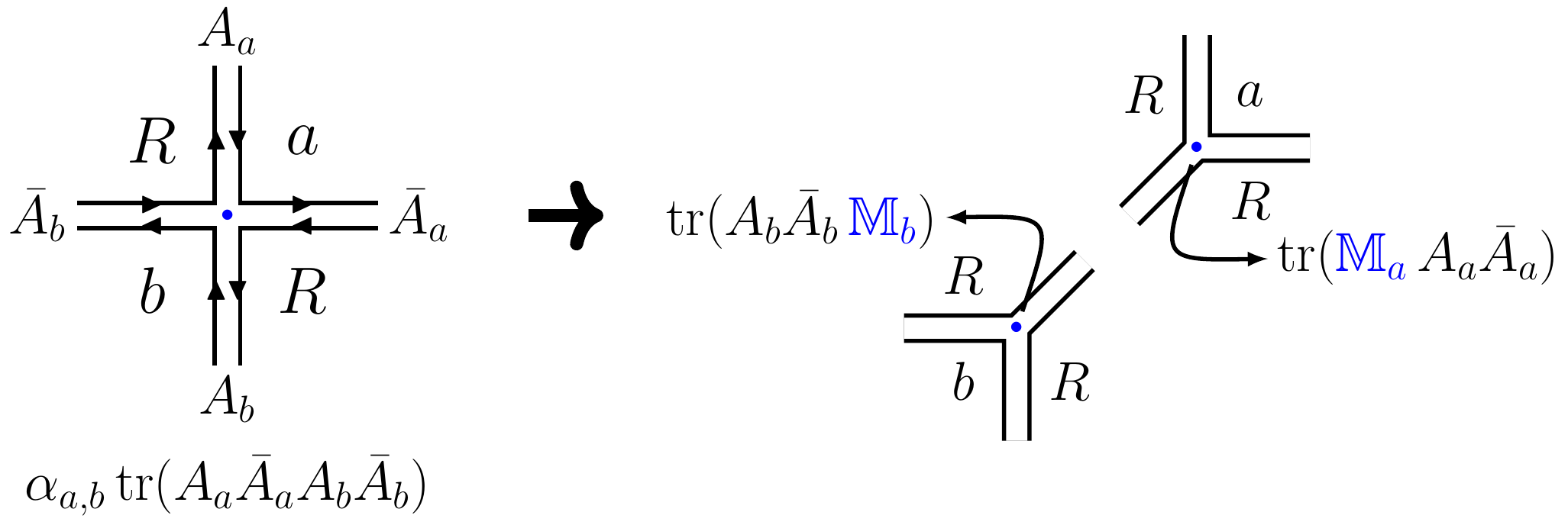}}
\caption{Splitting a BPS vertex by introducing hermitian matrices
$\mathbb{M}_{a}$ and $\mathbb{M}_{b}$, which satisfy $\langle \mathbb{M}_{a}\mathbb{M}_{b}\rangle=\alpha_{a,b}$.}
\label{fig:ExtSplit}
\end{figure}

\begin{figure}[ht]
\centering
\resizebox{1.5\totalheight}{!}{\includegraphics[width=\textwidth]{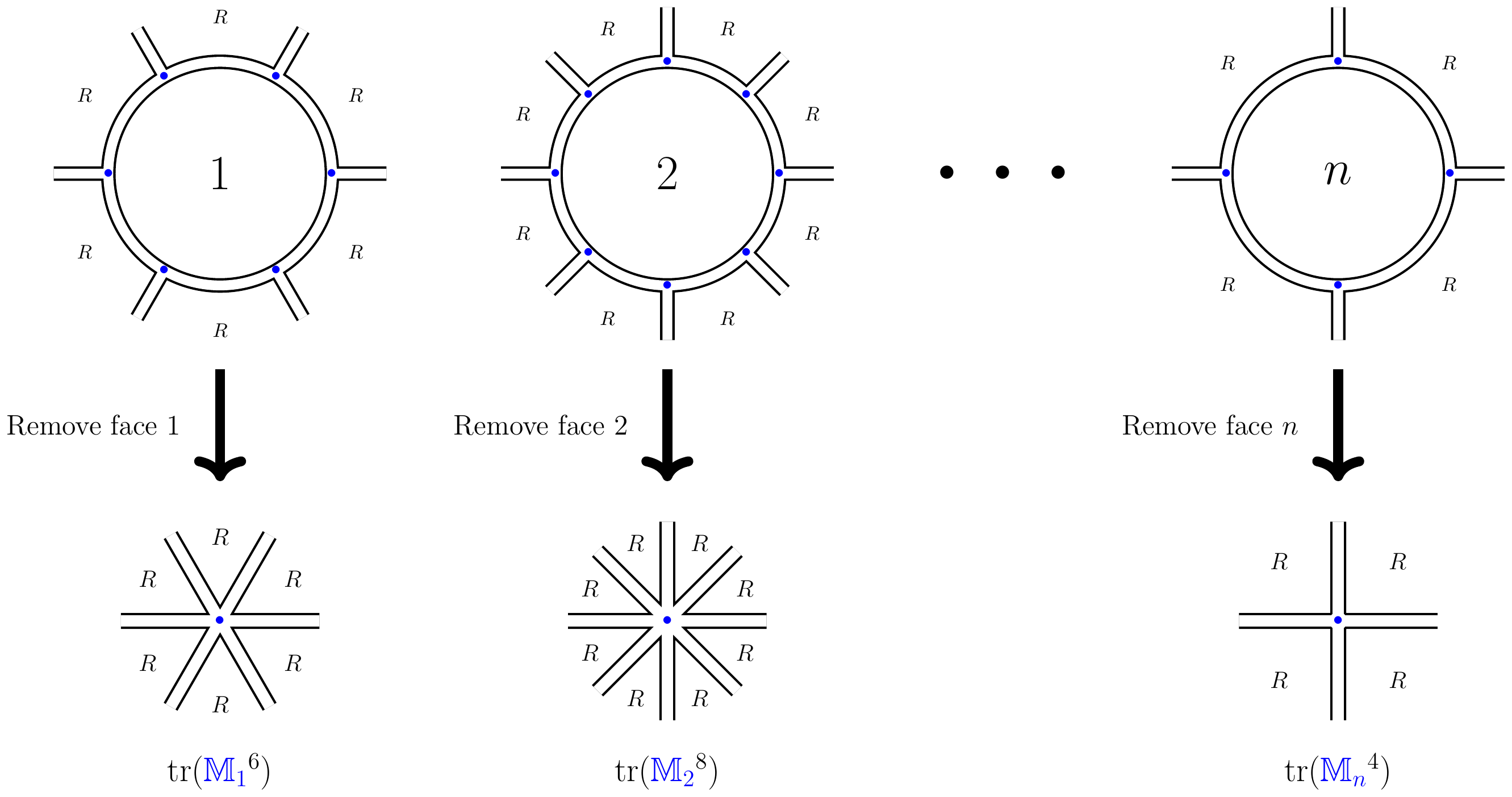}}
\caption{Integrating-out all faces except the reservoir $R$.}
\label{fig:ExtRemoveFace}
\end{figure}

We can simplify the correlators $\langle \cdots\rangle_{(n+1)\text{
faces}}$ in~\eqref{eq:ExtApartition} by performing integrating-in and
-out operations. First, we integrate-in to obtain three-valent vertices
where the faces lying between their edges are $(RRa)$ as shown in
figure \figref{fig:ExtSplit}.
Then, we can arrange all vertices of the form $(RRa)$ around the face
$a$, for all $a=1,\cdots,n$, as shown in \figref{fig:ExtRemoveFace}.

Finally, by removing these $n$ reference faces, we obtain $n$ effective
vertices, with the only remaining face being the reservoir~$R$. So the
partition $\mathbf{A}^{(E)}_{n}$ is now effectively
computed by a sum of one-face correlators:
\beq\label{eq:ExtAsimplified}
\mathbf{A}^{(\text{E})}_{n}(\alpha_{a,b})=\sum_{g=0}^{\infty}\,\sum_{i_{1}+\cdots + i_{n}=2(n+2g-1)}  \frac{\langle\tr(\mathbb{M}_{1}^{i_{1}})\cdots \tr(\mathbb{M}_{n}^{i_{n}})\rangle_{1\text{ face}}}{i_{1}!\cdots i_{n}!}\,=\,\left\langle\prod_{a=1}^{n}\tr\left(e^{\mathbb{M}_{a}}-\mathbb{I}\right)\right\rangle_{1\text{ face}}\,
\eeq
where all indices $i_{m}>0$, and the couplings appear in the Wick contractions as:
\beq
\langle \mathbb{M}_{a}\mathbb{M}_{b}\rangle \,=\,  \alpha_{a,b}
\,.
\eeq
We have computed this partition function~\eqref{eq:ExtAsimplified} for the
first few genera, and found a match against~\eqref{eq:ExtDSL} under the
replacement~\eqref{eq:ExtReplaceA}. We have further recognized that in
terms of couplings $\alpha_{a,b}$, the partition function evaluates to:
\beq
\mathbf{A}^{(\text{E})}_{n}(\alpha_{a,b})=\left\langle\prod_{a=1}^{n}\tr\left(e^{\mathbb{M}_{a}}-\mathbb{I}\right)\right\rangle_{1\text{ face}}\,=\,P_{n}(\alpha_{a,b}) \prod_{a=1}^{n}\frac{\sinh\left(\frac{1}{2}\sum_{b=1}^{n}\alpha_{a,b}\right)}{\frac{1}{2}\sum_{b=1}^{n}\alpha_{a,b}}
\,.
\eeq
The prefactor $P_{n}$ is a non-factorizable homogeneous polynomial of
degree $n-1$ in the couplings $\alpha_{a,b}$, with $a\neq b$, and is
independent of the genus. For $n=3$ and $n=4$, it is simply given by:
\begin{align}
P_{n=3}&= \alpha_{1,2}\,\alpha_{1,3} +\, \alpha_{1,2}\,\alpha_{2,3} + \,\alpha_{1,3}\, \alpha_{2,3}
\,,\\
P_{n=4}
&=\alpha_{1,2}\alpha_{3,4}\brk{\alpha_{1,3}+\alpha_{1,4}+\alpha_{2,3}+\alpha_{2,4}}
+ \alpha_{1,3}\alpha_{2,4}\brk{\alpha_{1,2}+\alpha_{1,4}+\alpha_{2,3}+\alpha_{3,4}}
\nn \\ & \quad
+ \alpha_{1,4}\alpha_{2,3}\brk{\alpha_{1,2}+\alpha_{1,3}+\alpha_{2,3}+\alpha_{2,4}}
+ \alpha_{1,2}\alpha_{1,3}\alpha_{1,4}
+ \alpha_{1,2}\alpha_{2,3}\alpha_{2,4}
\nn \\ & \quad
+ \alpha_{1,3}\alpha_{2,3}\alpha_{3,4}
+ \alpha_{1,4}\alpha_{2,4}\alpha_{3,4}
\,.
\end{align}
We have not found its closed form for generic $n$, although we know it
explicitly up to $n=6$. It is related to the planar contribution of
the DSL of the extremal correlator:
\begin{equation}
\lim_{\Nc\to\infty} E_{n} \overset{\mathrm{DSL}}{=} P_{n}(\alpha_{a,b})\big{|}_{\alpha_{a,b}\to J_{a}J_{b}}= (J_{1}+\cdots + J_{n})^{n-2}\,\,\prod_{i=1}^{n}J_{i}   \,=\, J_{R}^{n-2}\,\prod_{i=1}^{n}J_{i}
\,.
\end{equation}
%

\subsection{$n$-Point Cyclic Correlators In DSL}
\label{sec:bps-part-all-cyclic}

\begin{figure}[ht]
\centering
\resizebox{1\totalheight}{!}{\includegraphics[width=\textwidth]{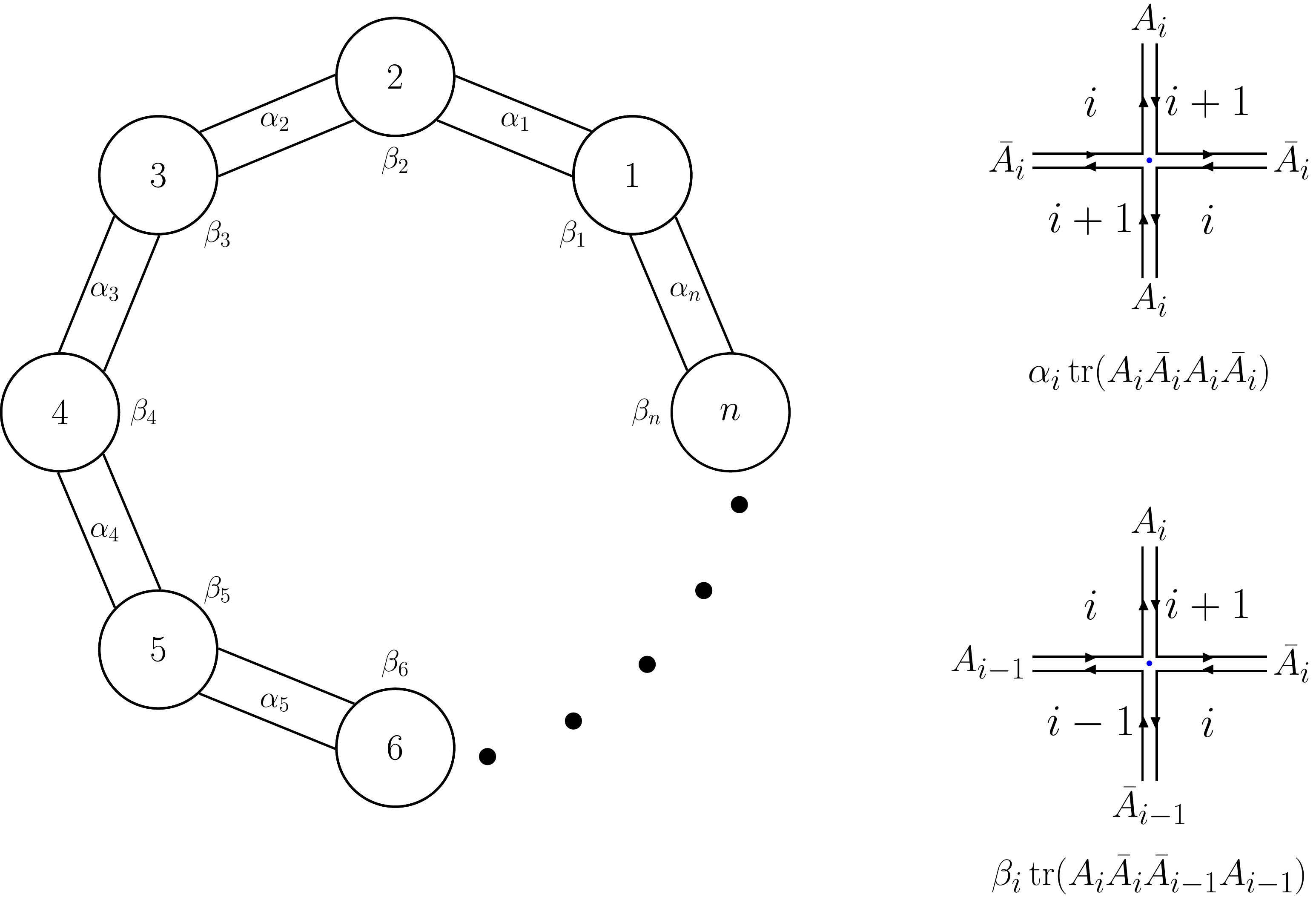}}
\caption{Cyclic correlator and four-valent vertices dual to the BPS squares $[i(i+1)i(i+1)]$ and $[i(i+1)i(i-1)]$.}
\label{fig:n-Cyclic}
\end{figure}

Finally, we consider the $n$-point cyclic correlator shown in
\figref{fig:n-Cyclic}. For $\mathcal{N}=4$ SYM, only the correlators
with $n\leq 6$ are realizable, while for higher $n$ the theory
does not admit enough R-charge polarizations to prevent other connections
that break the cyclic pattern.
Furthermore, as explained in \secref{sec:Conclusions}, see paragraph
before \eqref{eq:Euler5&6}, for $n>4$ only BPS
quadrangulations dominate.

The relevant matrix model to count these
quadrangulations has the action
\beq\label{eq:nCyclicAction}
S^{(\text{C})}_{n}=-\sum_{i=1}^{n}\tr(A_{i}\bar{A}_{i})\,+\,\sum_{i=1}^{n}\frac{\alpha_{i}}{2} \tr(A_{i}\,\bar{A}_{i}A_{i}\,\bar{A}_{i}) +\beta_{i}\,\tr(A_{i}\bar{A}_{i}\bar{A}_{i-1}A_{i-1})
\eeq
Based on direct computations of the relevant correlators of
four-valent vertices up to genus two, we predict the generalization
of~\eqref{eq:BPSlimitA} is:
\beq
\mathbf{A}^{(\text{C})}_{n}(\alpha_{i},\beta_{i}) =  \prod_{a=1}^{n}\beta_{i} \frac{\sinh\left(\frac{1}{2}\left(\beta_{i}+\alpha_{i}+\beta_{i+1}\right)\right)}{\frac{1}{2}\left(\beta_{i}+\alpha_{i}+\beta_{i+1}\right)}
\,,
\eeq
where $\mathbf{A}^{(\text{C})}_{n}$ is defined analogously
to~\eqref{eq:Apartition}, with the four-valent vertices
in~\eqref{eq:nCyclicAction}, and demanding now $n$ faces $(\langle
\cdots \rangle_{n\text{ faces}})$. The corresponding cyclic correlator
in the DSL is obtained by introducing the bridge lengths $k_{i}$ with
the replacement $\alpha_{i}\to {k_{i}^{2}}/{\Nc^{2}}$ and
$\beta_{i}\to {k_{i-1} k_{i}}/{\Nc}$ with $k_{0}\equiv k_{n}$.

%

\pdfbookmark[1]{\refname}{references}
\bibliographystyle{nb}
\bibliography{references}

\end{document}